\documentclass[longauth]{aaEC}

\usepackage{graphicx}
\usepackage{natbib}
\usepackage{scalerel}
\usepackage{comment}
\usepackage{float}
\usepackage{makecell}
\usepackage{booktabs}
\usepackage{lastpage}

\usepackage[table]{xcolor}

\usepackage{txfonts}

\usepackage[pdfencoding=auto,psdextra]{hyperref}
\hypersetup{
    colorlinks=true,
    linkcolor=blue,
    filecolor=magenta,      
    urlcolor=blue,
    citecolor=blue
}
\makeatletter
\renewcommand*\aa@pageof{, page \thepage{} of \pageref*{LastPage}}
\makeatother

\usepackage[utf8]{inputenc}

\usepackage[switch, modulo]{lineno}

\usepackage{euclid}

\begin{document}

\title{\Euclid: the first statistical census of dusty and massive objects in the ERO/Perseus field\thanks{This paper is published on behalf of the Euclid Consortium.}}    

\newcommand{\orcid}[1]{} 	   
\author{G.~Girardi\orcid{0009-0005-6156-4066}\thanks{\email{giorgia.girardi.1@phd.unipd.it}}\inst{\ref{aff1},\ref{aff2}}
\and A.~Grazian\orcid{0000-0002-5688-0663}\inst{\ref{aff2}}
\and G.~Rodighiero\orcid{0000-0002-9415-2296}\inst{\ref{aff1},\ref{aff2}}
\and L.~Bisigello\orcid{0000-0003-0492-4924}\inst{\ref{aff2}}
\and G.~Gandolfi\orcid{0000-0003-3248-5666}\inst{\ref{aff3},\ref{aff2}}
\and E.~Ba\~nados\orcid{0000-0002-2931-7824}\inst{\ref{aff4}}
\and S.~Belladitta\orcid{0000-0003-4747-4484}\inst{\ref{aff4},\ref{aff5}}
\and J.~R.~Weaver\orcid{0000-0003-1614-196X}\inst{\ref{aff6}}
\and S.~Eales\inst{\ref{aff7}}
\and C.~C.~Lovell\orcid{0000-0001-7964-5933}\inst{\ref{aff8}}
\and K.~I.~Caputi\orcid{0000-0001-8183-1460}\inst{\ref{aff9},\ref{aff10}}
\and A.~Enia\orcid{0000-0002-0200-2857}\inst{\ref{aff11},\ref{aff5}}
\and A.~Bianchetti\orcid{0009-0002-8916-3430}\inst{\ref{aff1},\ref{aff2}}
\and E.~Dalla~Bont\`a\orcid{0000-0001-9931-8681}\inst{\ref{aff2},\ref{aff12},\ref{aff3}}
\and T.~Saifollahi\orcid{0000-0002-9554-7660}\inst{\ref{aff13}}
\and A.~Vietri\orcid{0000-0003-4032-0853}\inst{\ref{aff1}}
\and N.~Aghanim\orcid{0000-0002-6688-8992}\inst{\ref{aff14}}
\and B.~Altieri\orcid{0000-0003-3936-0284}\inst{\ref{aff15}}
\and S.~Andreon\orcid{0000-0002-2041-8784}\inst{\ref{aff16}}
\and N.~Auricchio\orcid{0000-0003-4444-8651}\inst{\ref{aff5}}
\and H.~Aussel\orcid{0000-0002-1371-5705}\inst{\ref{aff17}}
\and C.~Baccigalupi\orcid{0000-0002-8211-1630}\inst{\ref{aff18},\ref{aff19},\ref{aff20},\ref{aff21}}
\and M.~Baldi\orcid{0000-0003-4145-1943}\inst{\ref{aff11},\ref{aff5},\ref{aff22}}
\and A.~Balestra\orcid{0000-0002-6967-261X}\inst{\ref{aff2}}
\and S.~Bardelli\orcid{0000-0002-8900-0298}\inst{\ref{aff5}}
\and P.~Battaglia\orcid{0000-0002-7337-5909}\inst{\ref{aff5}}
\and A.~Biviano\orcid{0000-0002-0857-0732}\inst{\ref{aff19},\ref{aff18}}
\and E.~Branchini\orcid{0000-0002-0808-6908}\inst{\ref{aff23},\ref{aff24},\ref{aff16}}
\and M.~Brescia\orcid{0000-0001-9506-5680}\inst{\ref{aff25},\ref{aff26}}
\and J.~Brinchmann\orcid{0000-0003-4359-8797}\inst{\ref{aff27},\ref{aff28}}
\and S.~Camera\orcid{0000-0003-3399-3574}\inst{\ref{aff29},\ref{aff30},\ref{aff31}}
\and G.~Ca\~nas-Herrera\orcid{0000-0003-2796-2149}\inst{\ref{aff32},\ref{aff33},\ref{aff34}}
\and V.~Capobianco\orcid{0000-0002-3309-7692}\inst{\ref{aff31}}
\and C.~Carbone\orcid{0000-0003-0125-3563}\inst{\ref{aff35}}
\and J.~Carretero\orcid{0000-0002-3130-0204}\inst{\ref{aff36},\ref{aff37}}
\and S.~Casas\orcid{0000-0002-4751-5138}\inst{\ref{aff38}}
\and M.~Castellano\orcid{0000-0001-9875-8263}\inst{\ref{aff39}}
\and G.~Castignani\orcid{0000-0001-6831-0687}\inst{\ref{aff5}}
\and S.~Cavuoti\orcid{0000-0002-3787-4196}\inst{\ref{aff26},\ref{aff40}}
\and K.~C.~Chambers\orcid{0000-0001-6965-7789}\inst{\ref{aff41}}
\and A.~Cimatti\inst{\ref{aff42}}
\and C.~Colodro-Conde\inst{\ref{aff43}}
\and G.~Congedo\orcid{0000-0003-2508-0046}\inst{\ref{aff44}}
\and C.~J.~Conselice\orcid{0000-0003-1949-7638}\inst{\ref{aff45}}
\and L.~Conversi\orcid{0000-0002-6710-8476}\inst{\ref{aff46},\ref{aff15}}
\and Y.~Copin\orcid{0000-0002-5317-7518}\inst{\ref{aff47}}
\and F.~Courbin\orcid{0000-0003-0758-6510}\inst{\ref{aff48},\ref{aff49}}
\and H.~M.~Courtois\orcid{0000-0003-0509-1776}\inst{\ref{aff50}}
\and M.~Cropper\orcid{0000-0003-4571-9468}\inst{\ref{aff51}}
\and A.~Da~Silva\orcid{0000-0002-6385-1609}\inst{\ref{aff52},\ref{aff53}}
\and H.~Degaudenzi\orcid{0000-0002-5887-6799}\inst{\ref{aff54}}
\and G.~De~Lucia\orcid{0000-0002-6220-9104}\inst{\ref{aff19}}
\and A.~M.~Di~Giorgio\orcid{0000-0002-4767-2360}\inst{\ref{aff55}}
\and H.~Dole\orcid{0000-0002-9767-3839}\inst{\ref{aff14}}
\and M.~Douspis\orcid{0000-0003-4203-3954}\inst{\ref{aff14}}
\and F.~Dubath\orcid{0000-0002-6533-2810}\inst{\ref{aff54}}
\and C.~A.~J.~Duncan\orcid{0009-0003-3573-0791}\inst{\ref{aff44},\ref{aff45}}
\and X.~Dupac\inst{\ref{aff15}}
\and S.~Dusini\orcid{0000-0002-1128-0664}\inst{\ref{aff56}}
\and S.~Escoffier\orcid{0000-0002-2847-7498}\inst{\ref{aff57}}
\and M.~Farina\orcid{0000-0002-3089-7846}\inst{\ref{aff55}}
\and R.~Farinelli\inst{\ref{aff5}}
\and F.~Faustini\orcid{0000-0001-6274-5145}\inst{\ref{aff39},\ref{aff58}}
\and S.~Ferriol\inst{\ref{aff47}}
\and S.~Fotopoulou\orcid{0000-0002-9686-254X}\inst{\ref{aff59}}
\and M.~Frailis\orcid{0000-0002-7400-2135}\inst{\ref{aff19}}
\and E.~Franceschi\orcid{0000-0002-0585-6591}\inst{\ref{aff5}}
\and M.~Fumana\orcid{0000-0001-6787-5950}\inst{\ref{aff35}}
\and S.~Galeotta\orcid{0000-0002-3748-5115}\inst{\ref{aff19}}
\and K.~George\orcid{0000-0002-1734-8455}\inst{\ref{aff60}}
\and B.~Gillis\orcid{0000-0002-4478-1270}\inst{\ref{aff44}}
\and C.~Giocoli\orcid{0000-0002-9590-7961}\inst{\ref{aff5},\ref{aff22}}
\and J.~Gracia-Carpio\inst{\ref{aff61}}
\and F.~Grupp\inst{\ref{aff61},\ref{aff60}}
\and S.~V.~H.~Haugan\orcid{0000-0001-9648-7260}\inst{\ref{aff62}}
\and J.~Hoar\inst{\ref{aff15}}
\and W.~Holmes\inst{\ref{aff63}}
\and I.~M.~Hook\orcid{0000-0002-2960-978X}\inst{\ref{aff64}}
\and F.~Hormuth\inst{\ref{aff65}}
\and A.~Hornstrup\orcid{0000-0002-3363-0936}\inst{\ref{aff66},\ref{aff67}}
\and P.~Hudelot\inst{\ref{aff68}}
\and K.~Jahnke\orcid{0000-0003-3804-2137}\inst{\ref{aff4}}
\and M.~Jhabvala\inst{\ref{aff69}}
\and E.~Keih\"anen\orcid{0000-0003-1804-7715}\inst{\ref{aff70}}
\and S.~Kermiche\orcid{0000-0002-0302-5735}\inst{\ref{aff57}}
\and A.~Kiessling\orcid{0000-0002-2590-1273}\inst{\ref{aff63}}
\and B.~Kubik\orcid{0009-0006-5823-4880}\inst{\ref{aff47}}
\and M.~K\"ummel\orcid{0000-0003-2791-2117}\inst{\ref{aff60}}
\and M.~Kunz\orcid{0000-0002-3052-7394}\inst{\ref{aff71}}
\and H.~Kurki-Suonio\orcid{0000-0002-4618-3063}\inst{\ref{aff72},\ref{aff73}}
\and A.~M.~C.~Le~Brun\orcid{0000-0002-0936-4594}\inst{\ref{aff74}}
\and D.~Le~Mignant\orcid{0000-0002-5339-5515}\inst{\ref{aff75}}
\and P.~Liebing\inst{\ref{aff51}}
\and S.~Ligori\orcid{0000-0003-4172-4606}\inst{\ref{aff31}}
\and P.~B.~Lilje\orcid{0000-0003-4324-7794}\inst{\ref{aff62}}
\and V.~Lindholm\orcid{0000-0003-2317-5471}\inst{\ref{aff72},\ref{aff73}}
\and I.~Lloro\orcid{0000-0001-5966-1434}\inst{\ref{aff76}}
\and G.~Mainetti\orcid{0000-0003-2384-2377}\inst{\ref{aff77}}
\and D.~Maino\inst{\ref{aff78},\ref{aff35},\ref{aff79}}
\and E.~Maiorano\orcid{0000-0003-2593-4355}\inst{\ref{aff5}}
\and O.~Mansutti\orcid{0000-0001-5758-4658}\inst{\ref{aff19}}
\and S.~Marcin\inst{\ref{aff80}}
\and O.~Marggraf\orcid{0000-0001-7242-3852}\inst{\ref{aff81}}
\and M.~Martinelli\orcid{0000-0002-6943-7732}\inst{\ref{aff39},\ref{aff82}}
\and N.~Martinet\orcid{0000-0003-2786-7790}\inst{\ref{aff75}}
\and F.~Marulli\orcid{0000-0002-8850-0303}\inst{\ref{aff83},\ref{aff5},\ref{aff22}}
\and R.~Massey\orcid{0000-0002-6085-3780}\inst{\ref{aff84}}
\and S.~Maurogordato\inst{\ref{aff85}}
\and E.~Medinaceli\orcid{0000-0002-4040-7783}\inst{\ref{aff5}}
\and S.~Mei\orcid{0000-0002-2849-559X}\inst{\ref{aff86},\ref{aff87}}
\and Y.~Mellier\inst{\ref{aff88},\ref{aff68}}
\and M.~Meneghetti\orcid{0000-0003-1225-7084}\inst{\ref{aff5},\ref{aff22}}
\and E.~Merlin\orcid{0000-0001-6870-8900}\inst{\ref{aff39}}
\and G.~Meylan\inst{\ref{aff89}}
\and A.~Mora\orcid{0000-0002-1922-8529}\inst{\ref{aff90}}
\and M.~Moresco\orcid{0000-0002-7616-7136}\inst{\ref{aff83},\ref{aff5}}
\and L.~Moscardini\orcid{0000-0002-3473-6716}\inst{\ref{aff83},\ref{aff5},\ref{aff22}}
\and R.~Nakajima\orcid{0009-0009-1213-7040}\inst{\ref{aff81}}
\and C.~Neissner\orcid{0000-0001-8524-4968}\inst{\ref{aff91},\ref{aff37}}
\and R.~C.~Nichol\orcid{0000-0003-0939-6518}\inst{\ref{aff92}}
\and S.-M.~Niemi\orcid{0009-0005-0247-0086}\inst{\ref{aff32}}
\and C.~Padilla\orcid{0000-0001-7951-0166}\inst{\ref{aff91}}
\and S.~Paltani\orcid{0000-0002-8108-9179}\inst{\ref{aff54}}
\and F.~Pasian\orcid{0000-0002-4869-3227}\inst{\ref{aff19}}
\and K.~Pedersen\inst{\ref{aff93}}
\and W.~J.~Percival\orcid{0000-0002-0644-5727}\inst{\ref{aff94},\ref{aff95},\ref{aff96}}
\and V.~Pettorino\inst{\ref{aff32}}
\and G.~Polenta\orcid{0000-0003-4067-9196}\inst{\ref{aff58}}
\and M.~Poncet\inst{\ref{aff97}}
\and L.~A.~Popa\inst{\ref{aff98}}
\and L.~Pozzetti\orcid{0000-0001-7085-0412}\inst{\ref{aff5}}
\and F.~Raison\orcid{0000-0002-7819-6918}\inst{\ref{aff61}}
\and R.~Rebolo\orcid{0000-0003-3767-7085}\inst{\ref{aff43},\ref{aff99},\ref{aff100}}
\and A.~Renzi\orcid{0000-0001-9856-1970}\inst{\ref{aff1},\ref{aff56}}
\and J.~Rhodes\orcid{0000-0002-4485-8549}\inst{\ref{aff63}}
\and G.~Riccio\inst{\ref{aff26}}
\and E.~Romelli\orcid{0000-0003-3069-9222}\inst{\ref{aff19}}
\and M.~Roncarelli\orcid{0000-0001-9587-7822}\inst{\ref{aff5}}
\and E.~Rossetti\orcid{0000-0003-0238-4047}\inst{\ref{aff11}}
\and B.~Rusholme\orcid{0000-0001-7648-4142}\inst{\ref{aff101}}
\and R.~Saglia\orcid{0000-0003-0378-7032}\inst{\ref{aff60},\ref{aff61}}
\and Z.~Sakr\orcid{0000-0002-4823-3757}\inst{\ref{aff102},\ref{aff103},\ref{aff104}}
\and D.~Sapone\orcid{0000-0001-7089-4503}\inst{\ref{aff105}}
\and B.~Sartoris\orcid{0000-0003-1337-5269}\inst{\ref{aff60},\ref{aff19}}
\and J.~A.~Schewtschenko\orcid{0000-0002-4913-6393}\inst{\ref{aff44}}
\and P.~Schneider\orcid{0000-0001-8561-2679}\inst{\ref{aff81}}
\and T.~Schrabback\orcid{0000-0002-6987-7834}\inst{\ref{aff106}}
\and A.~Secroun\orcid{0000-0003-0505-3710}\inst{\ref{aff57}}
\and G.~Seidel\orcid{0000-0003-2907-353X}\inst{\ref{aff4}}
\and M.~Seiffert\orcid{0000-0002-7536-9393}\inst{\ref{aff63}}
\and S.~Serrano\orcid{0000-0002-0211-2861}\inst{\ref{aff107},\ref{aff108},\ref{aff109}}
\and P.~Simon\inst{\ref{aff81}}
\and C.~Sirignano\orcid{0000-0002-0995-7146}\inst{\ref{aff1},\ref{aff56}}
\and G.~Sirri\orcid{0000-0003-2626-2853}\inst{\ref{aff22}}
\and L.~Stanco\orcid{0000-0002-9706-5104}\inst{\ref{aff56}}
\and J.~Steinwagner\orcid{0000-0001-7443-1047}\inst{\ref{aff61}}
\and P.~Tallada-Cresp\'{i}\orcid{0000-0002-1336-8328}\inst{\ref{aff36},\ref{aff37}}
\and D.~Tavagnacco\orcid{0000-0001-7475-9894}\inst{\ref{aff19}}
\and A.~N.~Taylor\inst{\ref{aff44}}
\and I.~Tereno\orcid{0000-0002-4537-6218}\inst{\ref{aff52},\ref{aff110}}
\and R.~Toledo-Moreo\orcid{0000-0002-2997-4859}\inst{\ref{aff111}}
\and F.~Torradeflot\orcid{0000-0003-1160-1517}\inst{\ref{aff37},\ref{aff36}}
\and I.~Tutusaus\orcid{0000-0002-3199-0399}\inst{\ref{aff103}}
\and L.~Valenziano\orcid{0000-0002-1170-0104}\inst{\ref{aff5},\ref{aff112}}
\and J.~Valiviita\orcid{0000-0001-6225-3693}\inst{\ref{aff72},\ref{aff73}}
\and T.~Vassallo\orcid{0000-0001-6512-6358}\inst{\ref{aff60},\ref{aff19}}
\and G.~Verdoes~Kleijn\orcid{0000-0001-5803-2580}\inst{\ref{aff9}}
\and A.~Veropalumbo\orcid{0000-0003-2387-1194}\inst{\ref{aff16},\ref{aff24},\ref{aff23}}
\and Y.~Wang\orcid{0000-0002-4749-2984}\inst{\ref{aff113}}
\and J.~Weller\orcid{0000-0002-8282-2010}\inst{\ref{aff60},\ref{aff61}}
\and G.~Zamorani\orcid{0000-0002-2318-301X}\inst{\ref{aff5}}
\and F.~M.~Zerbi\inst{\ref{aff16}}
\and E.~Zucca\orcid{0000-0002-5845-8132}\inst{\ref{aff5}}
\and M.~Bolzonella\orcid{0000-0003-3278-4607}\inst{\ref{aff5}}
\and C.~Burigana\orcid{0000-0002-3005-5796}\inst{\ref{aff114},\ref{aff112}}
\and L.~Gabarra\orcid{0000-0002-8486-8856}\inst{\ref{aff115}}
\and J.~Mart\'{i}n-Fleitas\orcid{0000-0002-8594-569X}\inst{\ref{aff90}}
\and V.~Scottez\inst{\ref{aff88},\ref{aff116}}}
										   
\institute{Dipartimento di Fisica e Astronomia "G. Galilei", Universit\`a di Padova, Via Marzolo 8, 35131 Padova, Italy\label{aff1}
\and
INAF-Osservatorio Astronomico di Padova, Via dell'Osservatorio 5, 35122 Padova, Italy\label{aff2}
\and
Dipartimento di Fisica e Astronomia ``G. Galilei", Universit\`a di Padova, Vicolo dell'Osservatorio 3, 35122 Padova, Italy\label{aff3}
\and
Max-Planck-Institut f\"ur Astronomie, K\"onigstuhl 17, 69117 Heidelberg, Germany\label{aff4}
\and
INAF-Osservatorio di Astrofisica e Scienza dello Spazio di Bologna, Via Piero Gobetti 93/3, 40129 Bologna, Italy\label{aff5}
\and
Department of Astronomy, University of Massachusetts, Amherst, MA 01003, USA\label{aff6}
\and
School of Physics and Astronomy, Cardiff University, The Parade, Cardiff, CF24 3AA, UK\label{aff7}
\and
Institute of Cosmology and Gravitation, University of Portsmouth, Portsmouth PO1 3FX, UK\label{aff8}
\and
Kapteyn Astronomical Institute, University of Groningen, PO Box 800, 9700 AV Groningen, The Netherlands\label{aff9}
\and
Cosmic Dawn Center (DAWN)\label{aff10}
\and
Dipartimento di Fisica e Astronomia, Universit\`a di Bologna, Via Gobetti 93/2, 40129 Bologna, Italy\label{aff11}
\and
Jeremiah Horrocks Institute, University of Central Lancashire, Preston, PR1 2HE, UK\label{aff12}
\and
Universit\'e de Strasbourg, CNRS, Observatoire astronomique de Strasbourg, UMR 7550, 67000 Strasbourg, France\label{aff13}
\and
Universit\'e Paris-Saclay, CNRS, Institut d'astrophysique spatiale, 91405, Orsay, France\label{aff14}
\and
ESAC/ESA, Camino Bajo del Castillo, s/n., Urb. Villafranca del Castillo, 28692 Villanueva de la Ca\~nada, Madrid, Spain\label{aff15}
\and
INAF-Osservatorio Astronomico di Brera, Via Brera 28, 20122 Milano, Italy\label{aff16}
\and
Universit\'e Paris-Saclay, Universit\'e Paris Cit\'e, CEA, CNRS, AIM, 91191, Gif-sur-Yvette, France\label{aff17}
\and
IFPU, Institute for Fundamental Physics of the Universe, via Beirut 2, 34151 Trieste, Italy\label{aff18}
\and
INAF-Osservatorio Astronomico di Trieste, Via G. B. Tiepolo 11, 34143 Trieste, Italy\label{aff19}
\and
INFN, Sezione di Trieste, Via Valerio 2, 34127 Trieste TS, Italy\label{aff20}
\and
SISSA, International School for Advanced Studies, Via Bonomea 265, 34136 Trieste TS, Italy\label{aff21}
\and
INFN-Sezione di Bologna, Viale Berti Pichat 6/2, 40127 Bologna, Italy\label{aff22}
\and
Dipartimento di Fisica, Universit\`a di Genova, Via Dodecaneso 33, 16146, Genova, Italy\label{aff23}
\and
INFN-Sezione di Genova, Via Dodecaneso 33, 16146, Genova, Italy\label{aff24}
\and
Department of Physics "E. Pancini", University Federico II, Via Cinthia 6, 80126, Napoli, Italy\label{aff25}
\and
INAF-Osservatorio Astronomico di Capodimonte, Via Moiariello 16, 80131 Napoli, Italy\label{aff26}
\and
Instituto de Astrof\'isica e Ci\^encias do Espa\c{c}o, Universidade do Porto, CAUP, Rua das Estrelas, PT4150-762 Porto, Portugal\label{aff27}
\and
Faculdade de Ci\^encias da Universidade do Porto, Rua do Campo de Alegre, 4150-007 Porto, Portugal\label{aff28}
\and
Dipartimento di Fisica, Universit\`a degli Studi di Torino, Via P. Giuria 1, 10125 Torino, Italy\label{aff29}
\and
INFN-Sezione di Torino, Via P. Giuria 1, 10125 Torino, Italy\label{aff30}
\and
INAF-Osservatorio Astrofisico di Torino, Via Osservatorio 20, 10025 Pino Torinese (TO), Italy\label{aff31}
\and
European Space Agency/ESTEC, Keplerlaan 1, 2201 AZ Noordwijk, The Netherlands\label{aff32}
\and
Institute Lorentz, Leiden University, Niels Bohrweg 2, 2333 CA Leiden, The Netherlands\label{aff33}
\and
Leiden Observatory, Leiden University, Einsteinweg 55, 2333 CC Leiden, The Netherlands\label{aff34}
\and
INAF-IASF Milano, Via Alfonso Corti 12, 20133 Milano, Italy\label{aff35}
\and
Centro de Investigaciones Energ\'eticas, Medioambientales y Tecnol\'ogicas (CIEMAT), Avenida Complutense 40, 28040 Madrid, Spain\label{aff36}
\and
Port d'Informaci\'{o} Cient\'{i}fica, Campus UAB, C. Albareda s/n, 08193 Bellaterra (Barcelona), Spain\label{aff37}
\and
Institute for Theoretical Particle Physics and Cosmology (TTK), RWTH Aachen University, 52056 Aachen, Germany\label{aff38}
\and
INAF-Osservatorio Astronomico di Roma, Via Frascati 33, 00078 Monteporzio Catone, Italy\label{aff39}
\and
INFN section of Naples, Via Cinthia 6, 80126, Napoli, Italy\label{aff40}
\and
Institute for Astronomy, University of Hawaii, 2680 Woodlawn Drive, Honolulu, HI 96822, USA\label{aff41}
\and
Dipartimento di Fisica e Astronomia "Augusto Righi" - Alma Mater Studiorum Universit\`a di Bologna, Viale Berti Pichat 6/2, 40127 Bologna, Italy\label{aff42}
\and
Instituto de Astrof\'{\i}sica de Canarias, V\'{\i}a L\'actea, 38205 La Laguna, Tenerife, Spain\label{aff43}
\and
Institute for Astronomy, University of Edinburgh, Royal Observatory, Blackford Hill, Edinburgh EH9 3HJ, UK\label{aff44}
\and
Jodrell Bank Centre for Astrophysics, Department of Physics and Astronomy, University of Manchester, Oxford Road, Manchester M13 9PL, UK\label{aff45}
\and
European Space Agency/ESRIN, Largo Galileo Galilei 1, 00044 Frascati, Roma, Italy\label{aff46}
\and
Universit\'e Claude Bernard Lyon 1, CNRS/IN2P3, IP2I Lyon, UMR 5822, Villeurbanne, F-69100, France\label{aff47}
\and
Institut de Ci\`{e}ncies del Cosmos (ICCUB), Universitat de Barcelona (IEEC-UB), Mart\'{i} i Franqu\`{e}s 1, 08028 Barcelona, Spain\label{aff48}
\and
Instituci\'o Catalana de Recerca i Estudis Avan\c{c}ats (ICREA), Passeig de Llu\'{\i}s Companys 23, 08010 Barcelona, Spain\label{aff49}
\and
UCB Lyon 1, CNRS/IN2P3, IUF, IP2I Lyon, 4 rue Enrico Fermi, 69622 Villeurbanne, France\label{aff50}
\and
Mullard Space Science Laboratory, University College London, Holmbury St Mary, Dorking, Surrey RH5 6NT, UK\label{aff51}
\and
Departamento de F\'isica, Faculdade de Ci\^encias, Universidade de Lisboa, Edif\'icio C8, Campo Grande, PT1749-016 Lisboa, Portugal\label{aff52}
\and
Instituto de Astrof\'isica e Ci\^encias do Espa\c{c}o, Faculdade de Ci\^encias, Universidade de Lisboa, Campo Grande, 1749-016 Lisboa, Portugal\label{aff53}
\and
Department of Astronomy, University of Geneva, ch. d'Ecogia 16, 1290 Versoix, Switzerland\label{aff54}
\and
INAF-Istituto di Astrofisica e Planetologia Spaziali, via del Fosso del Cavaliere, 100, 00100 Roma, Italy\label{aff55}
\and
INFN-Padova, Via Marzolo 8, 35131 Padova, Italy\label{aff56}
\and
Aix-Marseille Universit\'e, CNRS/IN2P3, CPPM, Marseille, France\label{aff57}
\and
Space Science Data Center, Italian Space Agency, via del Politecnico snc, 00133 Roma, Italy\label{aff58}
\and
School of Physics, HH Wills Physics Laboratory, University of Bristol, Tyndall Avenue, Bristol, BS8 1TL, UK\label{aff59}
\and
Universit\"ats-Sternwarte M\"unchen, Fakult\"at f\"ur Physik, Ludwig-Maximilians-Universit\"at M\"unchen, Scheinerstrasse 1, 81679 M\"unchen, Germany\label{aff60}
\and
Max Planck Institute for Extraterrestrial Physics, Giessenbachstr. 1, 85748 Garching, Germany\label{aff61}
\and
Institute of Theoretical Astrophysics, University of Oslo, P.O. Box 1029 Blindern, 0315 Oslo, Norway\label{aff62}
\and
Jet Propulsion Laboratory, California Institute of Technology, 4800 Oak Grove Drive, Pasadena, CA, 91109, USA\label{aff63}
\and
Department of Physics, Lancaster University, Lancaster, LA1 4YB, UK\label{aff64}
\and
Felix Hormuth Engineering, Goethestr. 17, 69181 Leimen, Germany\label{aff65}
\and
Technical University of Denmark, Elektrovej 327, 2800 Kgs. Lyngby, Denmark\label{aff66}
\and
Cosmic Dawn Center (DAWN), Denmark\label{aff67}
\and
Institut d'Astrophysique de Paris, UMR 7095, CNRS, and Sorbonne Universit\'e, 98 bis boulevard Arago, 75014 Paris, France\label{aff68}
\and
NASA Goddard Space Flight Center, Greenbelt, MD 20771, USA\label{aff69}
\and
Department of Physics and Helsinki Institute of Physics, Gustaf H\"allstr\"omin katu 2, 00014 University of Helsinki, Finland\label{aff70}
\and
Universit\'e de Gen\`eve, D\'epartement de Physique Th\'eorique and Centre for Astroparticle Physics, 24 quai Ernest-Ansermet, CH-1211 Gen\`eve 4, Switzerland\label{aff71}
\and
Department of Physics, P.O. Box 64, 00014 University of Helsinki, Finland\label{aff72}
\and
Helsinki Institute of Physics, Gustaf H{\"a}llstr{\"o}min katu 2, University of Helsinki, Helsinki, Finland\label{aff73}
\and
Laboratoire d'etude de l'Univers et des phenomenes eXtremes, Observatoire de Paris, Universit\'e PSL, Sorbonne Universit\'e, CNRS, 92190 Meudon, France\label{aff74}
\and
Aix-Marseille Universit\'e, CNRS, CNES, LAM, Marseille, France\label{aff75}
\and
SKA Observatory, Jodrell Bank, Lower Withington, Macclesfield, Cheshire SK11 9FT, UK\label{aff76}
\and
Centre de Calcul de l'IN2P3/CNRS, 21 avenue Pierre de Coubertin 69627 Villeurbanne Cedex, France\label{aff77}
\and
Dipartimento di Fisica "Aldo Pontremoli", Universit\`a degli Studi di Milano, Via Celoria 16, 20133 Milano, Italy\label{aff78}
\and
INFN-Sezione di Milano, Via Celoria 16, 20133 Milano, Italy\label{aff79}
\and
University of Applied Sciences and Arts of Northwestern Switzerland, School of Computer Science, 5210 Windisch, Switzerland\label{aff80}
\and
Universit\"at Bonn, Argelander-Institut f\"ur Astronomie, Auf dem H\"ugel 71, 53121 Bonn, Germany\label{aff81}
\and
INFN-Sezione di Roma, Piazzale Aldo Moro, 2 - c/o Dipartimento di Fisica, Edificio G. Marconi, 00185 Roma, Italy\label{aff82}
\and
Dipartimento di Fisica e Astronomia "Augusto Righi" - Alma Mater Studiorum Universit\`a di Bologna, via Piero Gobetti 93/2, 40129 Bologna, Italy\label{aff83}
\and
Department of Physics, Institute for Computational Cosmology, Durham University, South Road, Durham, DH1 3LE, UK\label{aff84}
\and
Universit\'e C\^{o}te d'Azur, Observatoire de la C\^{o}te d'Azur, CNRS, Laboratoire Lagrange, Bd de l'Observatoire, CS 34229, 06304 Nice cedex 4, France\label{aff85}
\and
Universit\'e Paris Cit\'e, CNRS, Astroparticule et Cosmologie, 75013 Paris, France\label{aff86}
\and
CNRS-UCB International Research Laboratory, Centre Pierre Bin\'etruy, IRL2007, CPB-IN2P3, Berkeley, USA\label{aff87}
\and
Institut d'Astrophysique de Paris, 98bis Boulevard Arago, 75014, Paris, France\label{aff88}
\and
Institute of Physics, Laboratory of Astrophysics, Ecole Polytechnique F\'ed\'erale de Lausanne (EPFL), Observatoire de Sauverny, 1290 Versoix, Switzerland\label{aff89}
\and
Aurora Technology for European Space Agency (ESA), Camino bajo del Castillo, s/n, Urbanizacion Villafranca del Castillo, Villanueva de la Ca\~nada, 28692 Madrid, Spain\label{aff90}
\and
Institut de F\'{i}sica d'Altes Energies (IFAE), The Barcelona Institute of Science and Technology, Campus UAB, 08193 Bellaterra (Barcelona), Spain\label{aff91}
\and
School of Mathematics and Physics, University of Surrey, Guildford, Surrey, GU2 7XH, UK\label{aff92}
\and
DARK, Niels Bohr Institute, University of Copenhagen, Jagtvej 155, 2200 Copenhagen, Denmark\label{aff93}
\and
Waterloo Centre for Astrophysics, University of Waterloo, Waterloo, Ontario N2L 3G1, Canada\label{aff94}
\and
Department of Physics and Astronomy, University of Waterloo, Waterloo, Ontario N2L 3G1, Canada\label{aff95}
\and
Perimeter Institute for Theoretical Physics, Waterloo, Ontario N2L 2Y5, Canada\label{aff96}
\and
Centre National d'Etudes Spatiales -- Centre spatial de Toulouse, 18 avenue Edouard Belin, 31401 Toulouse Cedex 9, France\label{aff97}
\and
Institute of Space Science, Str. Atomistilor, nr. 409 M\u{a}gurele, Ilfov, 077125, Romania\label{aff98}
\and
Consejo Superior de Investigaciones Cientificas, Calle Serrano 117, 28006 Madrid, Spain\label{aff99}
\and
Universidad de La Laguna, Departamento de Astrof\'{\i}sica, 38206 La Laguna, Tenerife, Spain\label{aff100}
\and
Caltech/IPAC, 1200 E. California Blvd., Pasadena, CA 91125, USA\label{aff101}
\and
Institut f\"ur Theoretische Physik, University of Heidelberg, Philosophenweg 16, 69120 Heidelberg, Germany\label{aff102}
\and
Institut de Recherche en Astrophysique et Plan\'etologie (IRAP), Universit\'e de Toulouse, CNRS, UPS, CNES, 14 Av. Edouard Belin, 31400 Toulouse, France\label{aff103}
\and
Universit\'e St Joseph; Faculty of Sciences, Beirut, Lebanon\label{aff104}
\and
Departamento de F\'isica, FCFM, Universidad de Chile, Blanco Encalada 2008, Santiago, Chile\label{aff105}
\and
Universit\"at Innsbruck, Institut f\"ur Astro- und Teilchenphysik, Technikerstr. 25/8, 6020 Innsbruck, Austria\label{aff106}
\and
Institut d'Estudis Espacials de Catalunya (IEEC),  Edifici RDIT, Campus UPC, 08860 Castelldefels, Barcelona, Spain\label{aff107}
\and
Satlantis, University Science Park, Sede Bld 48940, Leioa-Bilbao, Spain\label{aff108}
\and
Institute of Space Sciences (ICE, CSIC), Campus UAB, Carrer de Can Magrans, s/n, 08193 Barcelona, Spain\label{aff109}
\and
Instituto de Astrof\'isica e Ci\^encias do Espa\c{c}o, Faculdade de Ci\^encias, Universidade de Lisboa, Tapada da Ajuda, 1349-018 Lisboa, Portugal\label{aff110}
\and
Universidad Polit\'ecnica de Cartagena, Departamento de Electr\'onica y Tecnolog\'ia de Computadoras,  Plaza del Hospital 1, 30202 Cartagena, Spain\label{aff111}
\and
INFN-Bologna, Via Irnerio 46, 40126 Bologna, Italy\label{aff112}
\and
Infrared Processing and Analysis Center, California Institute of Technology, Pasadena, CA 91125, USA\label{aff113}
\and
INAF, Istituto di Radioastronomia, Via Piero Gobetti 101, 40129 Bologna, Italy\label{aff114}
\and
Department of Physics, Oxford University, Keble Road, Oxford OX1 3RH, UK\label{aff115}
\and
ICL, Junia, Universit\'e Catholique de Lille, LITL, 59000 Lille, France\label{aff116}}    

\abstract{Our comprehension of the history of star formation at $z>3$ strongly relies on rest-frame UV observations. However, this selection systematically misses the most dusty and massive sources, resulting in an incomplete census at earlier times. Infrared facilities such as \textit{Spitzer} and the \textit{James Webb} Space Telescope have shed light on a hidden population lying at $z=3$--$6$, characterised by extreme red colours, named HIEROs (HST-to-IRAC extremely red objects), identified by the colour criterion $\HE - \mathrm{ch2} > 2.25$. Recently, \Euclid Early Release Observations (ERO) have opened the possibility to further study such objects, exploiting the comparison between \Euclid and ancillary \textit{Spitzer}/IRAC observations. The aim of this paper is to conduct an investigation on a small area of $232$\,arcmin$^2$ to investigate the effectiveness of this synergy in characterising this population. We utilize catalogues in the Perseus field across the VIS and NISP bands, supplemented by data from the four \textit{Spitzer} channels and several ground-based MegaCam bands (\textit{u}, \textit{g}, \textit{r}, ${\rm H}\,\alpha$, \textit{i}, and \textit{z}) already included in the ERO catalogue. We select $121$ HIEROs by applying the $\HE - \mathrm{ch2} > 2.25$ colour cut, clean this sample of globular clusters and brown dwarfs, and then inspect by eye the multiband cutouts of each source, ending with $42$ reliable HIEROs. Photometric redshifts and other physical properties of the final sample have been estimated using the SED-fitting software \texttt{Bagpipes}. From the $z_{\mathrm{phot}}$ and $M_*$ values, we compute the galaxy stellar mass function at $3.5<z<5.5$. When we exclude all galaxies that could host an AGN, or where the stellar masses might be overestimated, we still find that the high-mass end of the galaxy stellar mass function is similar to previous estimates, indicating that the true value could be even higher. This investigation highlights the importance of a deeper study of this still mysterious population, in particular, to assess its contribution to the cosmic star-formation rate density and its agreement with current galaxy evolution and formation models. Indeed, these early results show \Euclid's capabilities to push the boundaries of our understanding of obscured star formation across a wide range of epochs.}

\titlerunning{Statistical census of dusty and massive objects in the ERO/Perseus field} 
\authorrunning{Girardi et al.}

{\let\clearpage\relax
\maketitle}

\section{\label{sc:Intro}Introduction}

Addressing the process of stellar mass assembly over the age of the Universe is fundamental to our understanding of galaxy formation and evolution, offering critical insights into the processes driving the evolution of structures in the Universe. This topic becomes even more crucial during early epochs, at $z > 3$, where the build-up of massive galaxies remains a subject of active investigation. Resolving this fundamental issue remains a central objective of observational extragalactic astronomy, given its importance in validating or challenging the predictions of galaxy formation models \citep{santini2009star}.

An important technique for finding high-redshift `normal' star-forming galaxies has been to locate the Lyman break between the rest-frame ultraviolet (UV) and optical stellar emission \citep{steidel1993deep, steidel1995lyman, madau1996high, steidel1999lyman}. These `Lyman-break galaxies' are detected because they are much brighter on the long-wavelength side of the break than on the short-wavelength side \citep{bouwens2015uv,oesch2016remarkably}.
However, at $z > 3$ this selection method systematically misses the most dusty massive objects.

Observations with the \Spitzer, the Atacama Large Millimeter/submillimeter Array (ALMA), the \textit{James Webb} Space Telescope (JWST), and radio telescopes have discovered a population of dust-obscured massive galaxies missed by the Lyman-break technique \citep[e.g.,][]{wang2019dominant, gruppioni2020alpine,talia2021illuminating, enia2022new, gomez2023jwst, nelson2023jwst, perez2023ceers, gardner2023james, gentile2024illuminating, gottumukkala2024unveiling}. These galaxies have been called optically-dark galaxies (ODGs) or HST-dark galaxies, due to the fact that they are missed even in the deepest \textit{Hubble} Space Telescope observations. Their spectral energy distributions usually have very red slopes. 

The contribution of this population to the star-formation rate density (SFRD) may be as much as 10 times higher than the contribution of the galaxies found by the Lyman-break technique at $z \approx 4$, and may be even higher at $z=5$--$7$ \citep{wang2019dominant, rodighiero2023jwst, wang2024true, traina2024a3cosmos}. Their high stellar masses \citep{Caputi12, gottumukkala2024unveiling} mean that improved knowledge of this population is important for a better understanding of the evolution of the high-mass end of the galaxy stellar mass function in the early Universe.

\Euclid \citep{Laureijs11, EuclidSkyOverview}, which was launched by the European Space Agency (ESA) in July $2023$ to investigate the evolution of dark matter and energy in the Universe, is also well-suited to search for these galaxies. It is equipped with two instruments: the visible instrument \citep[VIS;][]{EuclidSkyVIS}, able to obtain high-resolution optical imaging in the \IE band, covering a wavelength range between $550$ and $900$\,nm; and the Near-Infrared Spectrometer and Photometer (NISP), which provides spectrophotometric observations in three NIR bands (\YE, \JE, and \HE) at 900--2000\,nm \citep{collaboration2022euclid, EuclidSkyNISP}. The deep \Euclid images of the Perseus cluster obtained during the Early Release Observations \citep[EROs;][]{EROcite, EROData}, coupled with \textit{Spitzer} observations of this cluster, allow us to search for these ODGs. This paper aims to test \Euclid's ability to search for such high-redshift dark galaxies, focusing on a particular type, the HST-to-IRAC extremely red objects, named HIEROs, as defined by \cite{wang2016infrared} using the colour criterion $\HE-\mathrm{ch2} > 2.25$ \citep[see also][for an introduction to this method]{Caputi12}. The goal of this work is to compute the galaxy stellar mass function (GSMF) and evaluate the contribution of these red galaxies to the overall mass function.

The paper is structured as follows. In Sect.~\ref{sc:Perseus} we report a brief overview of the images used in this paper. In Sect.~\ref{sc:Methods} we describe the procedure used to build our sample and to estimate the physical properties of the galaxies. We give the results in Sect.~\ref{sc:Results}, including our estimates of the GSMF. In Sect.~\ref{sc:Discussion}, we discuss the possible effects that could bias our results. Our conclusions are listed in Sect.~\ref{conclusions}. 

Throughout the paper, we assume a $\Lambda$CDM cosmology with parameters from \cite{Planck2016} and a \cite{chabrier2003galactic} initial mass function (IMF). All magnitudes presented in this work are reported in the AB magnitude system.

\section{\label{sc:Perseus} Multi-wavelength observations and photometry of the Perseus cluster}

The Perseus cluster (centred at RA: $\ra{03;19;48.1}$, Dec: $\ang{+41;30;54}$), also known as Abell $426$, is at a distance of 72\,Mpc, corresponding to $z = 0.0167$. It extends over more than 50\,Mpc and belongs to the Perseus-Pisces supercluster \citep{wegner1993survey, aguerri2020deep}. 
It has been observed by numerous X-ray telescopes \citep{lau2017physical, sanders2020measuring}, as well as at other wavelengths. 

\subsection{Euclid Early Release Observations} \label{ERO}
In this work, we have analysed the observations of the cluster that were made during the ERO programme\footnote{Euclid Early Release Observations. 2024, \url{https://doi.org/10.57780/esa-qmocze3}}\citep{EROPerseusOverview, EROPerseusDGs}.

The \IE images have a pixel size of $\ang{;;0 .1}$ and the NISP images have a pixel size of $\ang{;;0 .3}$. In the mosaics made from the images, the full width at half maximum (FWHM) values in the \IE, \YE, \JE, and \HE bands are $\ang{;;0 .16}$, $\ang{;;0 .48}$, $\ang{;;0 .49}$, and $\ang{;;0 .50}$, respectively \citep{EROPerseusOverview}. These FWHM values correspond to 56\,pc for VIS and 170\,pc for NISP at the distance of the Perseus cluster. We also used in our analysis observations of the cluster that were made with MegaCam at the Canada-France-Hawaii Telescope (CFHT) in the \textit{u}, \textit{g}, \textit{r}, \textit{i}, \textit{z}, and ${\rm H}\,\alpha$ ‘off’ filters \citep{EP-Zalesky}. 

We used the catalogues of the sources provided by the Euclid Consortium. Not all the sources detected in the \HE band were detected in the \IE band, given that these catalogues were obtained by running \texttt{SExtractor} \citep[Source Extractor;][]{bertin1996sextractor} separately on the \IE band and a combination of the three NISP images \citep{EROPerseusOverview}. We used the \texttt{MAG\_AUTO} measurements, automatically obtained by \texttt{SExtractor} within a Kron-like elliptical aperture, thus minimising the need for additional corrections. In fact, this measurement estimates the total flux by integrating the pixel values within an aperture that is adaptively scaled according to each source’s light distribution, thereby adjusting to its intrinsic morphology. We corrected for Galactic extinction using the formula $m_{\rm intrinsic} = m_{\rm observed} - C \, E(B-V)$ with $E(B-V)=0.16$ and the values of the parameters given in the \texttt{README} file and in Table~\ref{tab:mag_correction}.
The extinction was derived using the dust extinction curve of \citet{gordon2023one} and \citet{gordon2024dust_extinction}.

\begin{table}[htbp!]
    \centering
    \caption{Values applied to correct for Galactic extinction in the different filters.}

\addtolength{\tabcolsep}{-0.1em}
    \begin{tabular}{lrcc} 
    \hline\hline
    \noalign{\vskip 2pt}
         Filter   & \multicolumn{1}{c}{$\lambda_{\rm mean}$} & \textit{C} & $A(\lambda_{\rm mean})$ \\ 
        & \multicolumn{1}{c}{[\AA]} & &  [mag]\\ 
\noalign{\vskip 2pt}
         \hline
\noalign{\vskip 2pt}
         \texttt{CFHT/MegaCam.u} & $3692.7$ & $4.633$ & $0.7516$\\
         \texttt{CFHT/MegaCam.g} & $4824.0$ & $3.552$ &$0.5764$\\
         \texttt{CFHT/MegaCam.r} & $6425.4$ & $2.516$&$0.4081$\\
         \texttt{CFHT/Megaprime.Halpha} & $6590.6$ & $2.433$&$0.3947$\\
         \texttt{Euclid/VIS.vis} & $7324.8$ & $2.122$&$0.3443$\\
         \texttt{CFHT/MegaCam.i} & $7721.0$ & $1.919$&$0.3113$\\
         \texttt{CFHT/MegaCam.z} & $9004.8$ & $1.487$&$0.2413$\\
         \texttt{Euclid/NISP.Y} & $10\,898.7$ & $1.066$&$0.1730$\\
         \texttt{Euclid/NISP.J} & $13\,796.0$ & $0.726$&$0.1178$\\
         \texttt{Euclid/NISP.H} & $17\,877.9$ &$0.470$&$0.0763$\\
         \texttt{Spitzer/IRAC.ch1} & $35\,378.4$ & $0.136$ & $0.0221$\\
         \texttt{Spitzer/IRAC.ch2} & $44\,780.5$ & $0.098$ & $0.0159$\\
         \texttt{Spitzer/IRAC.ch3} & $56\,961.8$ & $0.082$ & $0.0133$\\
         \texttt{Spitzer/IRAC.ch4} & $77\,978.4$ &$0.087$ & $0.0142$\\
 \hline
    \end{tabular}
    
\label{tab:mag_correction}
\end{table} 

\subsection{\textit{Spitzer} imaging and catalogue extraction} \label{photom_analyses}
The Perseus cluster has been surveyed by the \textit{Spitzer} Space Telescope \citep{werner2004spitzer} in various programmes and observing modes, mostly aimed at studying the core of the cluster. We take advantage of the reduced and calibrated products available from the archive.
For this work, we downloaded the IRAC Post-BCD (Basic Calibrated Data) products, including observations in the four channels (from Programme IDs $3228$ and $80089$, P.Is. W.~Forman and D.~Sanders, respectively). 

\begin{figure*}
    \centering
   \includegraphics[width=1\linewidth]{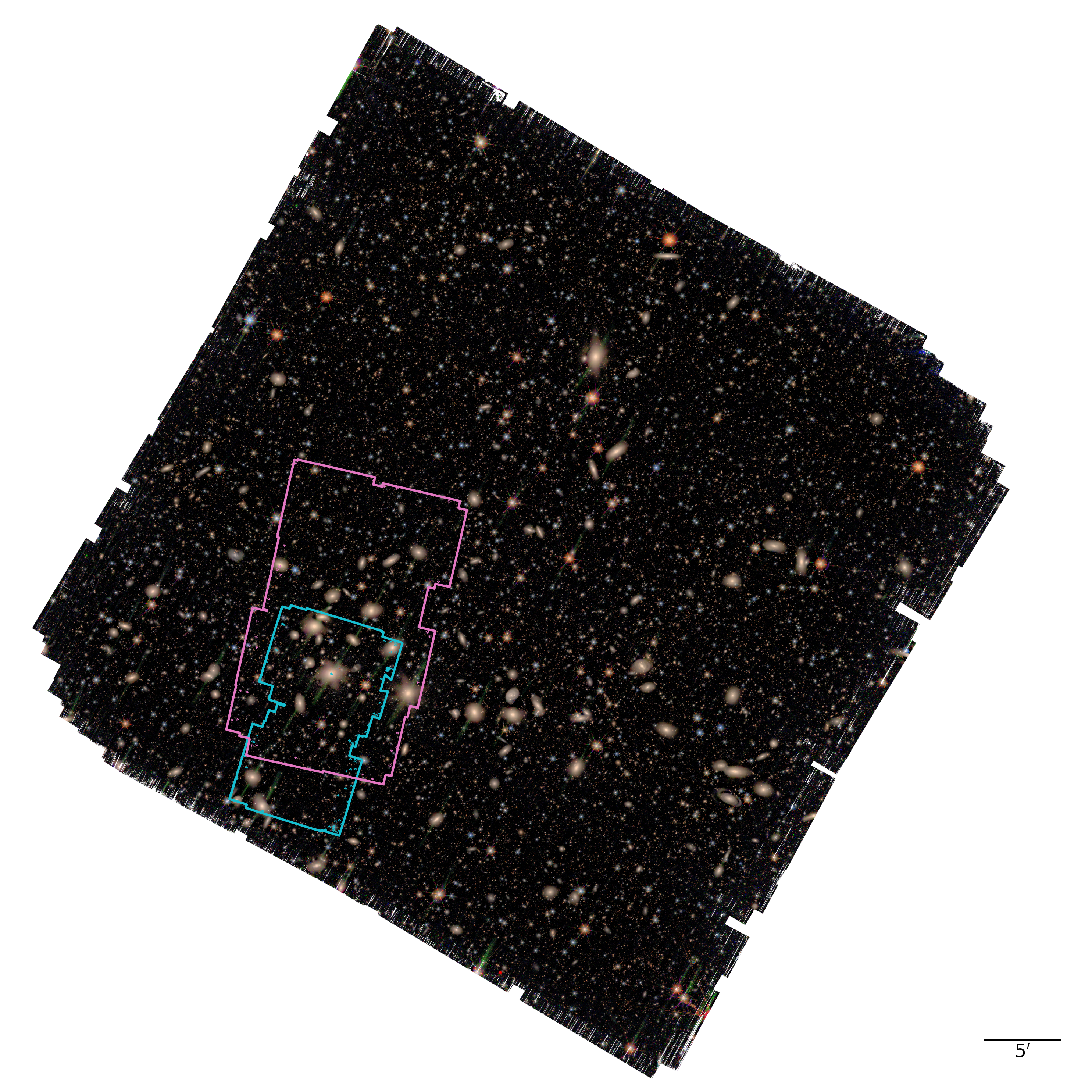}
      \caption{\Euclid RGB image of the Perseus cluster made by combining the images taken through the three NISP filters. The coloured solid lines show the extent of the two \textit{Spitzer}/IRAC data sets, with the cyan and magenta lines showing the extent of the first and second pointings, respectively. The images have been aligned with the \texttt{WCS} coordinate system, with north up and east to the left. In the low right corner, a scale bar representing $5$ arcmin is shown. } 
      \label{fig:perseus_field}
\end{figure*}

We used the two pointings that are within the area of the cluster imaged by \Euclid, which cover a total area of $232$\,arcmin$^2$. The four IRAC wavelengths are $3.6\,\micron$ (ch1), $4.5\,\micron$ (ch2), $5.7\,\micron$ (ch3), and $8\,\micron$ (ch4). For the first pointing there were only useful data in ch1, ch2, and ch3, while for the second pointing there were only useful data in ch1, ch2, and ch4. The layout of the \Euclid and \textit{Spitzer} images are shown in Fig.~\ref{fig:perseus_field}, where the difference in the sizes of the images provided by the two telescopes can be clearly seen. 

We used the \texttt{SExtractor} set-up described by \cite{moneti2022euclid} to detect sources in the \textit{Spitzer} images, with the only change being that we lowered the values of \texttt{DETECT\_THRESH and ANALYSIS\_THRESH} from $2$ to $1$ in order to find the faintest objects. The parameters we used are listed in Table~\ref{table:sextractor}. To convert the fluxes in the IRAC maps from MJy\,sr$^{-1}$ to AB magnitudes, we used the flux density for a zeroth magnitude star of $3631$\,Jy and a pixel size \ang{;;0.6}.\footnote{See \url{https://irsa.ipac.caltech.edu/data/SPITZER/docs/irac/iracinstrumenthandbook/19/} and \cite{moneti2022euclid}.}

 We used the $4.5\,\micron$ images as detection images, using \texttt{SExtractor} in dual mode to derive the fluxes in the other channels. In order for \texttt{SExtractor} to run in this mode, we had to convert the images in the different channels to a common size. For the first pointing, we converted the ch1 and ch3 images to the same size as the ch2 image using the \texttt{reproject} package.
 For the second pointing, we expanded the size of the ch1 image to match the size of the ch2 image by adding a blank boundary. As result, some of the sources detected in the ch2 image do not have observations in ch1.

\begin{table}[htbp!]
    \centering
\caption{Key parameters used for the \texttt{SExtractor} analysis of the \textit{Spitzer} images.}
\label{table:sextractor}
    \begin{tabular}{lc} 
    \hline\hline
    \noalign{\vskip 2pt}
    Parameter name & Value \\
    \hline
         \texttt{DETECT\_MINAREA}   & $5$\\ 
         \texttt{DETECT\_MAXAREA}   & $1000000$\\
         \texttt{DETECT\_THRESH}  & $1$\\ 
         \texttt{ANALYSIS\_THRESH}& $1$\\ 
         \texttt{FILTER\_NAME}      & gauss\_2.5\_5$\times$5.conv\\ 
         \texttt{DEBLEND\_NTHRESH}  & $32$\\ 
         \texttt{DEBLEND\_MINCONT}  & $0.00001$\\ 
 \texttt{MAG\_ZEROPOINT}    &$21.58$\\   
 \texttt{BACK\_SIZE}        &$32$\\ 
 \texttt{BACK\_FILTERSIZE}  &$3$\\ 
 \texttt{PIXEL\_SCALE}      &$0.60$\\
 \hline
    \end{tabular}    
\end{table} 

Although the Galactic extinction at these wavelengths is very small, we applied the same procedure as we used for the other images (see Sect. \ref{ERO}). However, we had to calculate our own corrections, since \textit{Spitzer} data are not included in the \Euclid ERO catalogues. We used the same $E(B-V)$ value as before and calculate the $C$ factors using the \texttt{dust\_extinction} Python package \citep{gordon2024dust_extinction}. The extinction values are listed in Table~\ref{tab:mag_correction}.

We used the \texttt{MAG\_AUTO} magnitudes obtained from \texttt{SExtractor}, as in the \textit{Euclid} case, ensuring consistency among the colours derived in the combined photometric catalogue (see Sect.~\ref{cross_match}). We considered all detections with signal-to-noise ratio (S/N) greater than $3$. If a galaxy was detected with a lower S/N, we gave it a magnitude limit corresponding to $3\,\sigma$.

\section{\label{sc:Methods} Methods}

\subsection{A cross-matched \textit{Spitzer}/\Euclid catalogue} \label{cross_match}
 
We matched the \textit{Spitzer} catalogue with the \Euclid NISP catalogue using \texttt{TOPCAT} \citep{taylor2005topcat} with a matching radius of $\ang{;;0.5}$. The radius chosen may seem small, considering that IRAC positions can substantially move due to the presence of blending. However, most sources presenting a blending problem in the IRAC images would have been discarded anyway during the visual check (see below). Therefore, we adopted this conservative choice to end up only with candidates having reliable photometry. After we had combined the \Euclid NISP and \texttt{Spitzer} catalogues, we used \texttt{TOPCAT} to match the sources to the VIS and MegaCam catalogues, using the same matching radius.

The \textit{Spitzer} catalogue from the first pointing contains $3294$ objects, whereas the matched catalogue contains $1939$ objects having both \Euclid NISP and \textit{Spitzer} ch2 detections. Thus, $1355$ objects do not have a \Euclid counterpart. For the second pointing, the initial \textit{Spitzer} catalogue has $5793$ objects, but only $3377$ also have a \Euclid NISP detection, so $2416$ lack a \Euclid counterpart.

\begin{table}[H]
    \begin{center}
    \caption{Magnitude limits ($1\,\sigma$) for the data used in this paper.}
     \label{tab:mag_lim}
    \begin{tabular}{lcccccc}
    \hline\hline
    \noalign{\vskip 2pt}
    \multicolumn{7}{c}{MegaCam} \\
        \noalign{\vskip 1pt}
    \hline
       Band&  $u$&  $g$&  $r$&  H$\alpha$&  $i$&  $z$\\
    
           \noalign{\vskip 1pt}
        Mag limit& $26.1$&  $26.9$&  $27.1$&  $24.8$&  $26.1$ &  $24.6$\\
        \hline
                   \noalign{\vskip 2pt}
    \multicolumn{7}{c}{\Euclid} \\
    \hline
     Band& & \IE&  \YE&  \JE& \HE &\\
    
                  \noalign{\vskip 1pt}
      Mag limit& & $28.4$&  $26.2$&  $26.1$& $26.3$ &\\
        \hline
                   \noalign{\vskip 2pt}
         \multicolumn{7}{c}{\textit{Spitzer}/IRAC} \\
    \hline
    Band&  & ch1&  ch2&  ch3&  ch4\\
   
               \noalign{\vskip 1pt}
        \makecell[l]{Mag limit of \\pointing $1$ }& & $22.1$&  $22.5$&  $21.1$&  \dots\\
    
                   \noalign{\vskip 1pt}
        \makecell[l]{Mag limit of\\pointing $2$}& & $22.3$&  $22.7$&  \dots&  $21.7$\\
        \hline
    \end{tabular}
    \end{center}
\end{table}

In Table~\ref{tab:mag_lim}, we report the magnitude limits (at $1\,\sigma$) in all bands. For \Euclid, we simply scaled the S/N values in the \Euclid catalogues to obtain the $1\,\sigma$ values. For \textit{Spitzer}, we determined the noise by placing 100 circular apertures randomly in background areas of the images, ensuring they did not overlap with sources by checking the segmentation maps provided by \texttt{SExtractor}. We used the Python package \texttt{photutils} \citep{larry_bradley_2024_10967176} to measure the fluxes in these apertures and the mean value of these measurements as our estimate of the noise.

\subsection{HIERO sample selection}
\begin{figure}
    \centering
    \includegraphics[width=1.\linewidth]{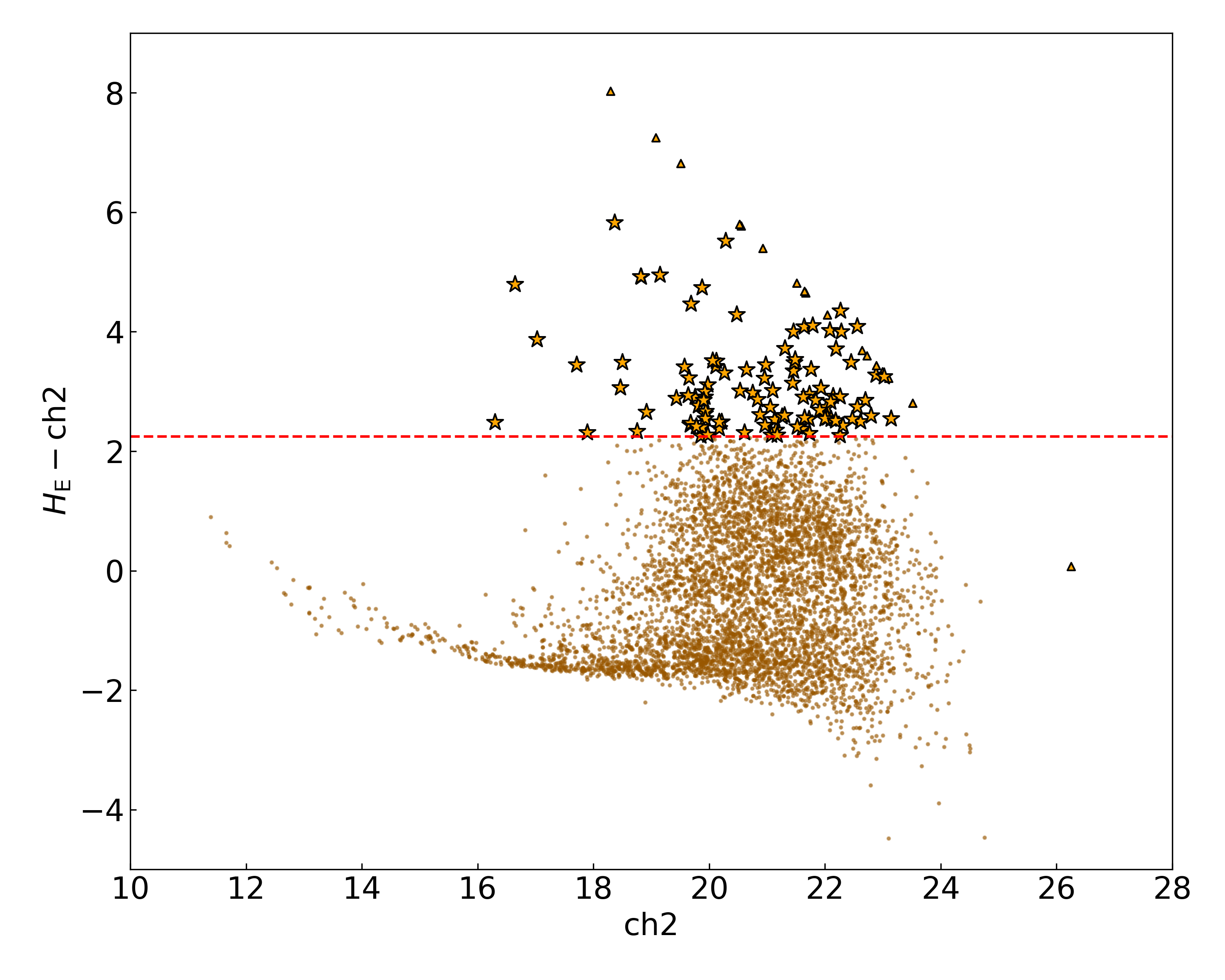}
    \caption{Colour-magnitude ($\HE-\mathrm{ch2}$ versus ch2) distribution of  the matched \Euclid and \textit{Spitzer} sample. The yellow star symbols show the objects that satisfy the HIERO colour criterion (shown as a dashed red line). The triangle symbols show lower limits for
    the \HE-undetected objects, calculated using the $1\,\sigma$ \HE magnitude limit. The other points show the parent sample.
      }
      \label{fig:color-selection}
\end{figure}

To identify the candidate HIEROs in the matched \Euclid$+$\textit{Spitzer} catalogue, we used the colour criterion of \cite{wang2016infrared}, $\HE - \mathrm{ch2} > 2.25$. This criterion provided us with an initial sample of $121$ objects ($50$ from pointing $1$ and $71$ from pointing $2$). Figure~\ref{fig:color-selection} shows the parent sample and the objects found with the colour cut. There were $18$ objects detected in the combined NISP image but that were not detected in the \HE image; we have shown these with a triangle symbol (using the $1\,\sigma$ \HE limit for clarity on the plot).

\begin{figure*}[hbtp!]
    \centering
   \includegraphics[width=0.9\linewidth, trim={0 63 0 42},clip, keepaspectratio]{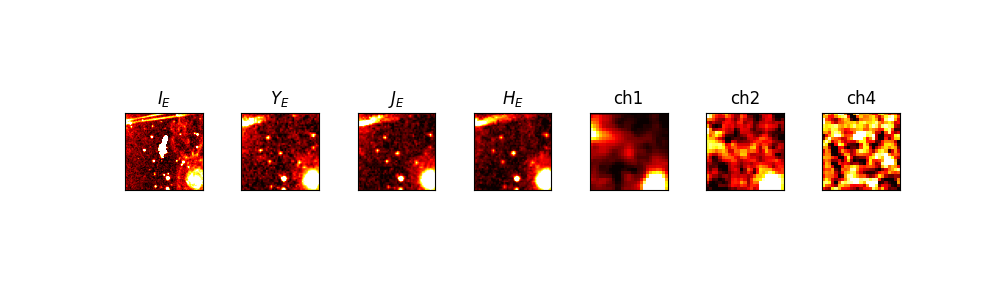}
    \caption{Example multi-wavelength cutout of a discarded object, due to an artefact in the VIS image. The cutout sizes are $15'' \times 15''$.}
     \label{fig:discarded_obj}
\end{figure*}

We performed a visual inspection of the \textit{Spitzer}/IRAC and \Euclid VIS and NISP images to spot objects that might be artefacts or noise features. We removed $69$ objects, reducing our sample to $52$. Figure~\ref{fig:discarded_obj} provides an example of the discarded objects.
In order to retain only reliable objects, we proceeded with a further series of tests (see Sects.~\ref{blending}, \ref{GC}, and \ref{BD}). 

\subsection{Blending contamination in the IRAC photometry} \label{blending}
Through the visual check on the \Euclid VIS and NISP$+$\textit{Spitzer} cutouts (see Fig.~\ref{fig:cutouts}) we identified a few objects that appeared to have several components in the VIS and NISP bands. We checked that the extraction executed in all the available bands treated these multi-component objects as a single source. This means that in all the bands the different components have been detected as a single source, therefore not invalidating the photometry by being a single source in some bands and two different sources in others (for a more detailed discussion of these objects, see Sect.~\ref{morphology}).

\subsection{Globular cluster contamination} \label{GC}
We removed possible globular clusters (GCs) by applying the selection criteria used for the Fornax cluster ERO data \citep{Saifollahi25}. The GCs in Fornax were selected as: $(1)$ being partially resolved and with small eccentricity; and $(2)$ having colours $0.0 < \IE - \YE < 0.8$, $ -0.3 < \YE - \JE < 0.5$, and $-0.3 < \JE - \HE < 0.5$. To be conservative, we decided to consider as GCs all objects that satisfied the colour conditions, independent of their size. Ultimately, nine candidates satisfied these conditions, reducing our sample to $43$ objects.

\subsection{Brown dwarf contamination} \label{BD}
We fitted the SEDs of the $43$ remaining objects with brown dwarf (BD) templates from \cite{burrows2006and}. We used the T dwarf models, which cover a temperature range from $700$\,K to $2200$\,K, gravities between $10^{4.5}$ and $10^{5.5}$\,cm\,s$^{-2}$, and metallicities between $\textrm{[Fe/H]} = -0.5$ and $0.5$. We then compared the $\chi^2$ of the fit with the
$\chi^2$ of the fit with the galaxy templates (see next section), removing any object with a better fit to a BD template. There was only one of these ($\textrm{ID}= 39$),
 giving us a final sample of $42$ objects.

\subsection{SED fitting analyses} \label{SEDfitting}
We fitted the SEDs of the 42 remaining objects with the \texttt{Bagpipes} package \citep{carnall2018inferring}. For all the candidates with no photometric detection in a specific band, we set the flux to $0$ and the error to $3\,\sigma$ (see Sect.~\ref{tab:mag_lim}). We adopted a \texttt{Bagpipes} set-up that maximises the exploration of the parameter space. We used a Calzetti law \citep{calzetti2000dust} for the dust extinction, a `delayed' star-formation history, and we included the nebular emission component, to allow for the effect of emission lines on the SED, which might otherwise lead to an overestimate of the stellar masses (Sect.~\ref{line_emitters}). The broad range used for the dust extinction value is essential for this particular type of galaxy, since they are expected to be very dusty. Table~\ref{tab:bagpipes} lists the parameters adopted as the input set up for the fitting code. 

\begin{table}[htbp!]
    \centering
     \caption{Main input parameters for \texttt{Bagpipes}.}
     \label{tab:bagpipes}
    \begin{tabular}{lc}
    \hline\hline
    \noalign{\vskip 2pt}
    Redshift & $[0.1, 15]$\\
    \hline
        \noalign{\vskip 1pt}
    \multicolumn{2}{c}{Dust} \\
    \hline
            \noalign{\vskip 1pt}
      Type   & Calzetti\\
       $A_V$ [mag]& $[0, 6]$\\
       \hline
               \noalign{\vskip 1pt}
       \multicolumn{2}{c}{Nebular emission} \\
       \hline
               \noalign{\vskip 1pt}
         $\logten$U& $[-4, -1]$\\
          \hline
                  \noalign{\vskip 1pt}
       \multicolumn{2}{c}{Delayed $\tau$ model} \\
       \hline
               \noalign{\vskip 1pt}
        Age [Gyr] & $[0.001, 15]$\\
        $\tau$ [Gyr]& $[0.01, 10]$\\
         Metallicity [$Z_\odot$]& $[0, 2.5]$\\
    Mass-formed [$\logten(M_*/M_\odot$)]& $[6, 12.5]$\\
    \hline
    \end{tabular}
\end{table}

After the first run of \texttt{Bagpipes}, we noticed that for two objects in the final sample, ID $23$ and $31$, the MegaCam fluxes were suspiciously high with respect to the others, in particular relative to the flux in the \IE band. For object $23$, the fluxes in the two MegaCam filters on either side of the \IE band, the ${\rm H}\,\alpha$ and $i$ bands, were, respectively, a factor of $1.3$ and $2.5$ higher than the \IE flux.
For object $31$, the fluxes in the ${\rm H}\,\alpha$ and $i$ filters were $30$ and $15$ times higher, respectively, than the \IE flux. To investigate the reasons behind these examples,
we visually checked the \Euclid VIS and NISP $+$ \textit{Spitzer} cutouts of these objects.
We concluded that the fluxes measured from the MegaCam images, which have lower resolution than the \Euclid images, are likely to be contaminated by nearby
objects visible in the \Euclid images. For these two objects, we performed a second run of \texttt{Bagpipes}, excluding the MegaCam data. We use the results of these fits in the rest of the paper.

\section{\label{sc:Results} Results}
 \begin{figure}
     \centering
     \includegraphics[width=1\linewidth, trim={0 25 0 25},clip, keepaspectratio]{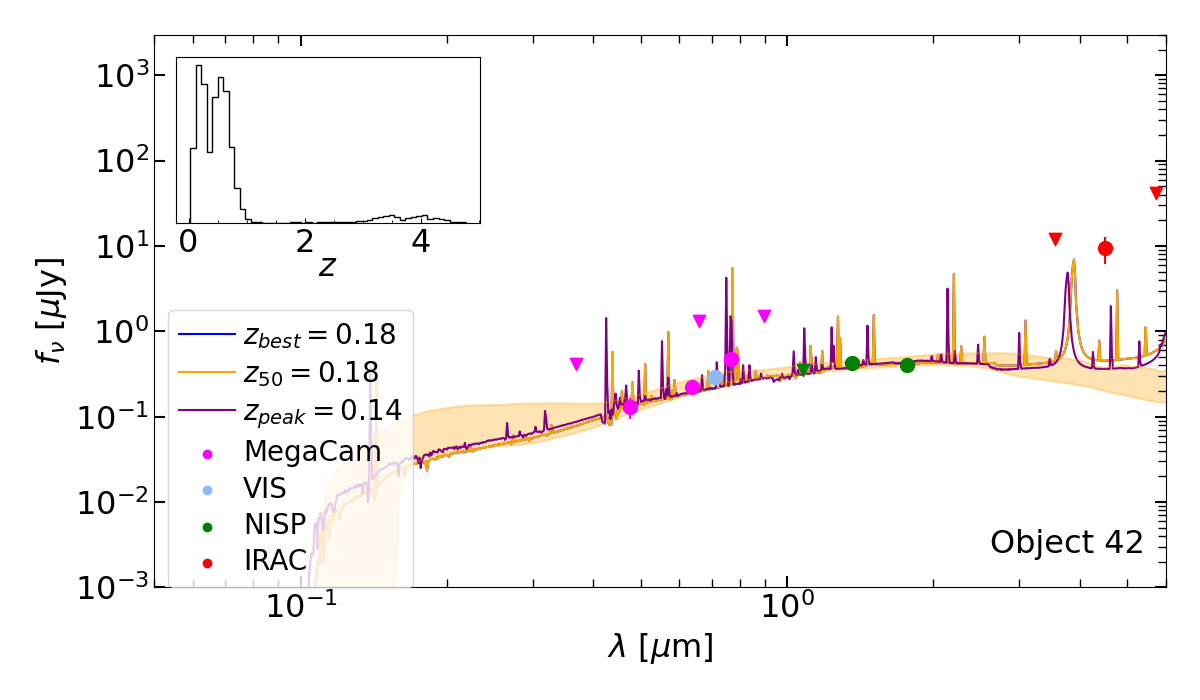}
     \includegraphics[width=1\linewidth, trim={0 25 0 25},clip, keepaspectratio]{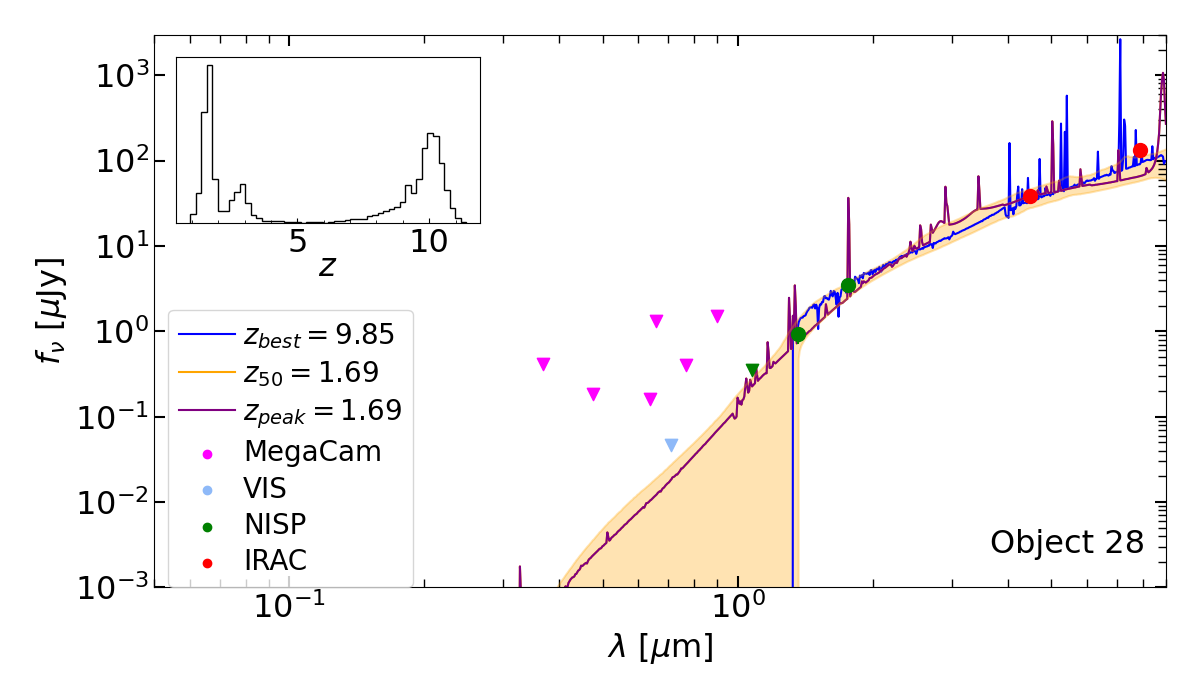}
     \includegraphics[width=1\linewidth, trim={0 25 0 25},clip, keepaspectratio]{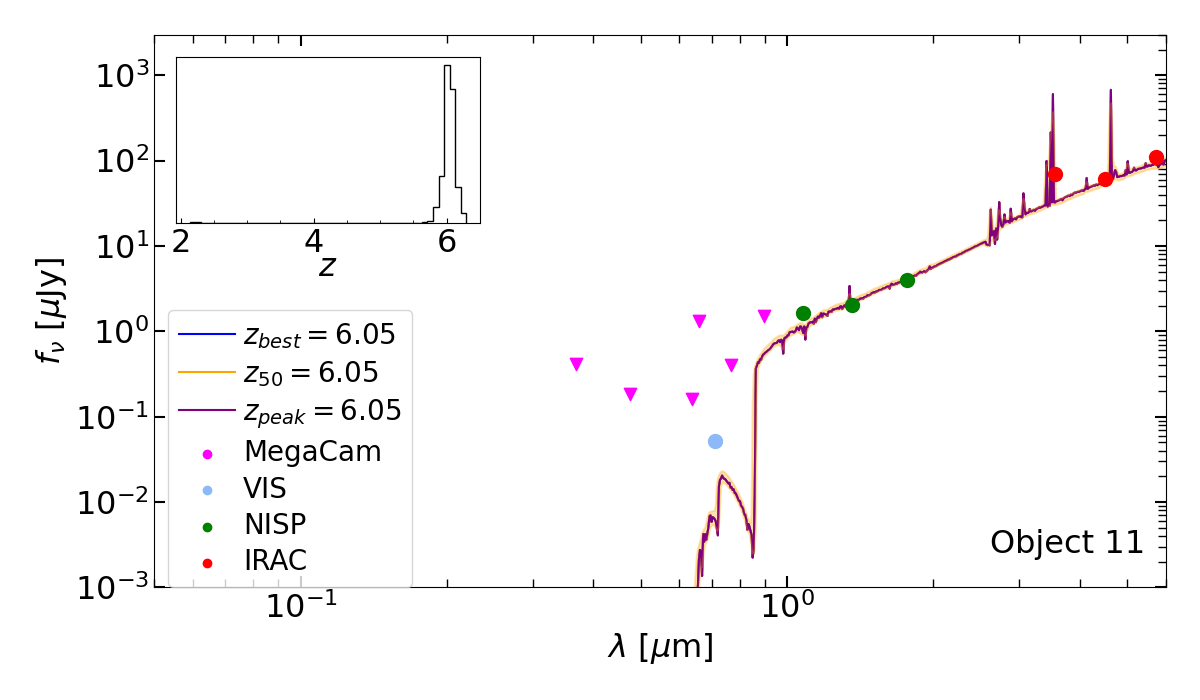}
     \caption{\texttt{Bagpipes} fits to the SEDs for three representative objects, IDs $42$, $28$, and $11$. The posterior probability distributions for the redshifts, PDF($z$), are shown as insets. Photometric detections are shown by the coloured circles and 3$\,\sigma$ upper limits are plotted as triangles. The coloured lines show the three different fits described in the text. The PDF for ID $42$ peaks at low redshift, while ID $28$ is bimodal and ID $11$ has a single high-redshift peak, making it representative of the objects that are likely to be HIEROs.}
     \label{fig:SED_fitting_examples}
 \end{figure}
 
\begin{figure}
    \centering
    \includegraphics[width=1\linewidth, trim={0 25 0 19},clip, keepaspectratio]{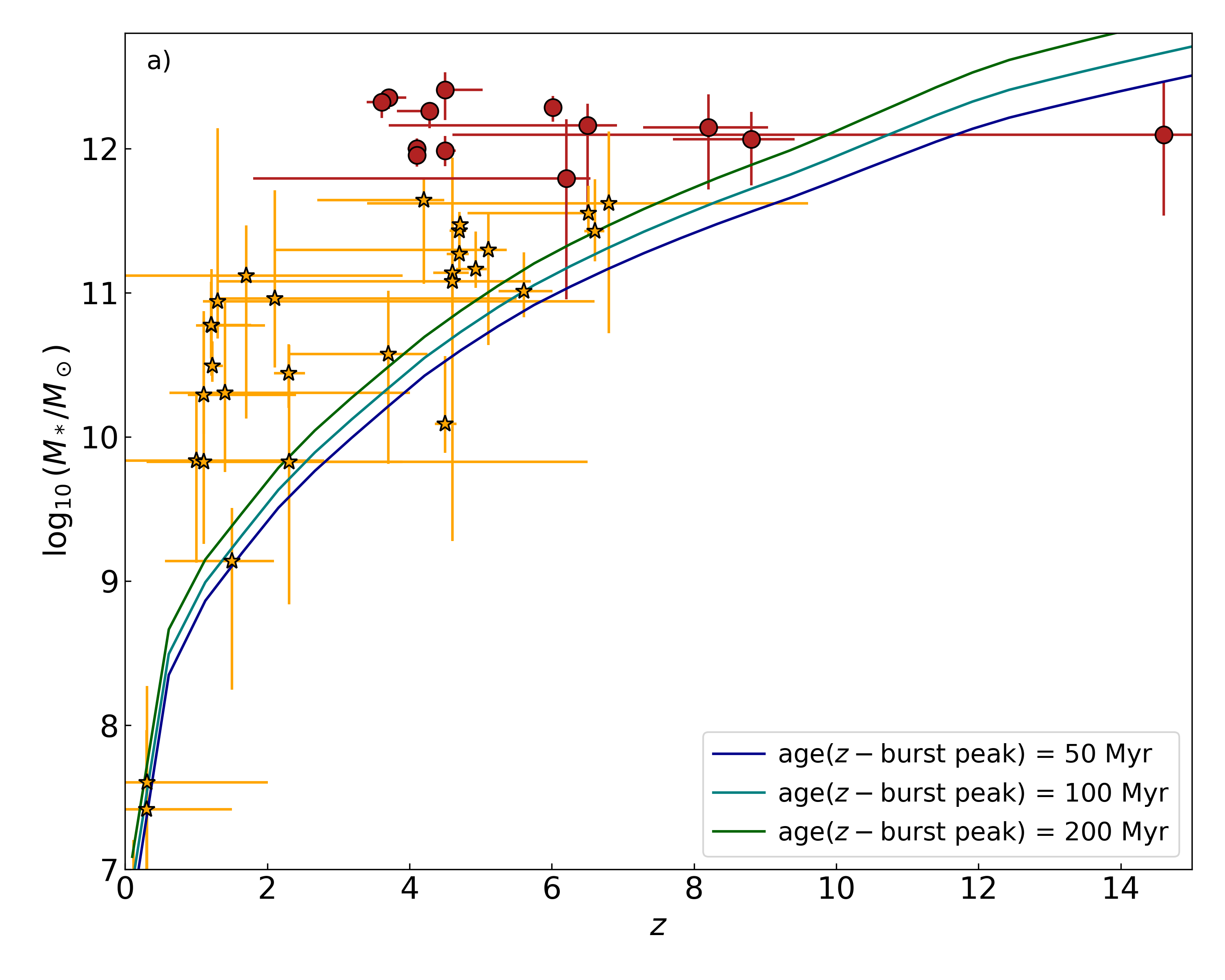}
    \includegraphics[width=1\linewidth, trim={0 25 0 19},clip, keepaspectratio]{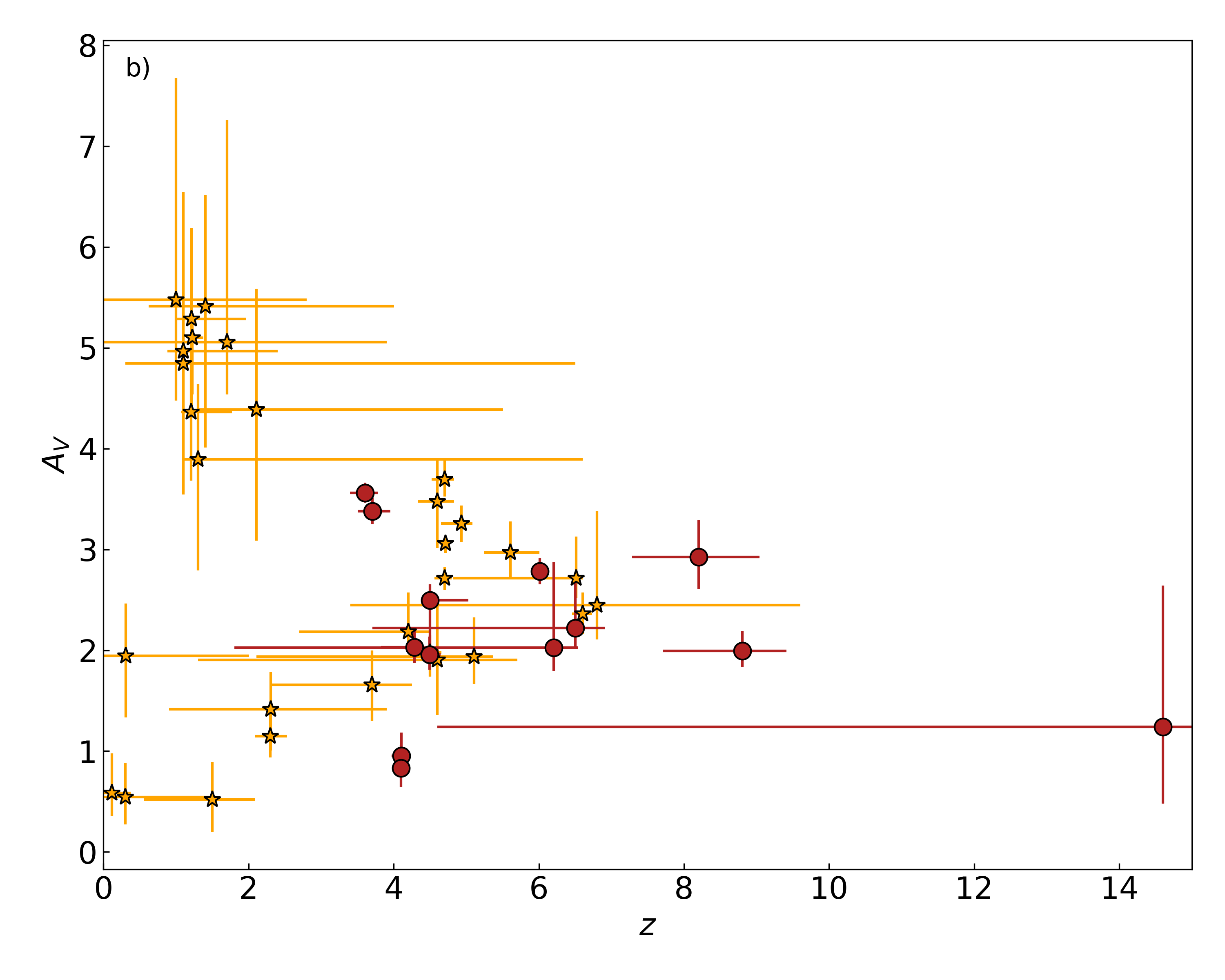}
    \includegraphics[width=1\linewidth, trim={0 7 0 7},clip, keepaspectratio]{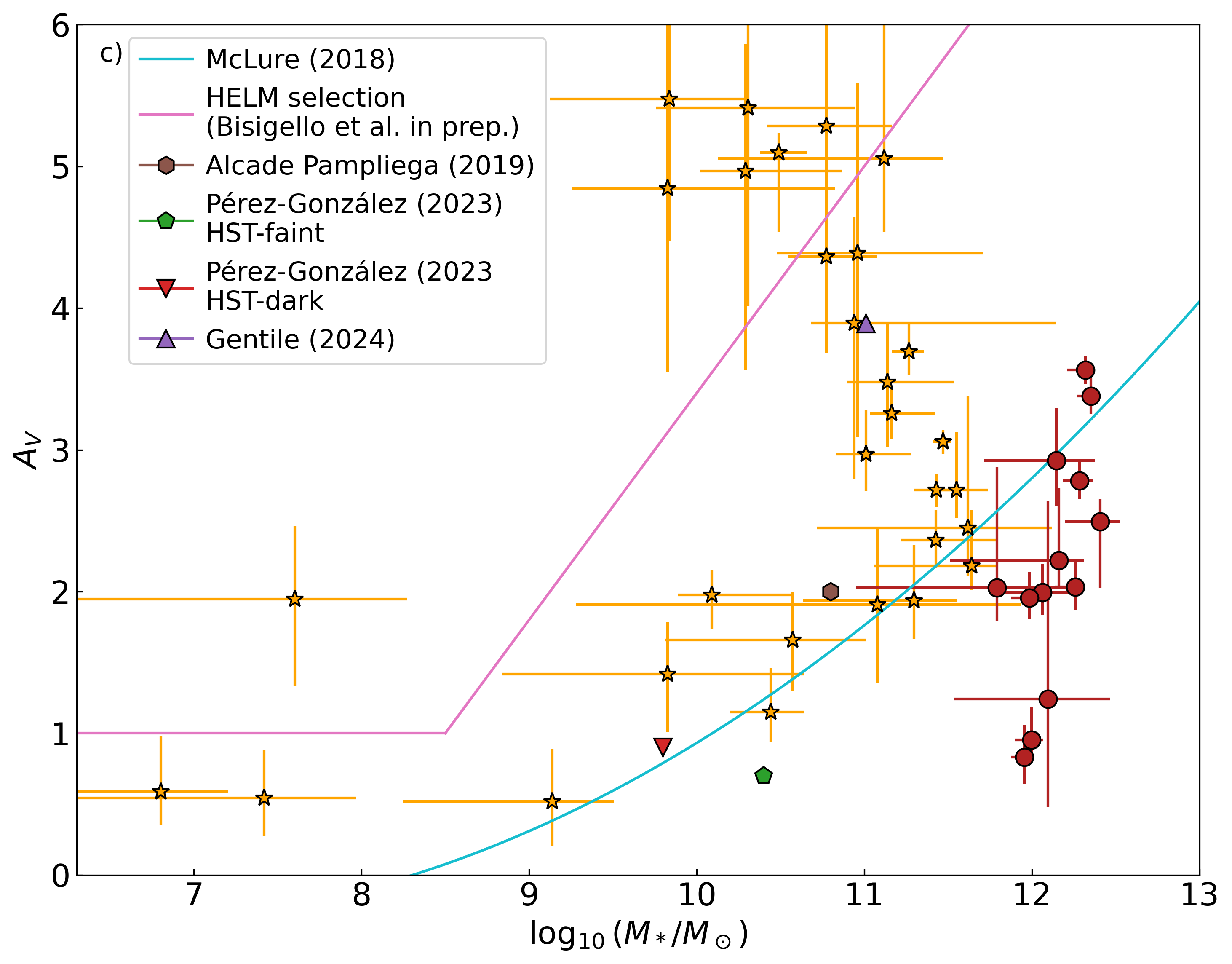}
    \caption{Properties of the HIEROs. Panel a) shows the redshift versus stellar mass. The red circles show the most massive candidates, with stellar mass greater than $10^{11.7} M_\odot$. These objects are from here on defined as `overmassive' and shown with the same symbol, i.e., red circles, in all other panels. The solid lines report the minimum observable stellar mass producing an IRAC ch2 magnitude of $22.7$. All three curves represent star-forming galaxies, the green with an age between the redshift epoch and the burst peak of $200$ Myr, the teal one of $100$ Myr, and the dark blue one of $50$ Myr. In panel b) the redshift versus dust attenuation distribution is reported. Panel c) shows the dust attenuation versus stellar mass. The cyan solid line shows the relation from \cite{mclure2018dust}, while the magenta line delimits the area identifying the so-called HELM galaxies \citep{bisigello2025spectroscopic}. Different symbols report values from previous studies, as indicated in the legend.}
    \label{fig:physical_param}
\end{figure}

\begin{figure}
    \centering
    \includegraphics[width=1.\linewidth, trim={0 5 0 7},clip, keepaspectratio]{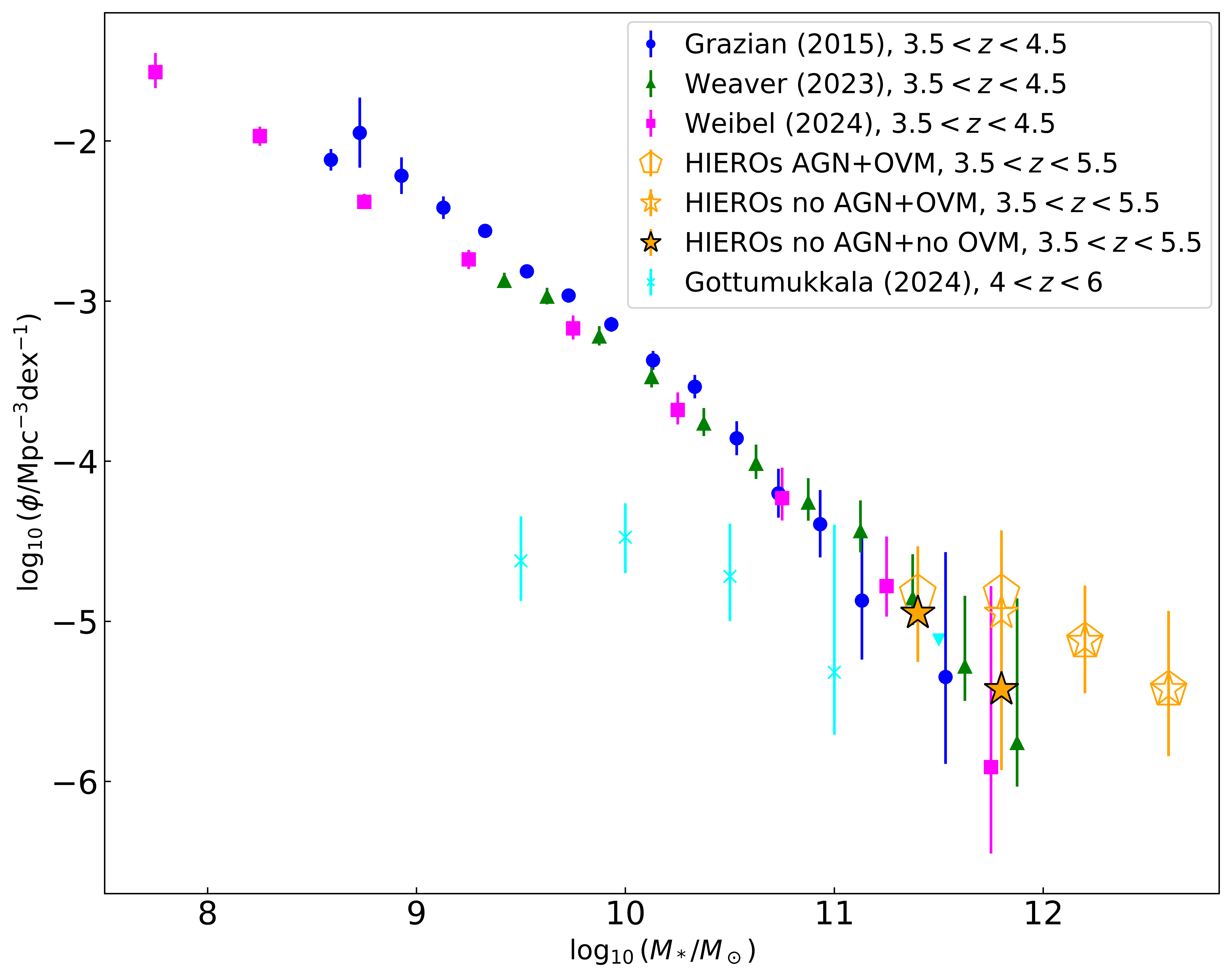} 
    \caption{Different galaxy stellar mass functions represented by orange open pentagons for the case with the complete original sample, orange open stars for the case where we remove the AGN candidates, and filled orange stars for the most conservative case, removing both the AGN candidates and the overmassive objects. The comparison is shown with respect to the values found by \cite{grazian2015galaxy}, reported in blue circles, \cite{weaver2023cosmos2020}, represented by green triangles, and \cite{weibel2024galaxy}, with magenta square symbols. The cyan cross symbols report the values found by \cite{gottumukkala2024unveiling}, who applied a similar selection on the JWST/CEERS data set.}
    \label{fig:final_MF}
\end{figure}

Before proceeding with showing the outcome of our work, we want to stress that we rely on only a few photometric data points. Even if upper limits are taken into account in the SED fitting analyses, sometimes we have just three or four detections with S/N $> 3$. For this reason, we invite the reader, as we do, to be cautious about the photometric redshifts retrieved and the galaxies properties estimated.

\subsection{Main SED fitting categories} \label{main_SEDfitting_categories}
Figure~\ref{fig:SED_fitting_examples} shows an example of the main categories of fits. All the fits are shown in Fig.~\ref{fig:SED_fitting}.
In each plot, the photometry is shown by a circle if there is a detection in a band ($\mathrm{S/N} > 3$) and a triangle if there is only a upper limit (shown as a $3\,\sigma$ upper limit). We show three different fits. The blue solid line shows the fit with the lowest $\chi^2$. The orange solid line shows the fit chosen by \texttt{Bagpipes}, which is the fit for the median value of the posterior probability distribution. The purple line shows the best fit for the redshift value with the highest probability [the peak in the PDF($z$)]. If some of these lines are not visible, it means that they are hidden by the others. We decided to use this final fit in the rest of the paper because this avoids any problems caused by the effect on the physical parameters if there is a bimodal PDF($z$). For this reason, in the rest of this paper we use the peak in the PDF($z$) as the photometric redshift of the object and the physical properties estimated with this redshift, in particular the dust attenuation $A_V$ and the stellar mass $M_*$ values. 

There are three categories of fit shown in Fig.~\ref{fig:SED_fitting_examples}: a low-$z$ solution; a bimodal PDF($z$); and a typical HIERO candidate. 
The \texttt{Bagpipes} fit of ID $42$ suggests that the object is actually at low redshift. The fit for  ID $28$, on the other hand, shows a bimodal PDF($z$), with a low-$z$ solution and a very high-$z$ ($z > 7$) one. As previously mentioned, we always selected the solution corresponding to the peak of the PDF, in this case the the low-$z$ solution. Finally, object ID $=11$ is an example of a typical HIERO source. The PDF($z$) shows a clear peak at $z \approx 6$, with all three fits in agreement that the object is at high redshift. We also find HIEROs at lower redshift ($z=3$--$4$), with all the fits being shown in Appendix~\ref{apdx:A}.

 \subsection{Physical properties}
 The physical properties derived from the \texttt{Bagpipes} fits are given in Table~\ref{tab:bagpipes_results} for all $42$ objects in the sample. We show in Fig.~\ref{fig:physical_param} the main physical properties as a function of redshift, in particular the dust extinction and stellar mass. We decide to not discuss the SFR estimates, since they are not reliable, given the lack of far-infrared data. The outlier point in the plots is for ID $=13$, which has an estimated redshift of $14.6$, although this result is uncertain due to the multi-peak behaviour of the redshift probability distribution (for further discussion regarding this source see Sect.~\ref{id13}).

The mean photometric redshift of the sample is approximately $4$. The top right panel of Fig.~\ref{fig:physical_param} shows that the majority of the objects have high values of dust extinction, with a mean value of $A_V = 2.8$, confirming that this kind of object is indeed dusty (to obtain these high values, it was necessary to allow the dust extinction parameter to cover a wide range of values, up to $A_V = 6$). The trend is similar to the one found by \cite{bisigello2023delving} for fainter galaxies observed with JWST. The second and third panels of Fig.~\ref{fig:physical_param} show that the galaxies generally have more extinction than the relationship between extinction and stellar mass found for normal galaxies by \cite{mclure2018dust}. We also show the relationship for highly extincted low-mass (HELM) galaxies \citep{bisigello2025spectroscopic}. Seven of the galaxies in our sample have extinction values that are above this relationship, while $16$ galaxies have extinction values between the two relationships.

We can see that, as expected, stellar mass increases with redshift. The mean value of $\logten(M_*/M_\odot) = 10.9$ clearly shows that we indeed find very massive galaxies. We have indicated the most massive galaxies, defined as $\logten(M_*/M_\odot) > 11.7$, by the red circles in Fig.~\ref{fig:physical_param}. We discuss these galaxies in Sect.~\ref{MF}.

\subsection{Stellar mass functions of HIEROs at \texorpdfstring{$z=3.5$--$5.5$}{z=3.5--5.5}} \label{MF} 
The GSMFs in the Perseus field have been derived by adopting the best-fit photometric redshifts and stellar mass estimates given in Table~\ref{tab:bagpipes_results}. GSMFs have been calculated in a single redshift bin, spanning from $3.5$ to $5.5$, to achieve better statistics given the low number of data points. We have intentionally decided to focus the analysis on this redshift range to investigate the contributions of the HIERO population to the stellar mass assembly beyond cosmic noon.

The GSMFs have been derived following the technique described in \citet{fontana06}, \citet{,santini12}, and \citet{grazian2015galaxy}. In detail, for each galaxy with $3.5\le z_{\mathrm{phot}}\le 5.5$ and magnitude in ch2 above the completeness limit (i.e., $[\textrm{ch2}]=20.9$ AB), we derived the non-parametric GSMF as the sum of the inverse accessible volumes. 

The GSMFs have been limited to a stellar mass above $1.6 \times 10^{11} M_\odot$ at $3.5<z<5.5$, corresponding to the magnitude limit in ch2 of the Perseus field, as described in Sect.~\ref{cross_match}. The error bars for each point have been computed using the approach of \cite{gehrels86}, which is Poisson statistics with proper corrections for small number counts below $10$ objects. In order to take into account the uncertainties on the photometric redshift derivation or on the stellar mass estimate, we generated $10^4$ synthetic catalogues by randomly sampling every time for each observed object the redshift probability distribution function (PDF) and its associated stellar mass PDF produced by \texttt{Bagpipes}. We derived the GSMF for each synthetic catalogue, and then we computed the uncertainties of the $10^4$ random GSMFs for each bin in redshift and stellar mass. This error was summed in quadrature with the statistical error of the observed GSMFs, as described above.

We have also included the sampling variance in our error estimates, in order to take into account the finite volume sampled and to avoid underestimating the total error budget. This variance has been derived by adopting the Cosmic Variance Calculator\footnote{\url{https://www.ph.unimelb.edu.au/~mtrenti/cvc/CosmicVariance.html}} by \citet{trenti08}. We have assumed an area of $232$ arcmin$^2$, a mean redshift of $4.5$, with a redshift interval of $2.0$, i.e., the one adopted here for the GSMF calculation. The other free parameters in the computation are the number of objects ($42$), the halo filling factor\footnote{This is the average halo occupation fraction, i.e., the fraction of dark matter halos hosting HIERO galaxies} ($1.0$) and the completeness of the observations ($1.0$). We have assumed both a halo filling factor and a completeness fraction of $1.0$ in order to maximise the amount of sample variance. For lower values of the halo filling factor and/or of the completeness fraction, the calculated variance would be lower. The resulting relative sampling variance is $0.17$ dex, which has been summed in quadrature with the GSMF errors listed above.

Here we consider the results from \cite{forrest2024magaz3ne}, who pointed out that SED fitting tools are not able to properly reproduce the spectrum of red sources similar to those studied in this work. They concluded that stellar mass values exceeding $\log_{10}(M_*/M_\odot) > 11.7$ in the redshift range $3 < z < 4$, derived through SED fitting, are unreliable due to incorrect $z_{\mathrm{phot}}$ estimates. Briefly, they selected a sample of galaxies with photometric redshifts between $3$ and $4$ and stellar mass values above $10^{11.7} M_\odot$, which they followed up spectroscopically. They found that the spectroscopic redshifts did not agree with the photometric ones ($\Delta z > 0.5$), generally lying below $z_{\mathrm{spec}}=2.5$. 

Taking account of these results, we created a subsample of galaxies with the same stellar mass and redshift cuts, namely $\logten(M_*/M_\odot) > 11.7$ and $z > 3.5$. There were $13$ objects in this sample, with six of them showing a secondary peak at $z<3.5$ in the \texttt{Bagpipes} redshift distribution. These sources will be referred to as `overmassive' from now on. In the next section, we discuss possible effects that could produce errors in the GSMFs, such as the presence of active galactic nuclei (AGN). 

We computed the GSFMs in three different ways: (1) using the whole HIERO sample, including the overmassive sources; (2) removing the AGN candidates (see Sect. \ref{agn}), but keeping the overmassive sources; and (3) removing both the AGN candidates and the overmassive sources. The different results are shown in Fig.~\ref{fig:final_MF}. The figure shows that our estimates of the GSMF in the two highest mass bins are entirely due to the presence of the overmassive galaxies, which therefore vanish when these objects are excluded (i.e., the most conservative case, represented by the filled orange star). Our estimate of the GSMF in the lowest mass bin does not change very much when the AGN and overmassive galaxies are removed. However, our estimate of the GSMF in the second lowest mass bin changes significantly when the AGN and overmassive galaxies are removed.

Nevertheless, even in the most conservative case, when all the AGN and overmassive galaxies are removed, our data points are very similar to previous estimates of the high-mass end of the GSMF. Our estimates are likely to be lower limits because of our very conservative removal of all objects that might possibly be spurious, which suggests that previous estimates of the high-mass end of the GSMF may well be too low. Future \Euclid imaging will be crucial for investigating this question further. 

\section{\label{sc:Discussion} Discussion}

\subsection{Overestimates of stellar mass?}

In this section we explore effects that might lead to overestimates of the stellar masses and thus produce the overmassive galaxies found in the \texttt{Bagpipes} analysis.

\subsubsection{Contribution from line emission} \label{line_emitters}
\cite{papovich2023ceers} pointed out that for a galaxy at $z>4$ the absence of a JWST/MIRI detection at $5.6$ and $7.7\,\micron$ can lead to an estimate of the stellar mass that is too high by up to a factor of $10$. \cite{bisigello2019statistical} found the same problem for a mock sample of galaxies with data in the JWST NIRCam bands. The error is caused by the contribution of emission lines in broad-band filters, which produce very red colours.

In order to test how much our sample is affected by this issue, we selected only the sources with a clear detection ($\textrm{S/N} > 3$) in the longer wavelengths bands available, which are IRAC ch3 or ch4, depending on the pointing (as explained in Sect.~\ref{photom_analyses}). We also required the galaxies to also have detections in the other shorter wavelength IRAC bands, i.e., ch1 and ch2. Only $12$ sources are detected in all three IRAC bands. We repeated the SED fitting run reported in Sect.~\ref{SEDfitting}, this time excluding the measurements for the filter at the longest wavelength. 
The result was unexpected, with the opposite trend to that described in the aforementioned papers: we found that by removing the last data point, the stellar masses estimated are significantly lower than the case where all the data are considered in the fit. We found seven objects with a difference greater than $1$ dex, but in only one case was the stellar mass estimate higher. For the other six, the new stellar mass estimate was lower. The explanation for five of them was that the redshift estimate in the new run was lower, which led to a lower stellar mass estimate. For the remaining galaxy, the two redshift estimates were similar but the new stellar mass estimate was still lower.

To further investigate this result, we performed a second run in which we still excluded the photometry at the longest wavelength, but forced the redshift to be the same as in the original run. We found that for four objects the new stellar mass was still about $1$ dex below the original estimate, but for the rest the difference was only about $0.13$ dex. 

Therefore, we conclude that our stellar masses are not strongly overestimated because of bright nebular emission lines. The difference in the result with respect to the previously mentioned papers might be due to the different sample luminosities. In addition to that, since we used the \texttt{Bagpipes} configuration that allows the nebular parameter to cover a range from $-4$ to $-1$ (see Table~\ref{tab:bagpipes}), our results may be less affected by their presence.

\subsubsection{AGN contribution} \label{agn}
An additional source of uncertainty for the physical properties is the elusive presence of AGN. If an AGN is present and we are wrongly attributing its emission to stars, we will over-estimate the stellar mass of the galaxy.

To investigate this possibility, we adopted the criteria used inside the Euclid consortium to classify an object as an AGN. The criteria were developed by \cite{bisigello2021simulating} from the \texttt{SPRITZ} simulation. These criteria distinguish between AGN1, which are unobscured AGN, and All-AGN, which includes obscured AGN, unobscured AGN, and galaxies dominated by star-formation with a minor contributions by the AGN.

For the AGN1, the colour selection is $\IE - \HE < 1.1 \,\wedge\,u-z < 1.2 \,\wedge\, \IE - \HE < - 1.3\,(u - z)+ 1.9$, while for the All-AGN  the colour selection is $\IE -  \YE < - 0.9\,(u - r) + 0.8 \,\wedge\, u - r < 0.2$ \citep{bisigello2024euclid}. Only one object satisfies the criteria for both the categories (ID $=10$), while nine fall in the All-AGN class (ID $=1, 2, 4, 22, 26, 29, 36, 37,\mathrm{\;and\;}40$). However, this colour selection method is tuned with empirical constraints only up to $z \approx 2$, and therefore it is not reliable in our case because the majority of our sample is at $z>2$. For this reason, we decided to compute the calculation of the GSMFs both including and excluding the AGN candidates. 

\subsection{Visual morphology} \label{morphology}
Most of the objects have a compact shape in the VIS image. However, it is worth noting that a few objects have extended structure 
in the VIS image (the cutouts are shown in Fig.~\ref{fig:cutouts}). This could be explained by the presence of a single source composed of multiple components or with an irregular clumpy structure. Another interesting explanation would be the merger between dusty objects, such as the `Cosmic Whale' identified by \cite{rodighiero2024optically}.
This aspect will be explored more in future works.

\subsection{A candidate at \texorpdfstring{$z \sim 14$}{z=14}} \label{id13}
\begin{figure*}
    \centering
    \includegraphics[width=0.9\linewidth, trim={0 63 0 42},clip, keepaspectratio]{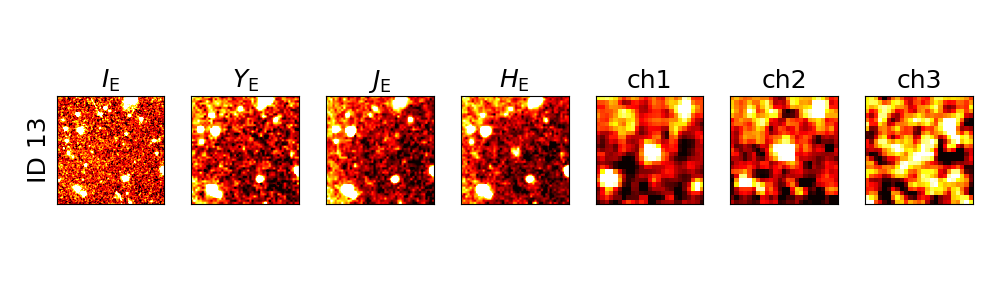}
    \caption{Multi-wavelength cutouts of the source ID $=13$, with an estimated $z\approx14$. The available images are the \Euclid bands \IE, \YE, \JE, and \HE and the \textit{Spitzer} channels ch1, ch2, and ch3. The cutouts size is $15'' \times 15''$.}
    \label{fig:ID13_cutout}
\end{figure*}

\begin{figure*}
    \centering
    \includegraphics[width=0.49\linewidth]{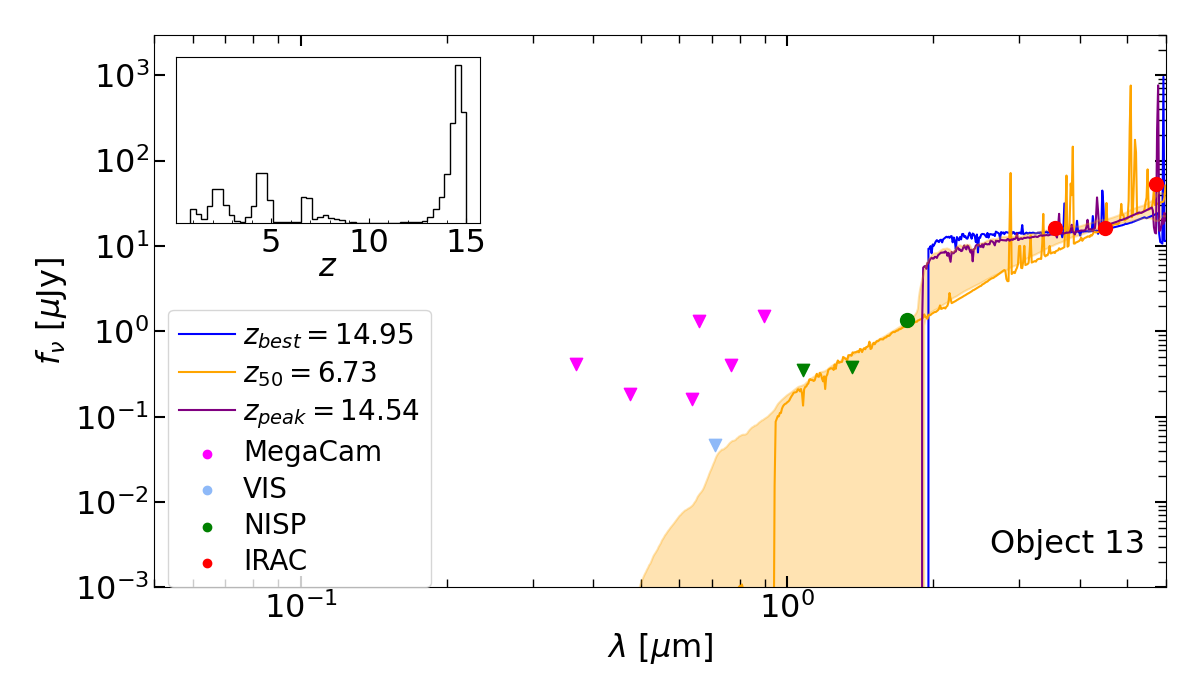}
    \includegraphics[width=0.49\linewidth]{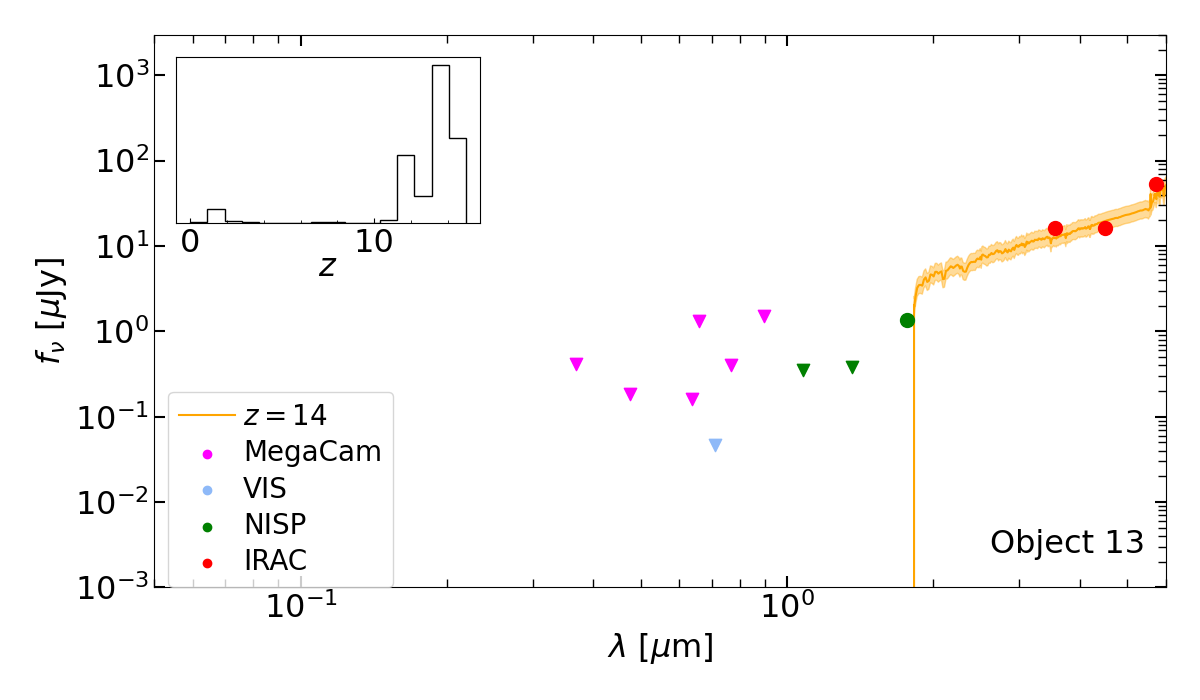}
    \caption{SED fits for ID $13$. Both panels show photometric data points, both detections (circles) and upper limits (triangles), along with the PDF($z$) displayed as an inset. The first panel shows the different SED fitting models found by \texttt{Bagpipes}, while the second panel shows the fit found by \texttt{CIGALE}.} 
    \label{fig:ID13}
\end{figure*}

\begin{table}[htbp!]
    \centering
    \caption{Input models and main parameters for the \texttt{CIGALE} code.}
    \label{tab:cigale}
    \begin{tabular}{lr}
    \hline\hline
    \noalign{\vskip 2pt}
    \multicolumn{2}{c}{\texttt{sfh\_delayed}} \\
    $\tau$ (main) [Myr] & $150, 200, 300, 500, 700, 1000,$ \\
    & $1500, 3000, 5000, 10000$ \\
    Age (main) [Myr]& $200, 300, 500, 700, 1000,$ \\
    & $1500, 3000, 5000, 10000$ \\
    \hline
        \noalign{\vskip 1pt}
    \multicolumn{2}{c}{\texttt{bc03}} \\
      IMF   & Chabrier\\
       Metallicity & $0.02$\\
       \hline
               \noalign{\vskip 1pt}
       \multicolumn{2}{c}{\texttt{nebular}} \\
         $\logten$U& $-3$\\
         Emission & True \\
       \hline
               \noalign{\vskip 1pt}
       \multicolumn{2}{c}{\texttt{dustatt\_modified\_starburst}} \\
        $E_{BV}$ lines [mag]& $0, 0.5, 1, 1.5, 2, 2.5, 3, 3.5, 4, 4.5$\\
        $E_{BV}$ factor& $0.44$\\
        $R_V$ & $3.1$ \\
       \hline
               \noalign{\vskip 1pt}
       \multicolumn{2}{c}{\texttt{fritz2006}} \\
        $r_{\rm ratio}$ & $10, 60, 150$\\
        $\tau$ & $0.1, 1, 6$\\
        $\beta$ & $-0.5$ \\
        $\gamma$ & $0, 4$ \\
        Opening angle [deg]& $60, 100$ \\
        Disk type & \cite{schartmann2005towards} spectrum \\
        $\delta$ & $-0.36$ \\
        fracAGN & $0, 0.1, 0.25, 0.5$ \\
        \hline
                \noalign{\vskip 1pt}
        \multicolumn{2}{c}{\texttt{redshifting}} \\
        Redshift & $0, 1, 2, \dots, 15$\\
    \hline
    \end{tabular}
    \tablefoot{The models used are: a delayed SFH with optional exponential burst; \cite{bruzual2003stellar} single stellar population; a continuum and line nebular emission model; a modified \cite{calzetti2000dust} dust attenuation law; AGN models from \cite{fritz2006revisiting}; and a redshifting model that also includes the IGM from \cite{meiksin2006colour}.}
\end{table}

As previously mentioned, with our SED fitting analyses we found a candidate galaxy that appears to be at $z\simeq14$ (ID $=13$). Its multi-wavelength cutout is shown in Fig.~\ref{fig:ID13_cutout}. In Fig.~\ref{fig:ID13} we show the different \texttt{Bagpipes} fits as explained above, along with the PDF($z$). Both the fit with the lowest
$\chi^2$ and the one corresponding to the highest peak in the PDF($z$) give a photometric redshift of around $14.6$, whereas the one from the median of the posterior probability distribution predicts the redshift corresponding to the minor peak in the PDF($z$) at $z=6.7$.

The stellar mass value found is $\logten(M_*/M_\odot)=12.1$, along with a dust extinction of about $A_V =1.24$ mag (Table~\ref{tab:bagpipes_results}). The stellar mass value puts this source among the overmassive objects, while not having a particularly high value for dust extinction, which makes us suspicious of this result.
However, we also fit the SED with \texttt{CIGALE} \citep{boquien2019cigale}, which includes an AGN component in the fit (although we note that this galaxy does not satisfy the \Euclid AGN criteria). The parameters used in this fit are listed in Table \ref{tab:cigale}. The fit is shown in the second panel of Fig. \ref{fig:ID13}. The best redshift estimate is also about $14$, strengthening the possibility that this galaxy is genuinely at a very high redshift, although the estimate of the stellar mass is now even higher, $1.23 \times 10^{13}$\,$M_\odot$. The PDF($z$) produced by \texttt{CIGALE} also presents a peak at lower redshift ($z < 1$). 

As mentioned before, the few photometric detections available are not enough to robustly constrain the photometric redshift of this source. In addition to this, the probability to find an object at this redshift in an area of about $232$ arcmin$^2$ is less than $10^{-4}$, based on the current estimates of the luminosity function \citep[LF;][]{Finkelstein2024}. Hence, this object is likely to be at a lower $z$ than the estimate found through the SED fitting procedures. We think that this object remains of great interest, but spectroscopic observations will be necessary to distinguish between the different possibilities and to determine the true $z$ value.

\section{Conclusions} \label{conclusions}
In this work we have used \textit{Spitzer} and \Euclid observations \citep{EROPerseusOverview} of the Perseus cluster to search for HIEROs, dusty galaxies in the redshift range $3<z<6$. The \textit{Spitzer} selection allows us to be certain about the non-spurious nature of these detections, thus avoiding any problem with artefacts in the \Euclid images. 

In the selection of our final sample we have been conservative, both applying a visual check and removing the possible contaminants to assure the robustness of the candidates. Therefore, the number of objects identified should be considered a lower limit. We performed a similar study using the $63\,\mathrm{deg}^2$ area from the first Quick Data Release \citep[Q1;][]{Q1cite}, identifying an initial sample of approximately $30\,000$ HIEROs candidates, which was reduced to $3870$ after visual inspection and data cleaning \citep{Q1-SP016}. Despite the smaller area in the present work, the observations are deeper, allowing for more robust results. In particular, we were able to examine the SED fittings individually, which was not feasible in the Q1 study due to the sample size. This allowed us to identify distinct categories among the HIEROs candidates. Thanks to the improved data quality, we were also able to compute the GSMF for the final sample of $42$ objects, which was not possible in the Q1 analysis. The GSMF was calculated in a redshift range between $3.5$ and $5.5$ and is shown in Fig.~\ref{fig:final_MF}. The presence of objects with excessive stellar masses was discussed in Sect.~\ref{sc:Discussion}, where we evaluated different possible explanations. The findings in the work by \cite{forrest2024magaz3ne} have led us to consider different scenarios, where we keep or remove possible biases. Even under the most conservative assumptions, our results remain consistent with the interpretation that we are probing the massive end of the GSMF. This may hint that previous GSMF estimates could have been somewhat underestimated, particularly if such objects were not included in the analyses.

Further improvements are expected with the upcoming first data release (DR1), which will cover the already observed Euclid Deep Fields at significantly greater depth, allowing for tighter constraints in the SED fitting. The rarity of these objects makes it difficult to study them with JWST, having only $0.6$ deg$^2$ as the largest contiguous programme \citep{casey2023cosmos}, hence their observations require wide fields of view, possible only with the synergy between \Euclid and \textit{Spitzer}.

This study case clearly points to the necessity of delving deeper in the analyses of these sources to be able to correctly characterise them. Above all, the possible unreliability of the photometric redshift estimates retrieved calls attention to the importance of follow-up observations, especially in regards of obtaining spectroscopic data. 

\begin{acknowledgements}
\AckERO  
\AckEC  
The Cosmic Variance Calculator was written by M. Trenti \& M. Stiavelli with support from NASA JWST grant NAG5-12458.
The research activities described in this paper were carried out with contribution of the Next Generation EU funds within the National Recovery and Resilience Plan (PNRR), Mission 4 -- Education and Research, Component 2 -- From Research to Business (M4C2), Investment Line 3.1 -- Strengthening and creation of Research Infrastructures, Project IR0000034 -- “STILES -- Strengthening the Italian Leadership in ELT and SKA”. 
The results obtained in this paper are based on observations obtained with MegaPrime/MegaCam, a joint project of CFHT and CEA/DAPNIA, at the Canada-France-Hawaii Telescope (CFHT) which is operated by the National Research Council (NRC) of Canada, the Institut National des Science de l'Univers of the Centre National de la Recherche Scientifique (CNRS) of France, and the University of Hawaii. The observations at the Canada-France-Hawaii Telescope were performed with care and respect from the summit of Maunakea which is a significant cultural and historic site. 

\end{acknowledgements}

\bibliography{mybib}

\begin{appendix}

  \onecolumn 
\section{Image cutouts and SED fits} \label{apdx:A}

Figure~\ref{fig:cutouts} shows cutouts of all our $42$ sources. For each object, we present both \Euclid and \textit{Spitzer} images. The SED fittings obtained by \texttt{Bagpipes} of all our sample are reported in Fig.~\ref{fig:SED_fitting}, along with the main physical quantities estimates retrieved, reported in Table~\ref{tab:bagpipes_results}.

\begin{figure*} 
    \centering
     \begin{minipage}{18cm}
        \centering
    \includegraphics[width=0.9\linewidth, trim={0 63 0 42},clip, keepaspectratio]{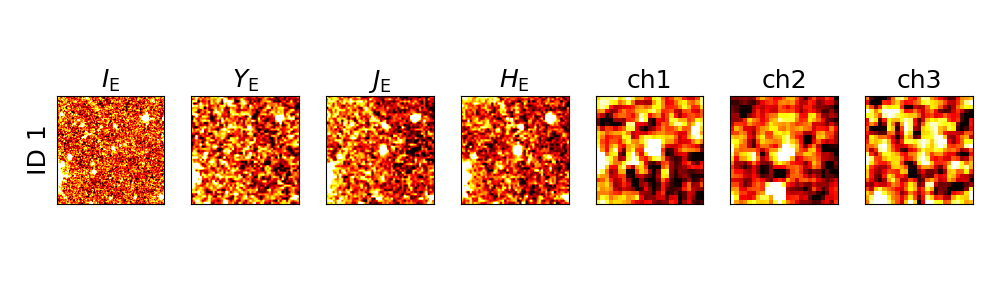}
    
    \end{minipage}
    
     \vspace{0.0cm}

     \begin{minipage}{18cm}
        \centering
         \includegraphics[width=0.9\linewidth, trim={0 63 0 63},clip, keepaspectratio]{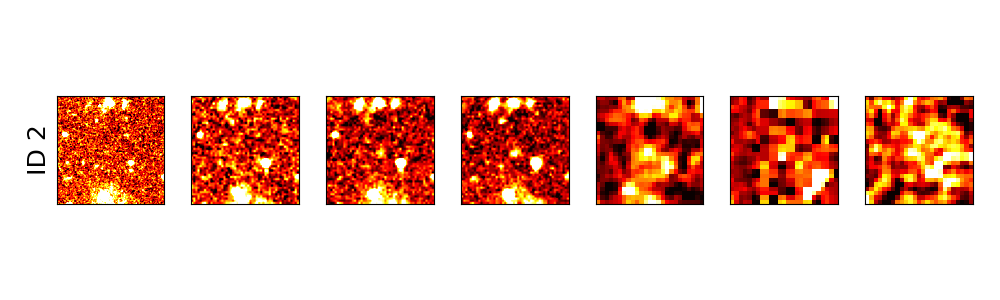}
    \end{minipage}

    \vspace{0.0cm}
     \begin{minipage}{18cm}
        \centering
         \includegraphics[width=0.9\linewidth, trim={0 63 0 63},clip, keepaspectratio]{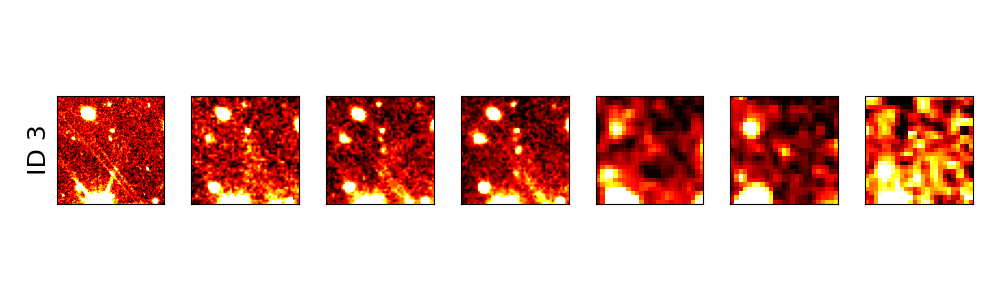}
    \end{minipage}

    \vspace{0.0cm} 

     \begin{minipage}{18cm}
        \centering
        \includegraphics[width=0.9\linewidth, trim={0 63 0 63},clip, keepaspectratio]{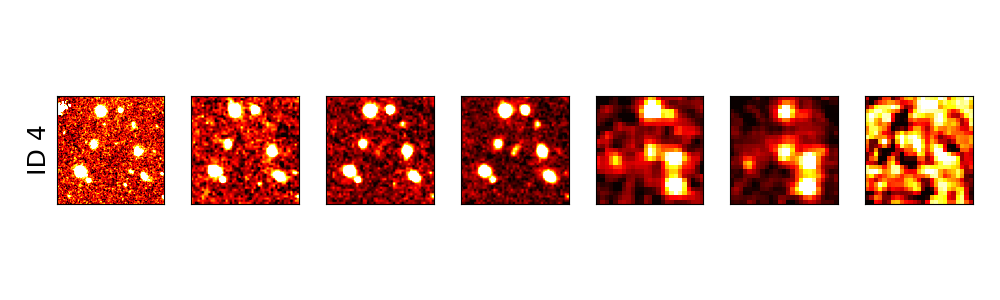}
\end{minipage}
\begin{minipage}{18cm}
        \centering
        \includegraphics[width=0.9\linewidth, trim={0 63 0 63},clip, keepaspectratio]{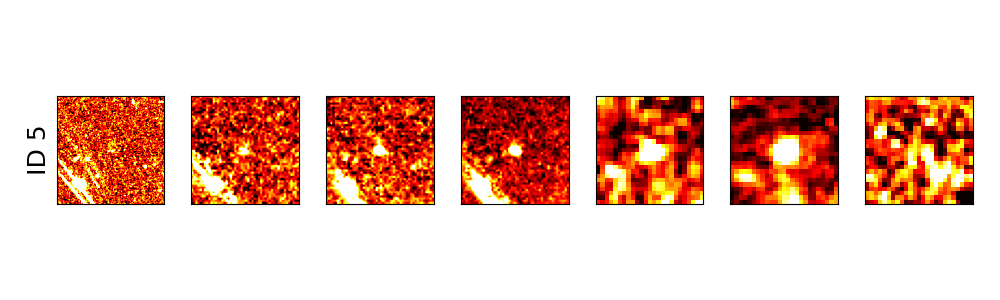}
\end{minipage}

    \vspace{0.0cm} 

     \begin{minipage}{18cm}
        \centering
        \includegraphics[width=0.9\linewidth, trim={0 63 0 63},clip, keepaspectratio]{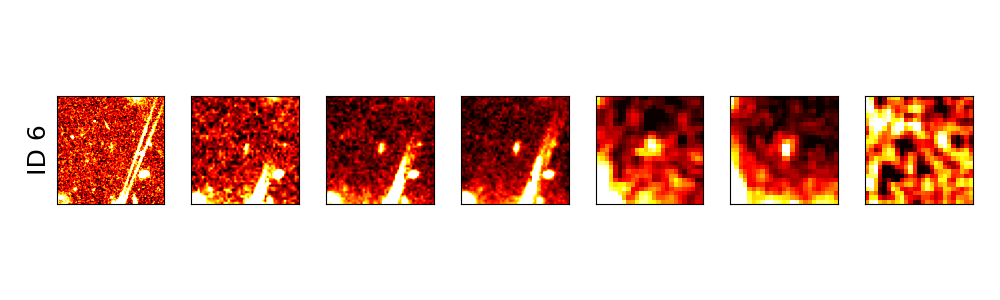}
\end{minipage}

    \vspace{0.0cm} 

     \begin{minipage}{18cm}
        \centering
        \includegraphics[width=0.9\linewidth, trim={0 63 0 63},clip, keepaspectratio]{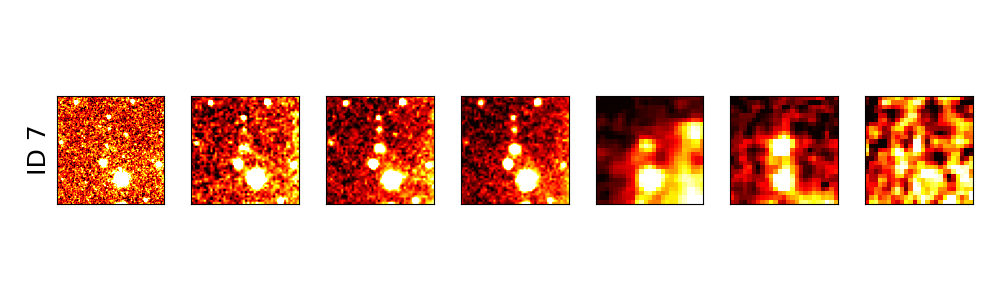}
\end{minipage}

    \vspace{0.0cm} 

     \begin{minipage}{18cm}
        \centering\includegraphics[width=0.9\linewidth, trim={0 63 0 63},clip, keepaspectratio]{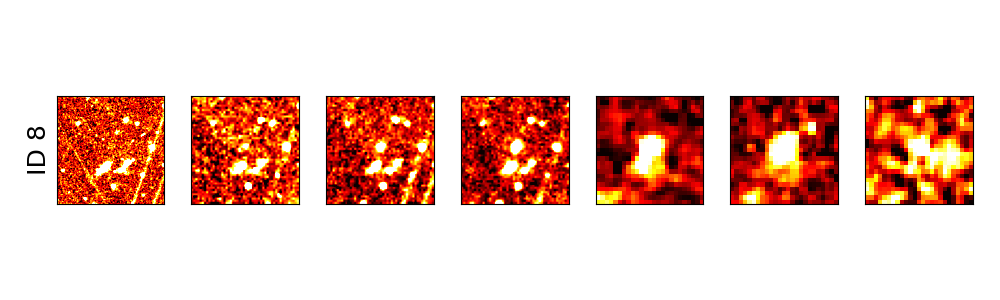}
\end{minipage}
     \begin{minipage}{18cm}
        \centering
        \includegraphics[width=0.9\linewidth, trim={0 63 0 63},clip, keepaspectratio]{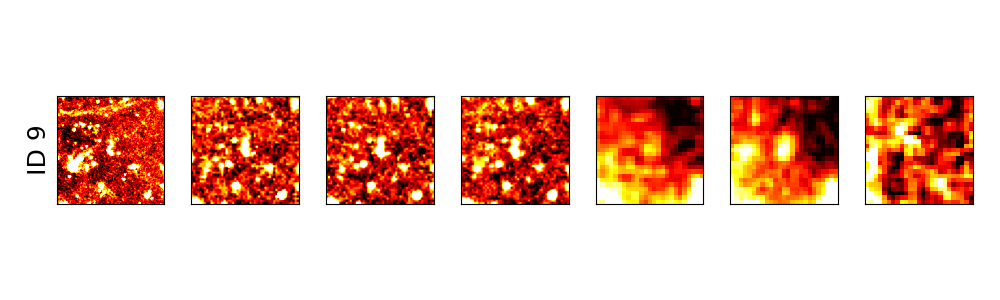}
\end{minipage}

    \vspace{0.0cm}

     \begin{minipage}{18cm}
        \centering
        \includegraphics[width=0.9\linewidth, trim={0 63 0 63},clip, keepaspectratio]{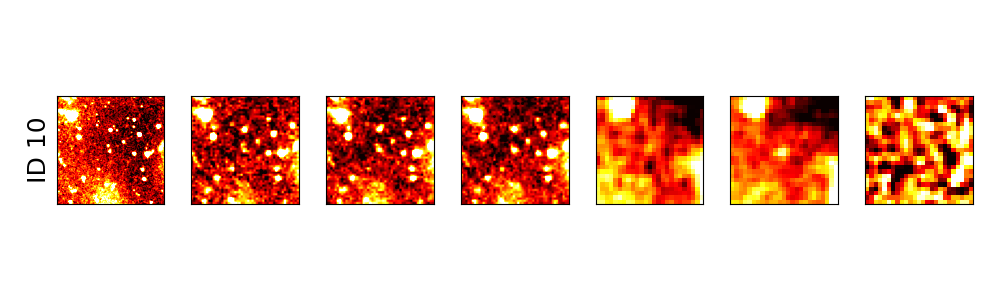}
\end{minipage}

    \vspace{0.0cm}

     \begin{minipage}{18cm}
        \centering
        \includegraphics[width=0.9\linewidth, trim={0 63 0 63},clip, keepaspectratio]{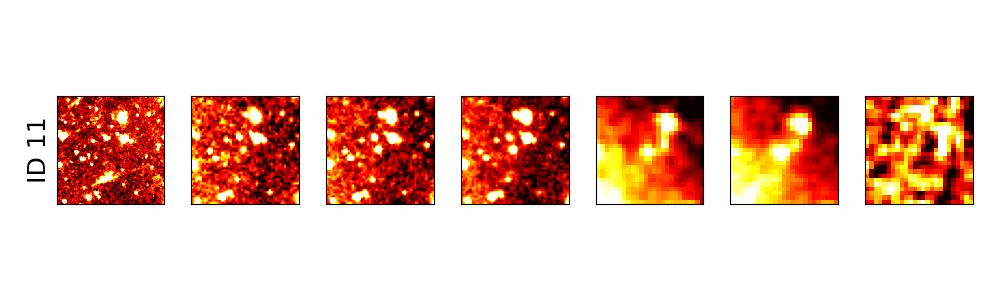}
\end{minipage}
\caption{Cutouts of all $42$ sources present in our final sample. ID $=13$ has already been shown in the main text. For IDs $ 1$ to $14$ and ID $=42, 43$, the sources belong to the first pointing, therefore having ch3 and not ch4. From ID $= 15$ until $43$, the objects are observed in ch4 and not in ch3; some of them are missing ch1 data because of the different image sizes (see the explanation in Sect.~\ref{photom_analyses}). These cutouts are $15'' \times 15''$.} 
 \label{fig:cutouts}
\end{figure*}
\begin{figure*}
\addtocounter{figure}{-1}
    \vspace{0.0cm} 

     \begin{minipage}{18cm}
        \centering
        \includegraphics[width=0.9\linewidth, trim={0 63 0 63},clip, keepaspectratio]{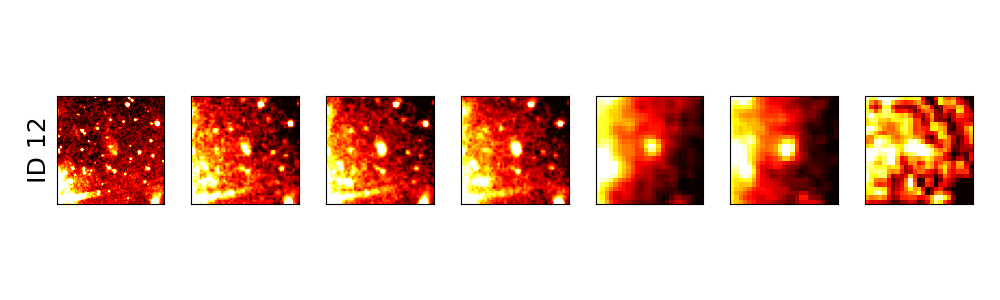}
\end{minipage}

    \vspace{0.0cm} 

     \begin{minipage}{18cm}
        \centering
        \includegraphics[width=0.9\linewidth, trim={0 63 0 63},clip, keepaspectratio]{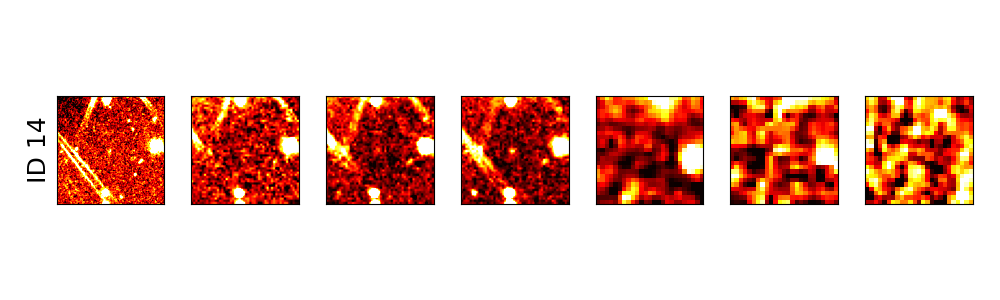}
\end{minipage}

    \vspace{0.0cm}

     \begin{minipage}{18cm}
        \centering
        \includegraphics[width=0.9\linewidth, trim={0 63 0 42},clip, keepaspectratio]{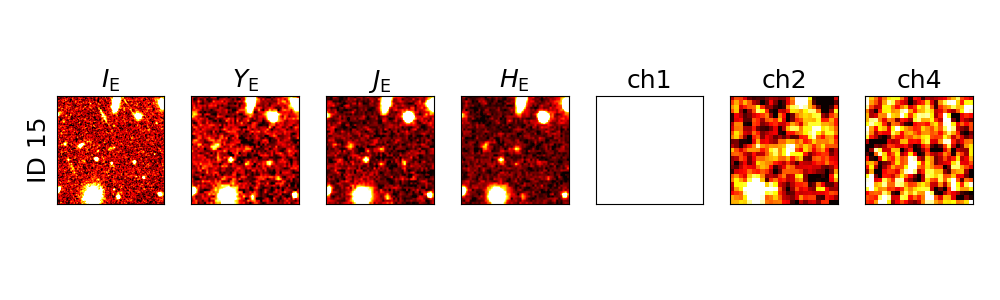}
\end{minipage}

    \vspace{0.0cm}

     \begin{minipage}{18cm}
        \centering
        \includegraphics[width=0.9\linewidth, trim={0 63 0 63},clip, keepaspectratio]{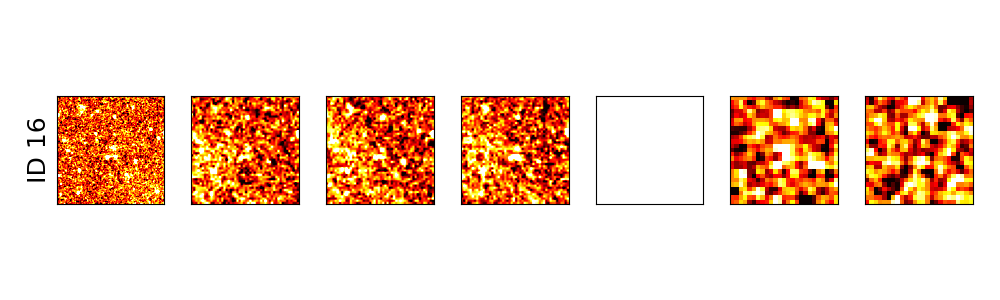}
\end{minipage}

    \vspace{0.0cm} 

     \begin{minipage}{18cm}
        \centering
        \includegraphics[width=0.9\linewidth, trim={0 63 0 63},clip, keepaspectratio]{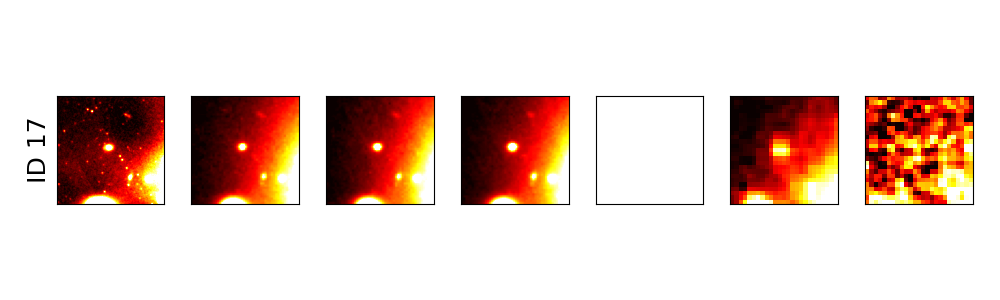}
\end{minipage}

     \begin{minipage}{18cm}
        \centering
        \includegraphics[width=0.9\linewidth, trim={0 63 0 63},clip, keepaspectratio]{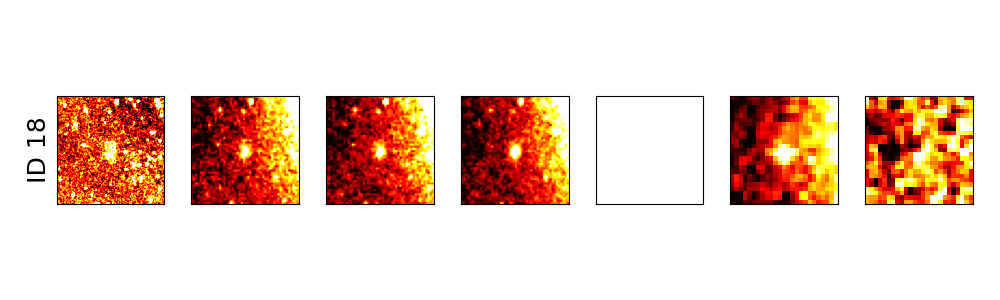}
\end{minipage}
    \vspace{0.0cm} 

     \begin{minipage}{18cm}
        \centering
        \includegraphics[width=0.9\linewidth, trim={0 63 0 63},clip, keepaspectratio]{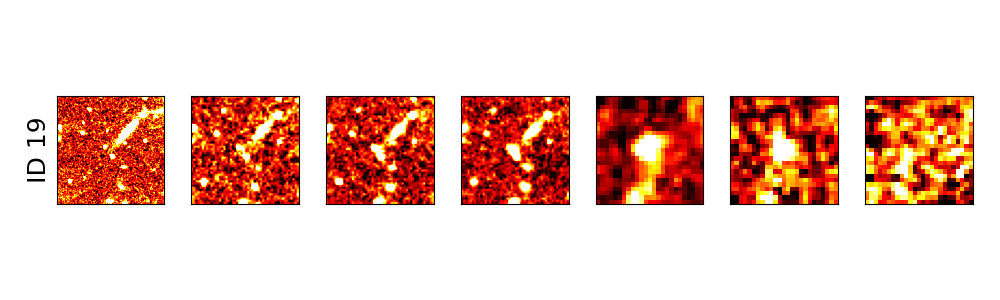}
\end{minipage}

    \vspace{0.0cm}

     \begin{minipage}{18cm}
        \centering
        \includegraphics[width=0.9\linewidth, trim={0 63 0 63},clip, keepaspectratio]{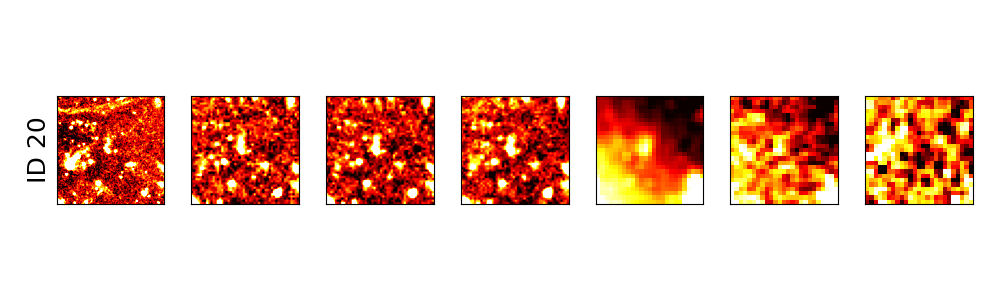}
\end{minipage}

    \vspace{0.0cm} 

     \begin{minipage}{18cm}
        \centering
        \includegraphics[width=0.9\linewidth, trim={0 63 0 63},clip, keepaspectratio]{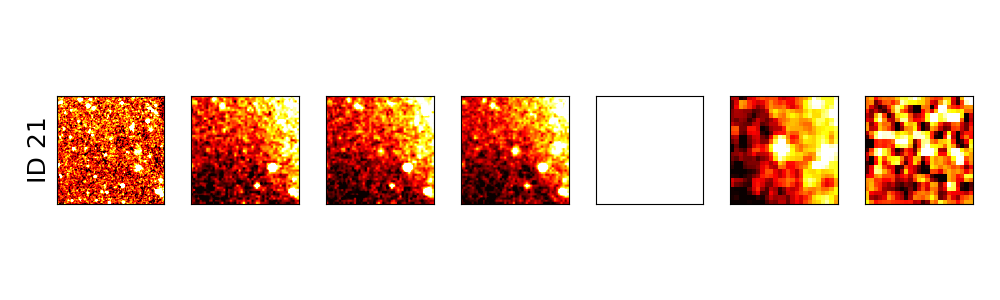}
\end{minipage}

    \vspace{0.0cm} 

     \begin{minipage}{18cm}
        \centering
        \includegraphics[width=0.9\linewidth, trim={0 63 0 63},clip, keepaspectratio]{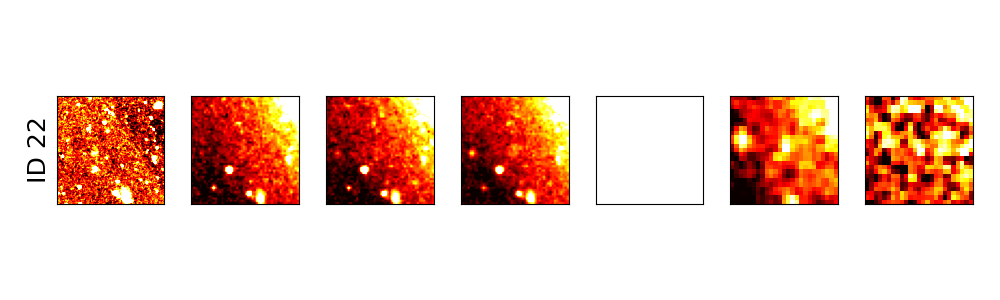}
\end{minipage}
    \vspace{0.0cm}

     \begin{minipage}{18cm}
        \centering
        \includegraphics[width=0.9\linewidth, trim={0 63 0 63},clip, keepaspectratio]{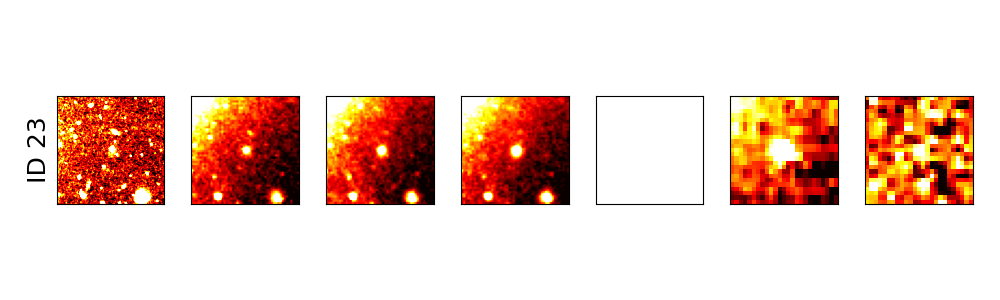}
\end{minipage}

    \vspace{0.0cm}
     \begin{minipage}{18cm}
        \centering
        \includegraphics[width=0.9\linewidth, trim={0 63 0 63},clip, keepaspectratio]{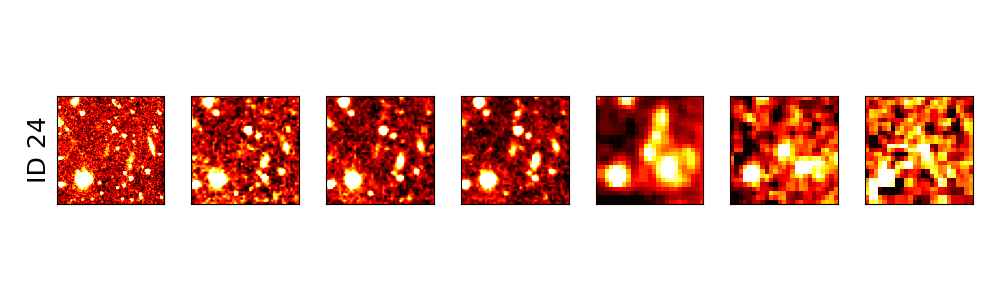}
\end{minipage}
\caption{Continued.} 
\end{figure*}
\begin{figure*}
\addtocounter{figure}{-1}
    \vspace{0.0cm}

     \begin{minipage}{18cm}
        \centering
        \includegraphics[width=0.9\linewidth, trim={0 63 0 63},clip, keepaspectratio]{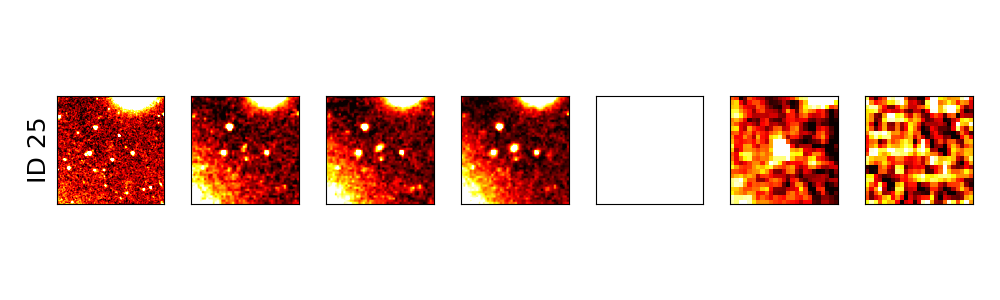}
\end{minipage}

    \vspace{0.0cm}

     \begin{minipage}{18cm}
        \centering
        \includegraphics[width=0.9\linewidth, trim={0 63 0 63},clip, keepaspectratio]{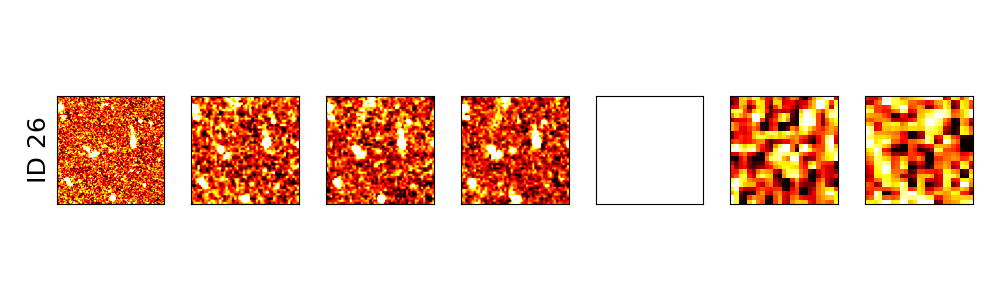}
\end{minipage}

    \vspace{0.0cm} 

     \begin{minipage}{18cm}
        \centering
        \includegraphics[width=0.9\linewidth, trim={0 63 0 63},clip, keepaspectratio]{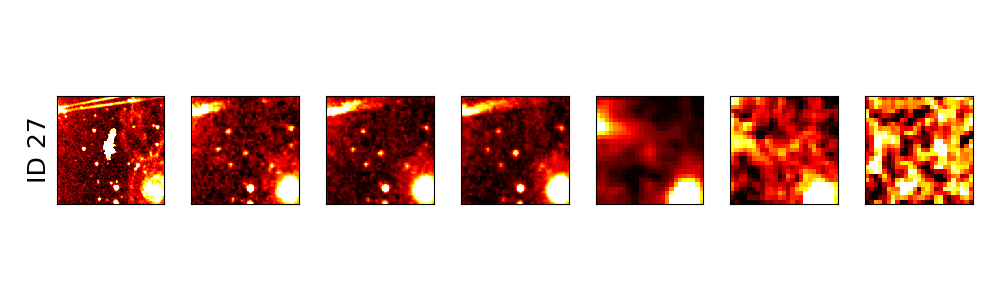}
\end{minipage}
    \vspace{0.0cm} 

     \begin{minipage}{18cm}
        \centering
        \includegraphics[width=0.9\linewidth, trim={0 63 0 63},clip, keepaspectratio]{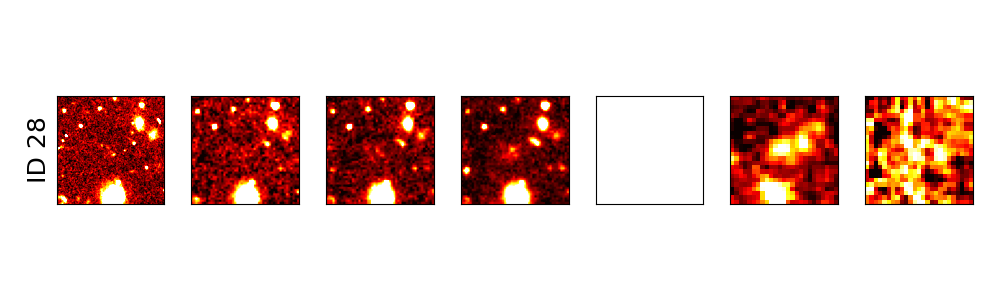}
\end{minipage}

     \begin{minipage}{18cm}
        \centering
        \includegraphics[width=0.9\linewidth, trim={0 63 0 63},clip, keepaspectratio]{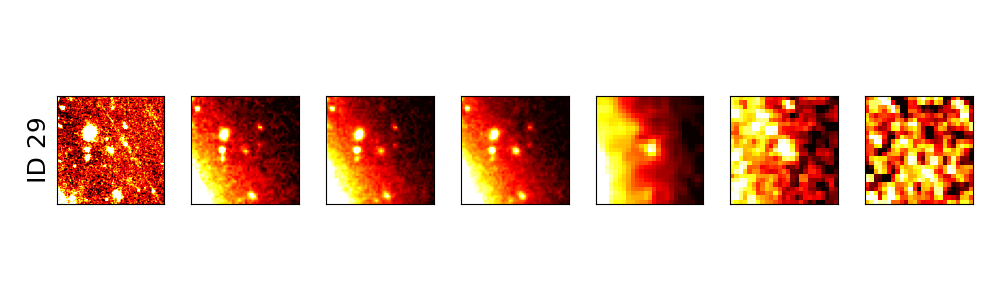}
\end{minipage}

    \vspace{0.0cm} 

     \begin{minipage}{18cm}
        \centering
        \includegraphics[width=0.9\linewidth, trim={0 63 0 63},clip, keepaspectratio]{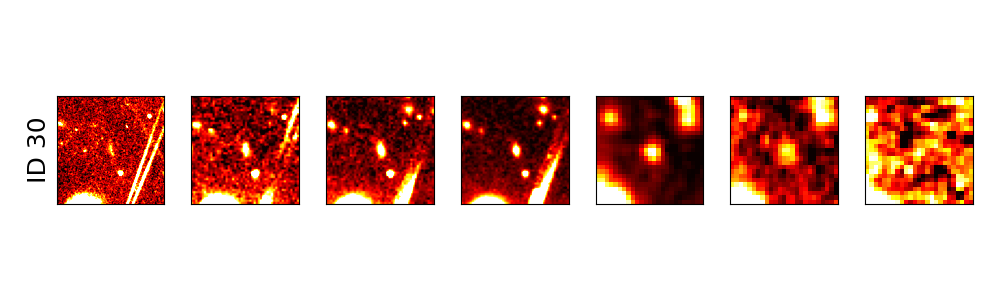}
\end{minipage}

    \vspace{0.0cm} 

     \begin{minipage}{18cm}
        \centering
        \includegraphics[width=0.9\linewidth, trim={0 63 0 63},clip, keepaspectratio]{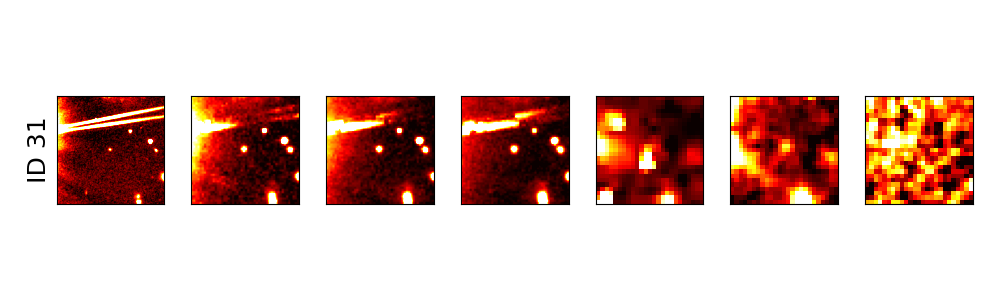}
\end{minipage}

    \vspace{0.0cm} 

     \begin{minipage}{18cm}
        \centering
        \includegraphics[width=0.9\linewidth, trim={0 63 0 63},clip, keepaspectratio]{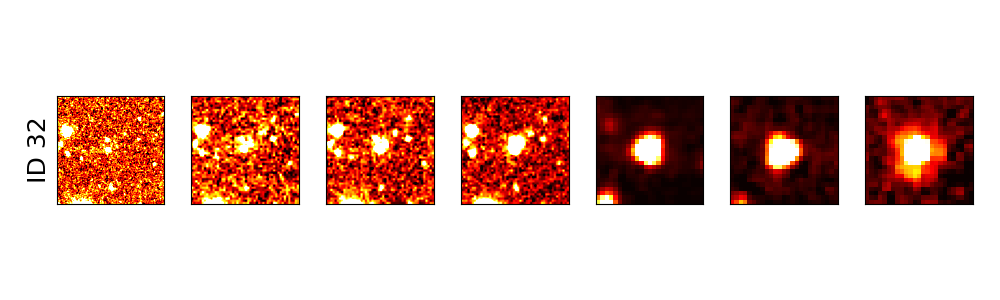}
\end{minipage}
    \vspace{0.0cm} 

     \begin{minipage}{18cm}
        \centering
        \includegraphics[width=0.9\linewidth, trim={0 63 0 63},clip, keepaspectratio]{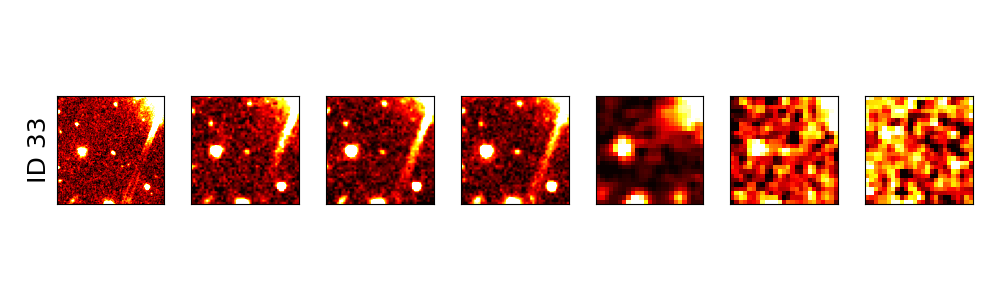}
\end{minipage}

    \vspace{0.0cm} 

     \begin{minipage}{18cm}
        \centering
        \includegraphics[width=0.9\linewidth, trim={0 63 0 63},clip, keepaspectratio]{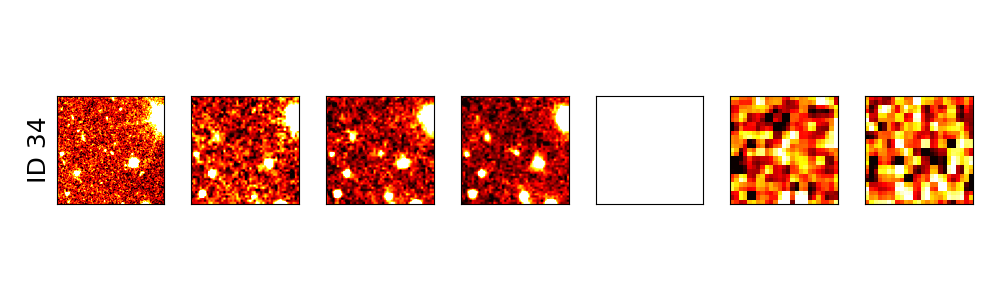}
\end{minipage}

    \vspace{0.0cm} 

     \begin{minipage}{18cm}
        \centering
        \includegraphics[width=0.9\linewidth, trim={0 63 0 63},clip, keepaspectratio]{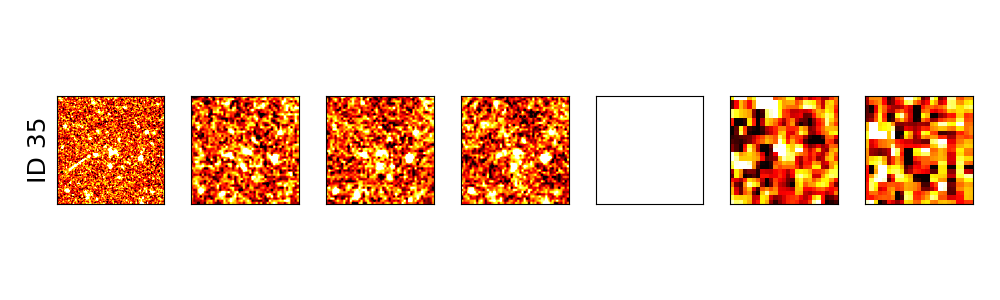}
\end{minipage}
\caption{Continued.} 
\end{figure*}
\begin{figure*}
\addtocounter{figure}{-1}
    \vspace{0.0cm} 

     \begin{minipage}{18cm}
        \centering
        \includegraphics[width=0.9\linewidth, trim={0 63 0 63},clip, keepaspectratio]{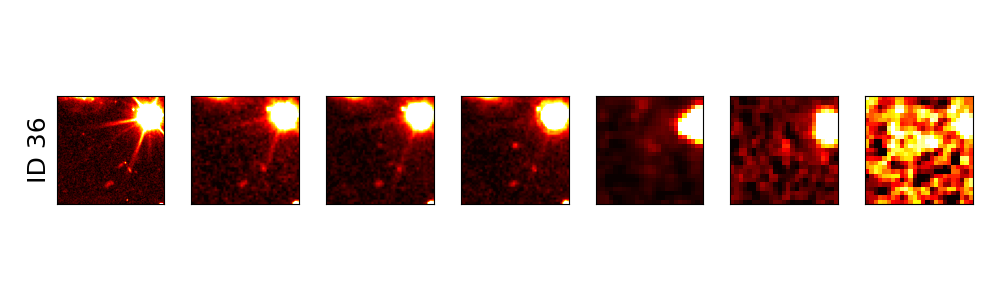}
\end{minipage}

    \vspace{0.0cm} 

     \begin{minipage}{18cm}
        \centering
        \includegraphics[width=0.9\linewidth, trim={0 63 0 63},clip, keepaspectratio]{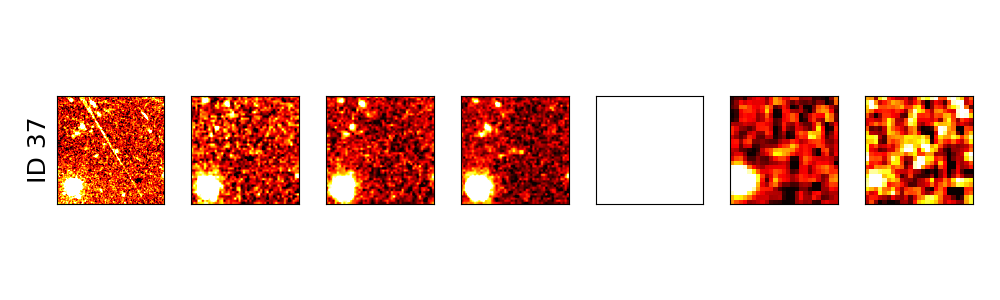}
\end{minipage}
\end{figure*}

\begin{figure*}
    \vspace{0.0cm}

     \begin{minipage}{18cm}
        \centering
        \includegraphics[width=0.9\linewidth, trim={0 63 0 63},clip, keepaspectratio]{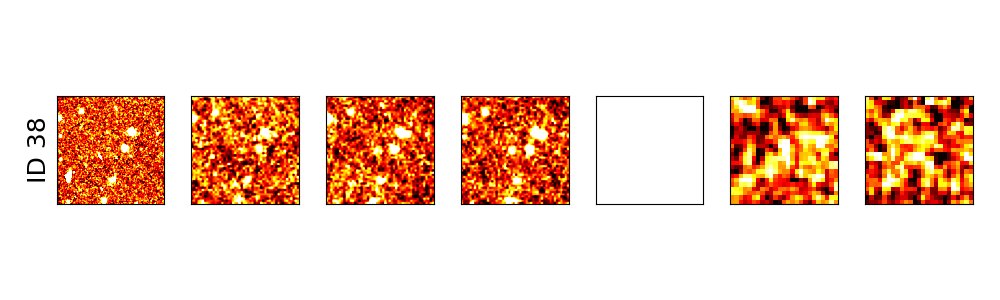}
\end{minipage}

\begin{minipage}{18cm}
        \centering
        \includegraphics[width=0.9\linewidth, trim={0 63 0 63},clip, keepaspectratio]{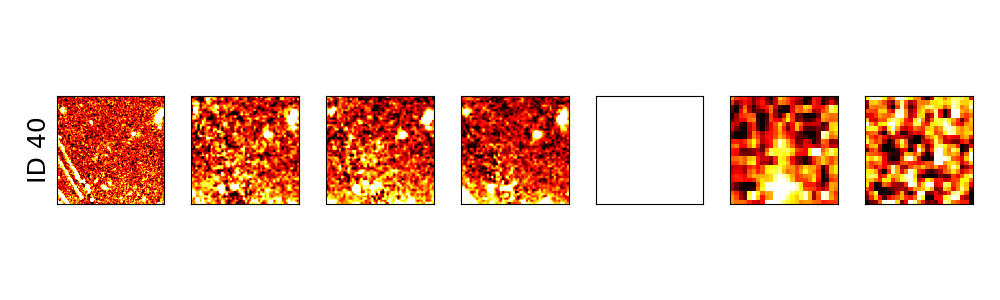}
\end{minipage}

    \vspace{0.0cm} 

     \begin{minipage}{18cm}
        \centering
        \includegraphics[width=0.9\linewidth, trim={0 63 0 63},clip, keepaspectratio]{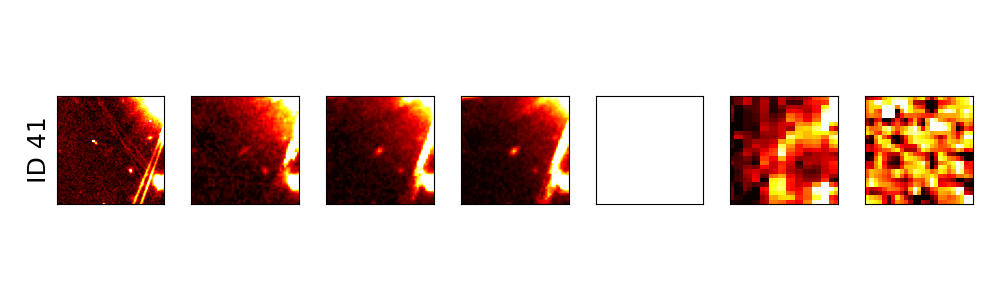}
\end{minipage}

    \vspace{0.0cm} 

     \begin{minipage}{18cm}
        \centering
        \includegraphics[width=0.9\linewidth, trim={0 63 0 42},clip, keepaspectratio]{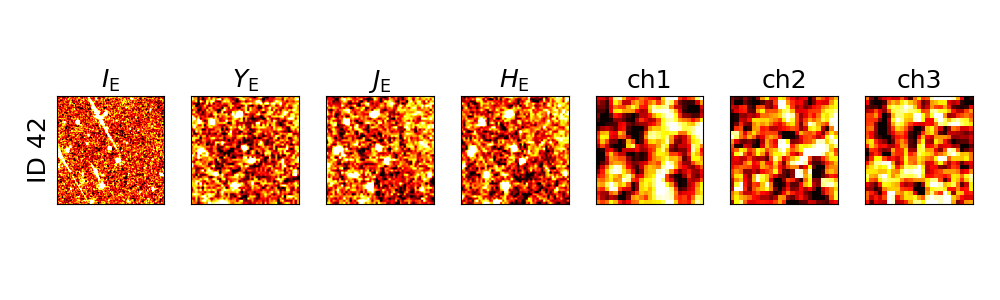}
\end{minipage}
    \vspace{0.0cm}

     \begin{minipage}{18cm}
        \centering
        \includegraphics[width=0.9\linewidth, trim={0 63 0 63},clip, keepaspectratio]{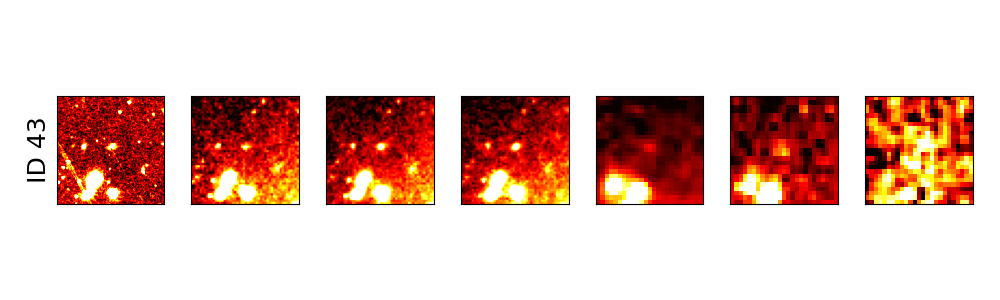}
\end{minipage}
\caption{Continued.}

\end{figure*}

\begin{figure*} 
    \centering
     \begin{minipage}{18cm}
        \centering
    \includegraphics[width=0.49\linewidth, trim={0 25 0 25},clip, keepaspectratio]{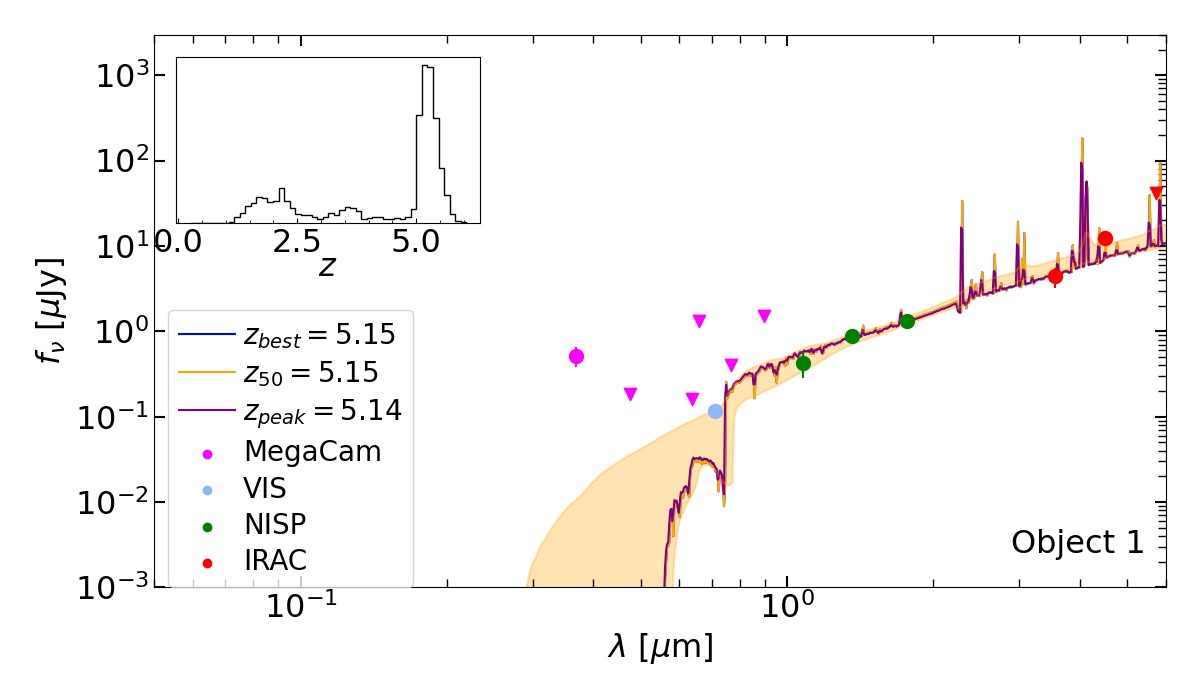}
    \includegraphics[width=0.49\linewidth, trim={0 25 0 25},clip, keepaspectratio]{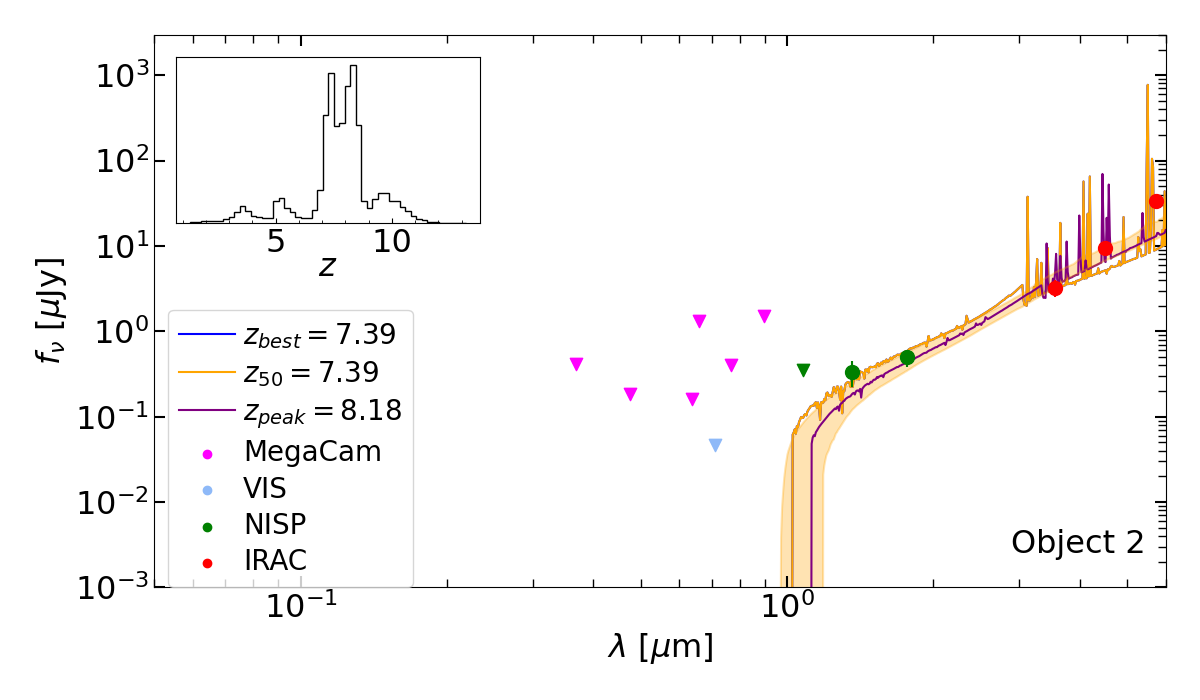}
 
    \end{minipage}
    
     \vspace{0.0cm}

     \begin{minipage}{18cm}
        \centering
         \includegraphics[width=0.49\linewidth, trim={0 25 0 25},clip, keepaspectratio]{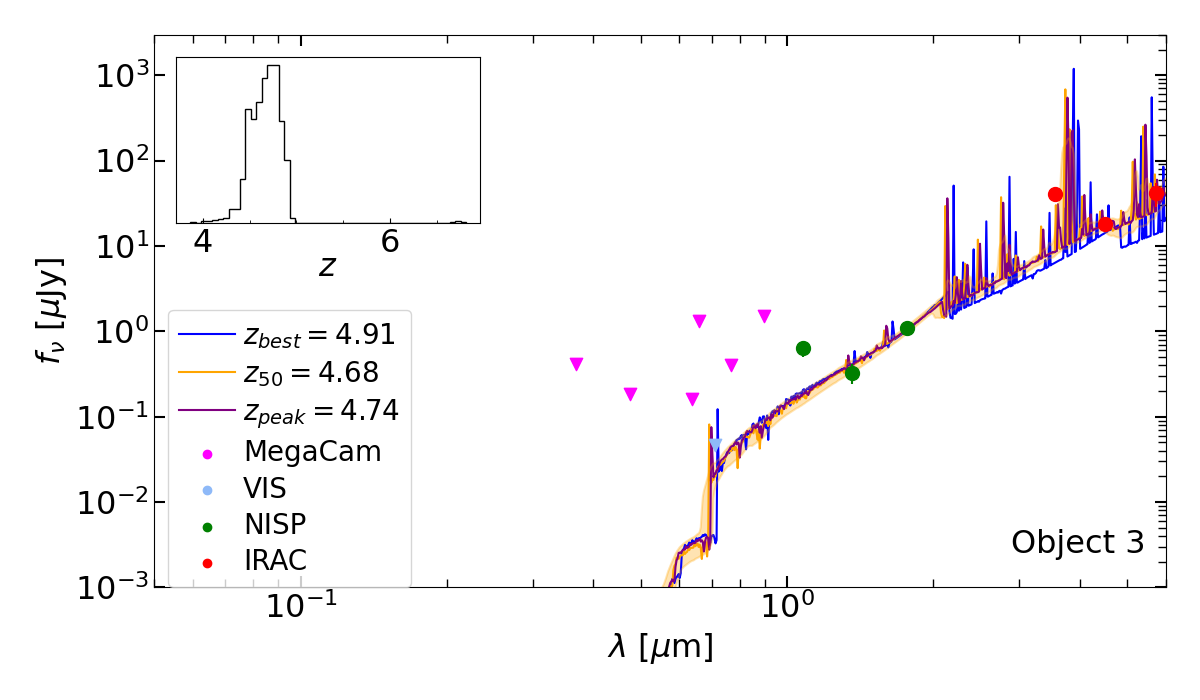}
         \includegraphics[width=0.49\linewidth, trim={0 25 0 25},clip, keepaspectratio]{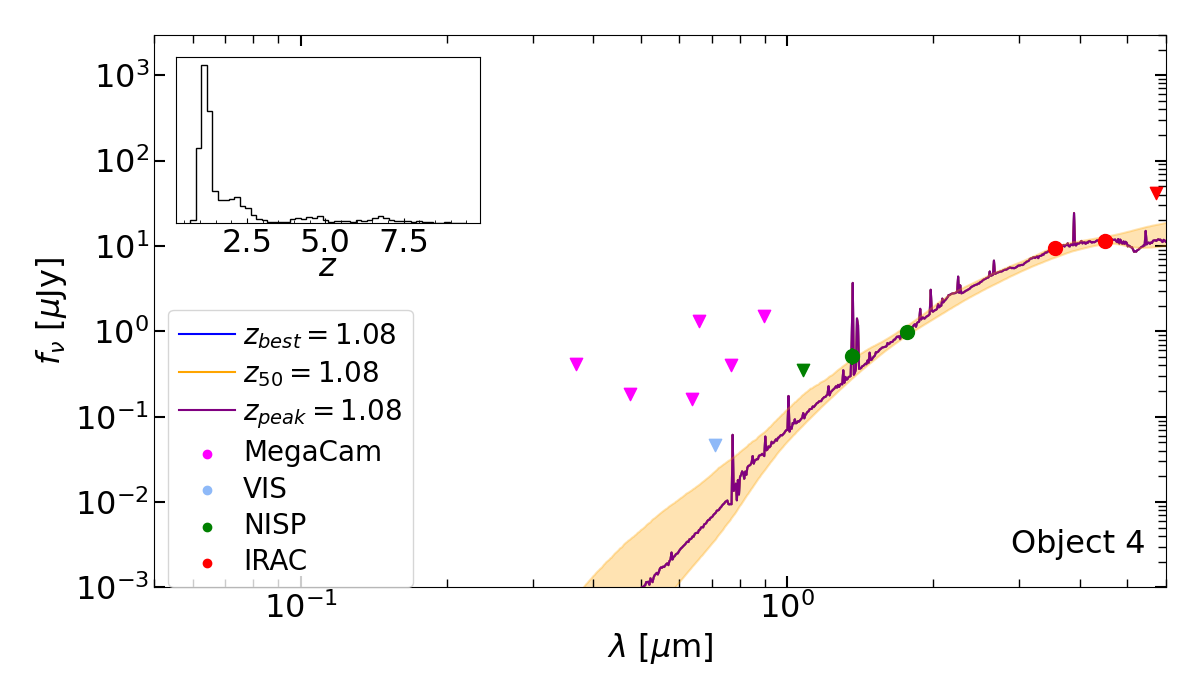}
    \end{minipage}

    \vspace{0.0cm}

     \begin{minipage}{18cm}
        \centering
         \includegraphics[width=0.49\linewidth, trim={0 25 0 25},clip, keepaspectratio]{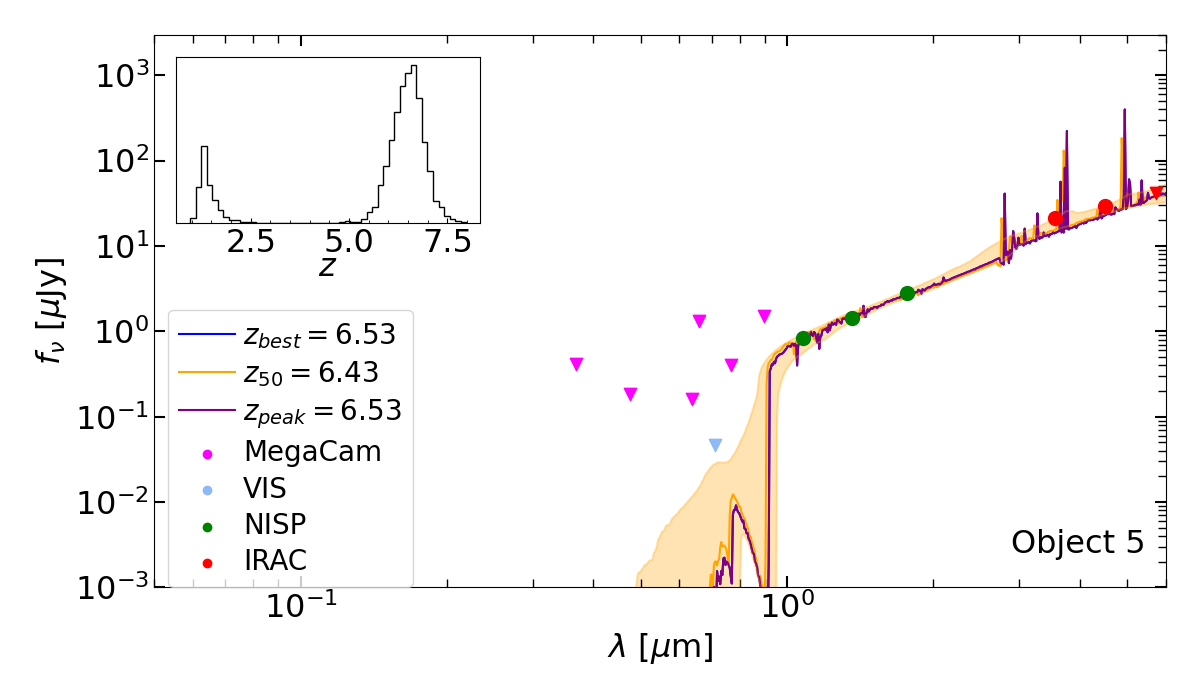}
         \includegraphics[width=0.49\linewidth, trim={0 25 0 25},clip, keepaspectratio]{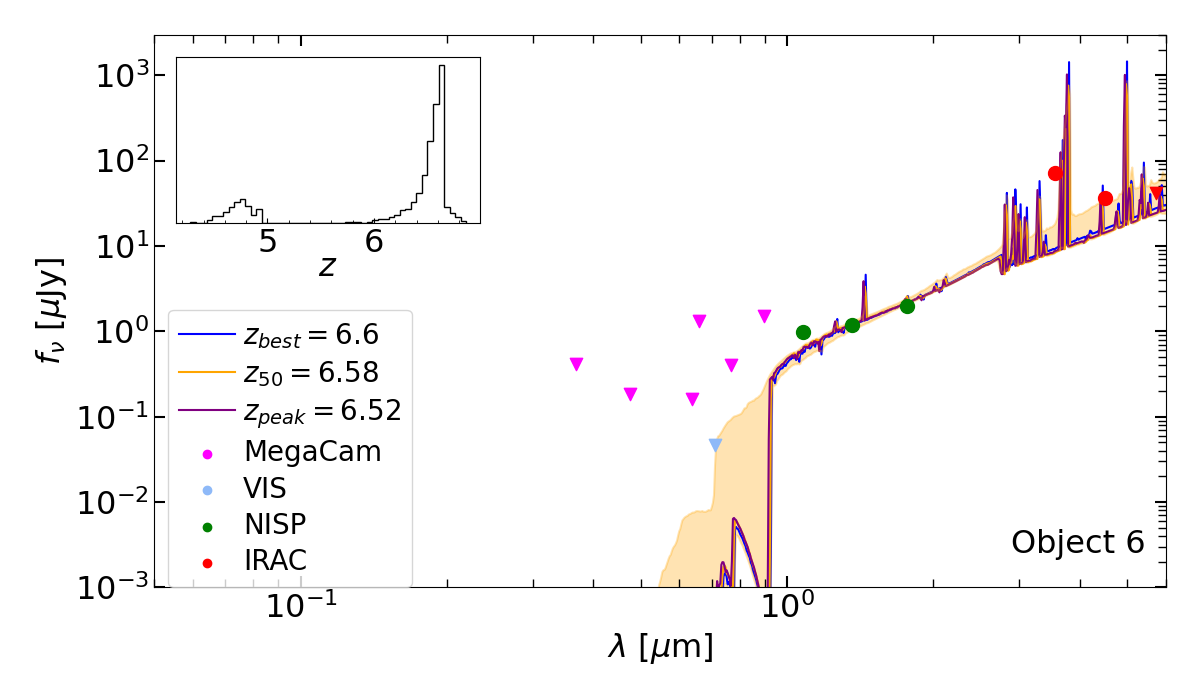}
    \end{minipage}
    \vspace{0.0cm} 

     \begin{minipage}{18cm}
        \centering
        \includegraphics[width=0.49\linewidth, trim={0 25 0 25},clip, keepaspectratio]{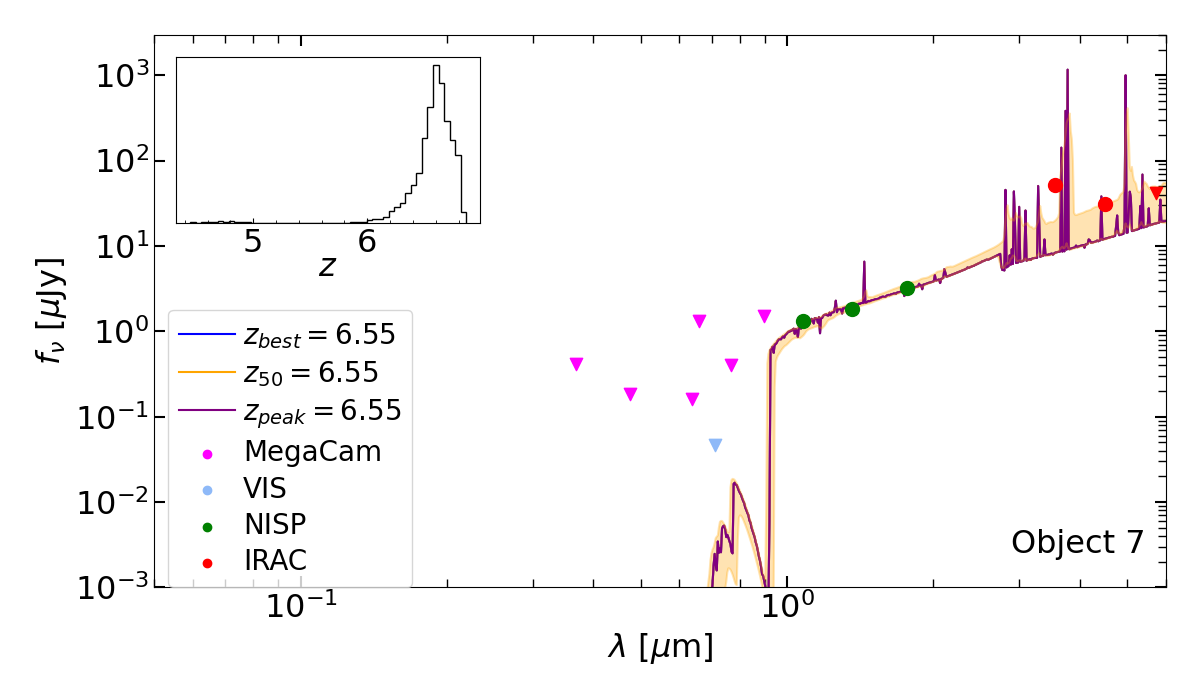}
         \includegraphics[width=0.49\linewidth, trim={0 25 0 25},clip, keepaspectratio]{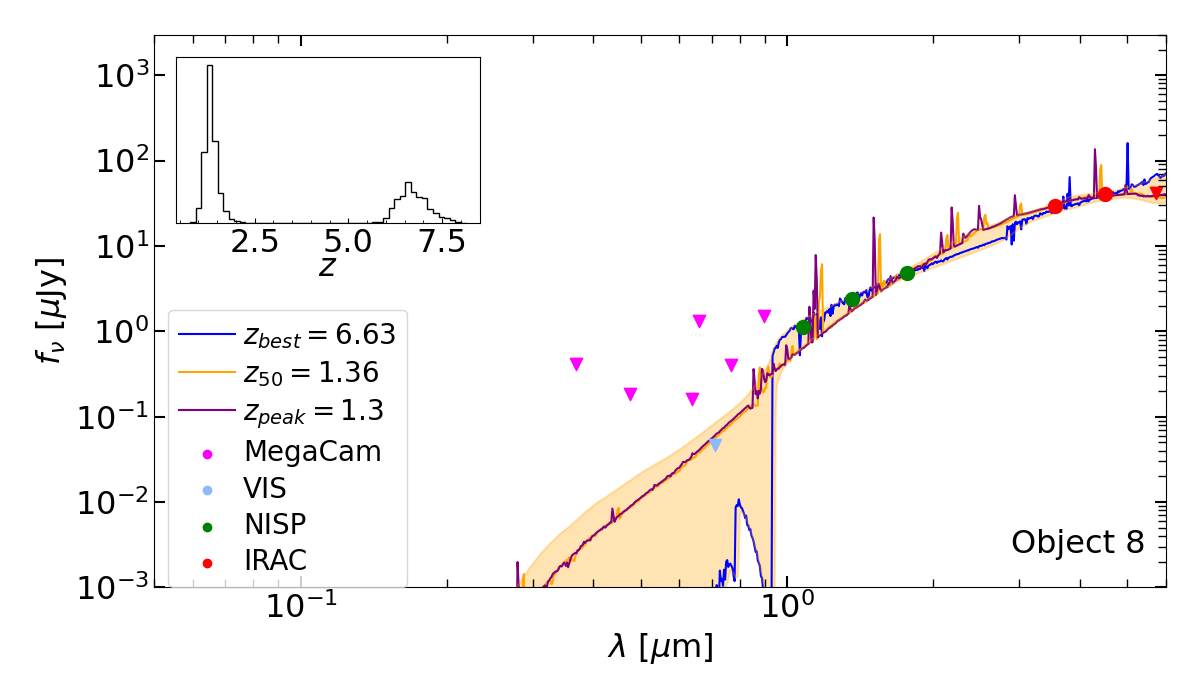}
\end{minipage}
\begin{minipage}{18cm}
        \centering
        \includegraphics[width=0.49\linewidth, trim={0 25 0 25},clip, keepaspectratio]{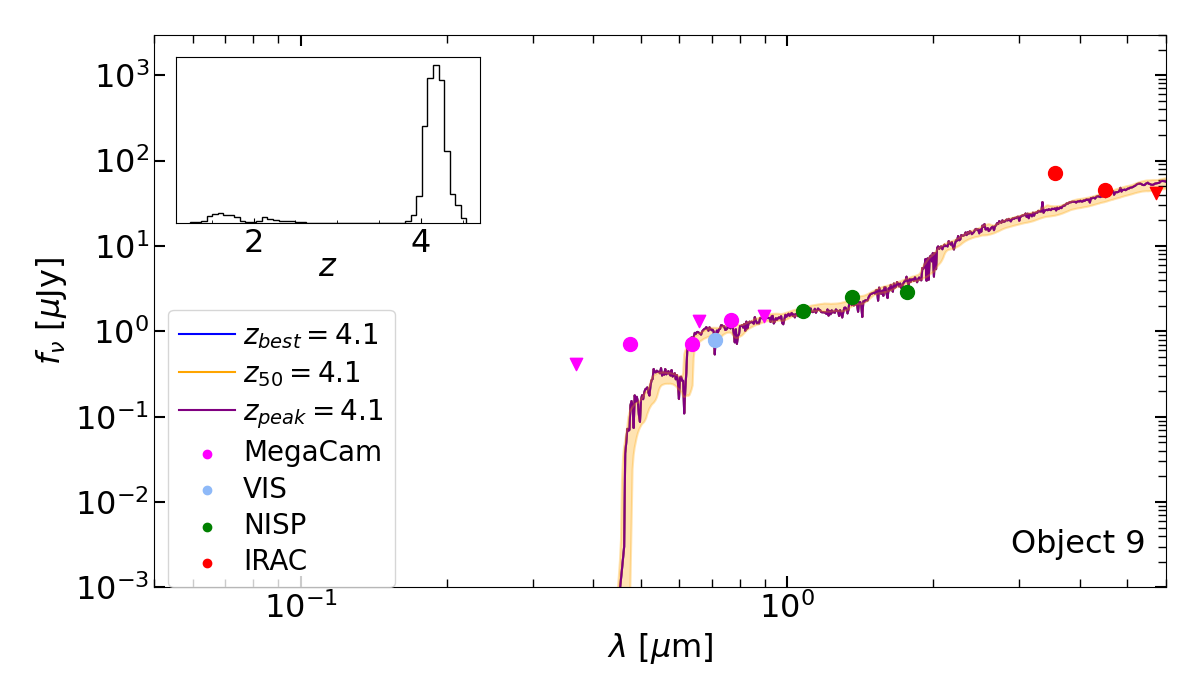}
         \includegraphics[width=0.49\linewidth, trim={0 25 0 25},clip, keepaspectratio]{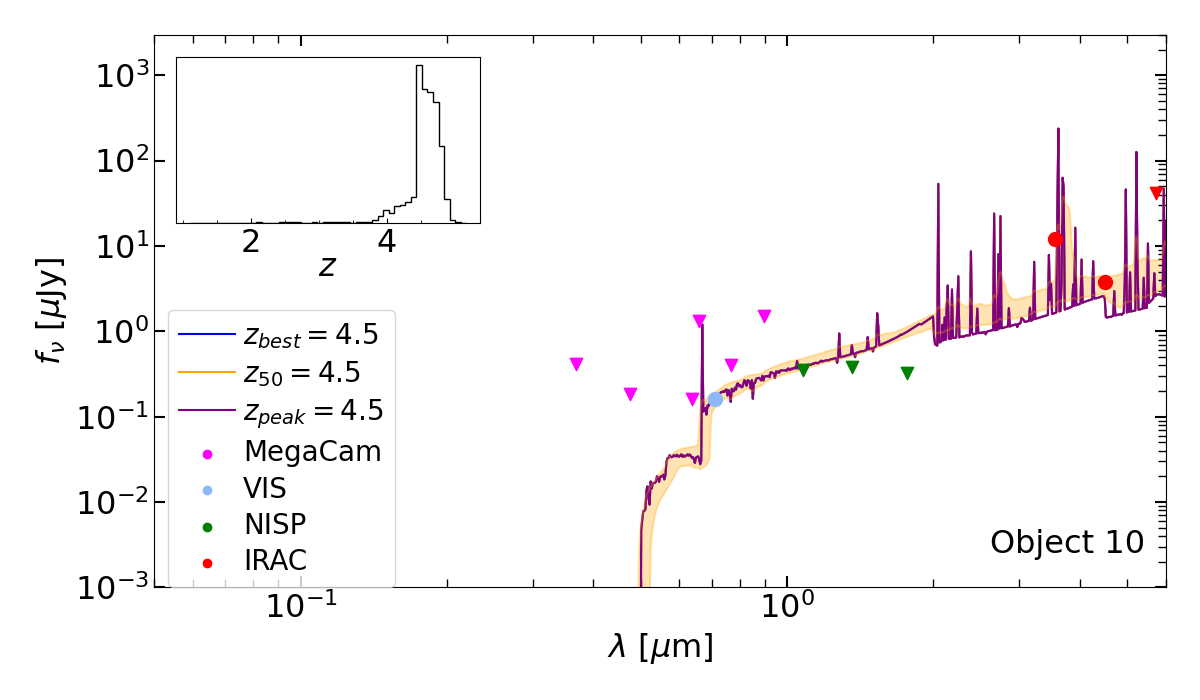}
\end{minipage}
 \caption{SED fits from \texttt{Bagpipes} in our main analyses. ID $11, 13, 28,$ and $42$ were already shown in the main text of the paper. The circles represent the photometric data points with $\textrm{S/N} > 3$, while the inverted triangles show upper limits at $3\,\sigma$. The different templates are explained in Sect.~\ref{main_SEDfitting_categories}. }

 \label{fig:SED_fitting}
\end{figure*}
\begin{figure*}
\addtocounter{figure}{-1}
   \begin{minipage}{18cm}
        \centering
        \includegraphics[width=0.49\linewidth, trim={0 25 0 25},clip, keepaspectratio]{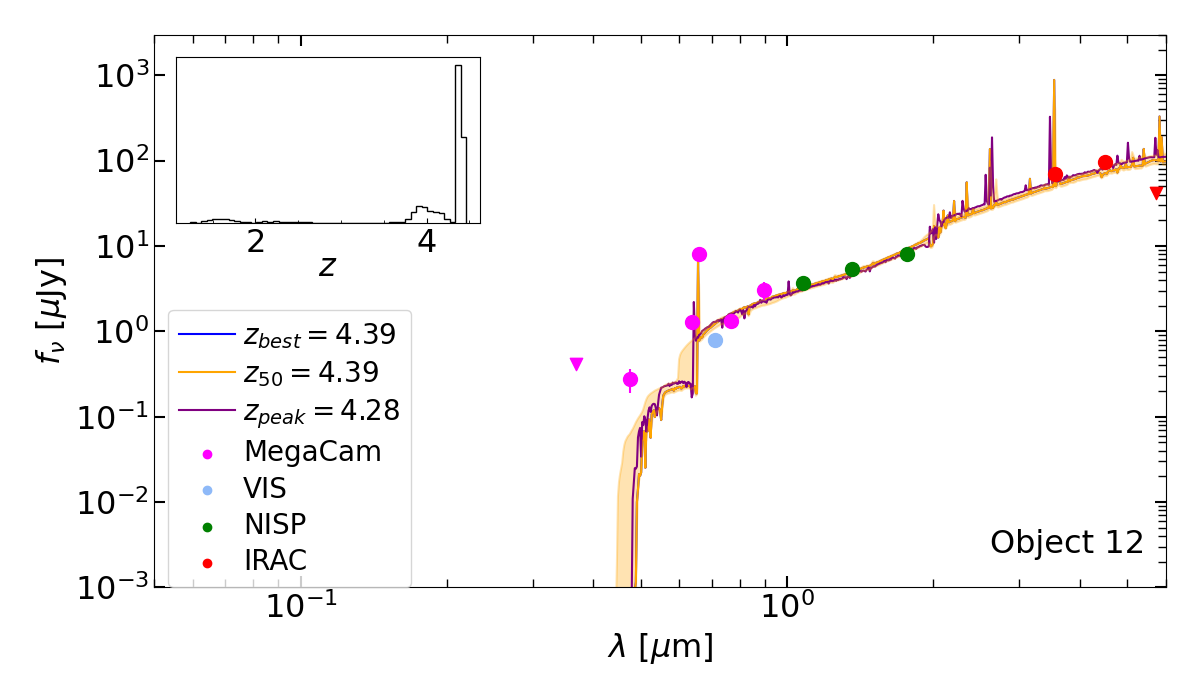}
         \includegraphics[width=0.49\linewidth, trim={0 25 0 25},clip, keepaspectratio]{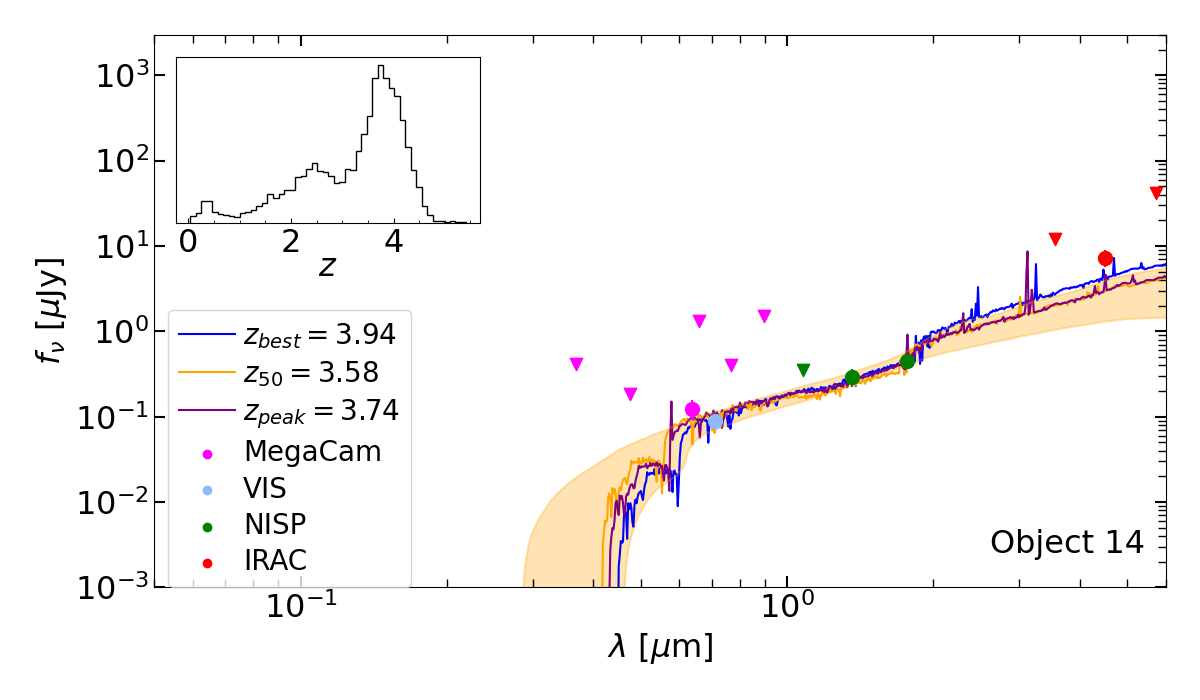}
\end{minipage}

    \vspace{0.0cm}

     \begin{minipage}{18cm}
         \includegraphics[width=0.49\linewidth, trim={0 25 0 25},clip, keepaspectratio]{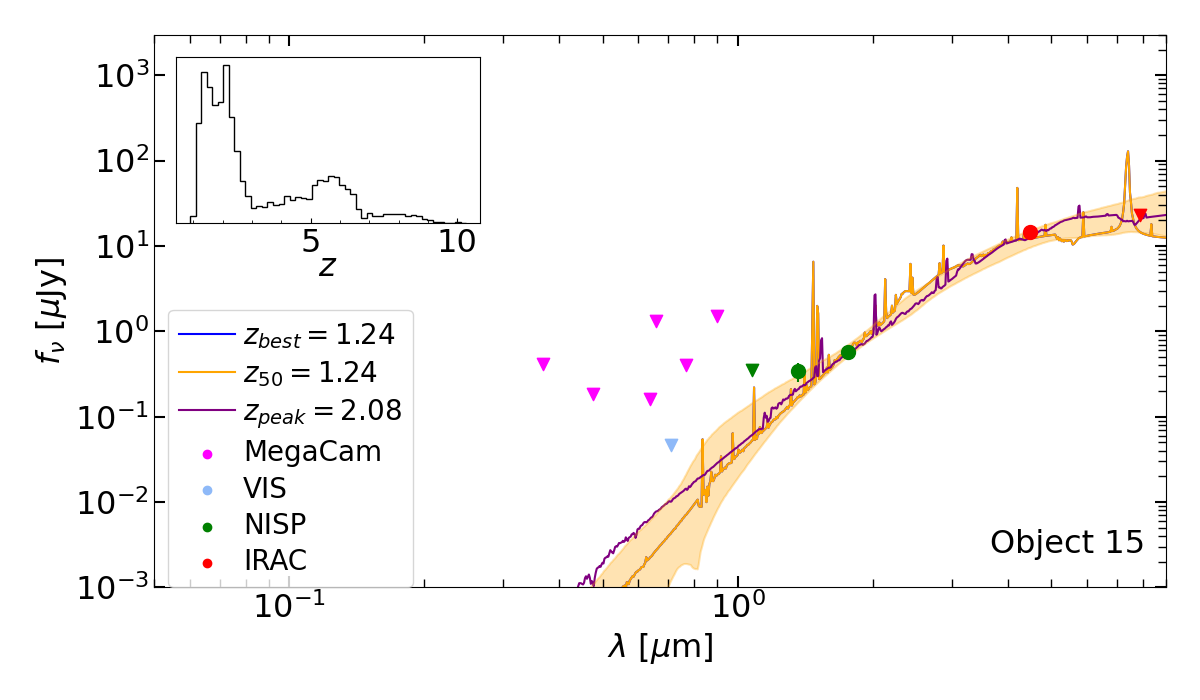}
         \includegraphics[width=0.49\linewidth, trim={0 25 0 25},clip, keepaspectratio]{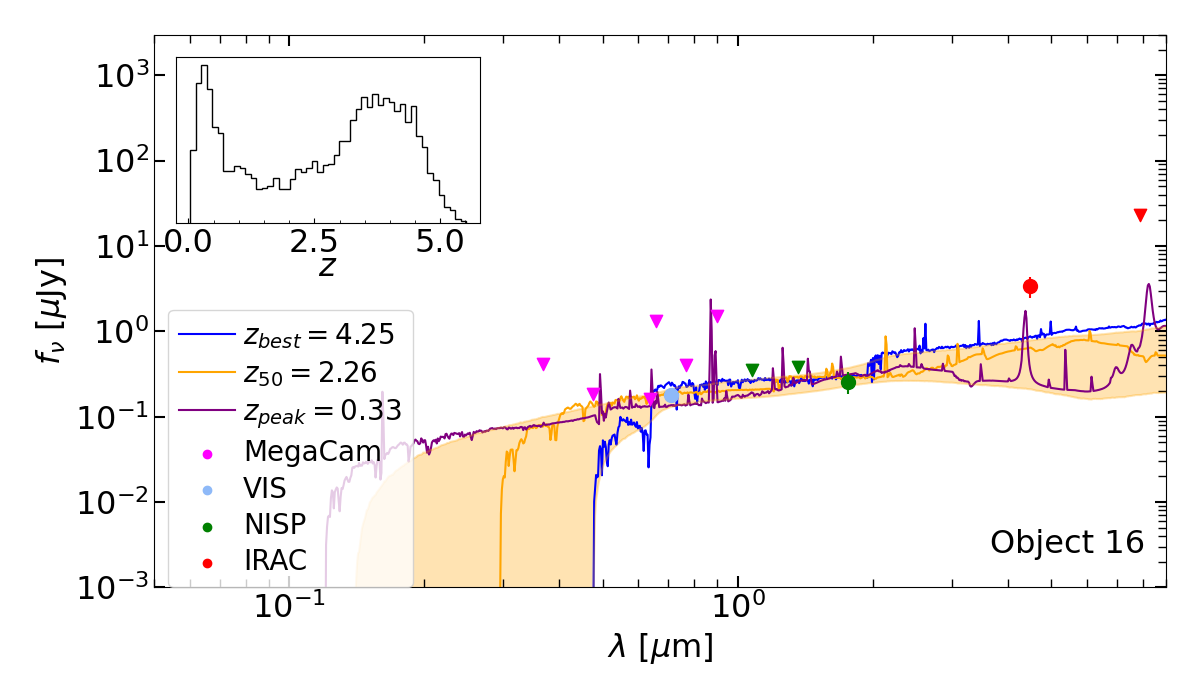}
\end{minipage}

     \begin{minipage}{18cm}
        \centering
         \includegraphics[width=0.49\linewidth, trim={0 25 0 25},clip, keepaspectratio]{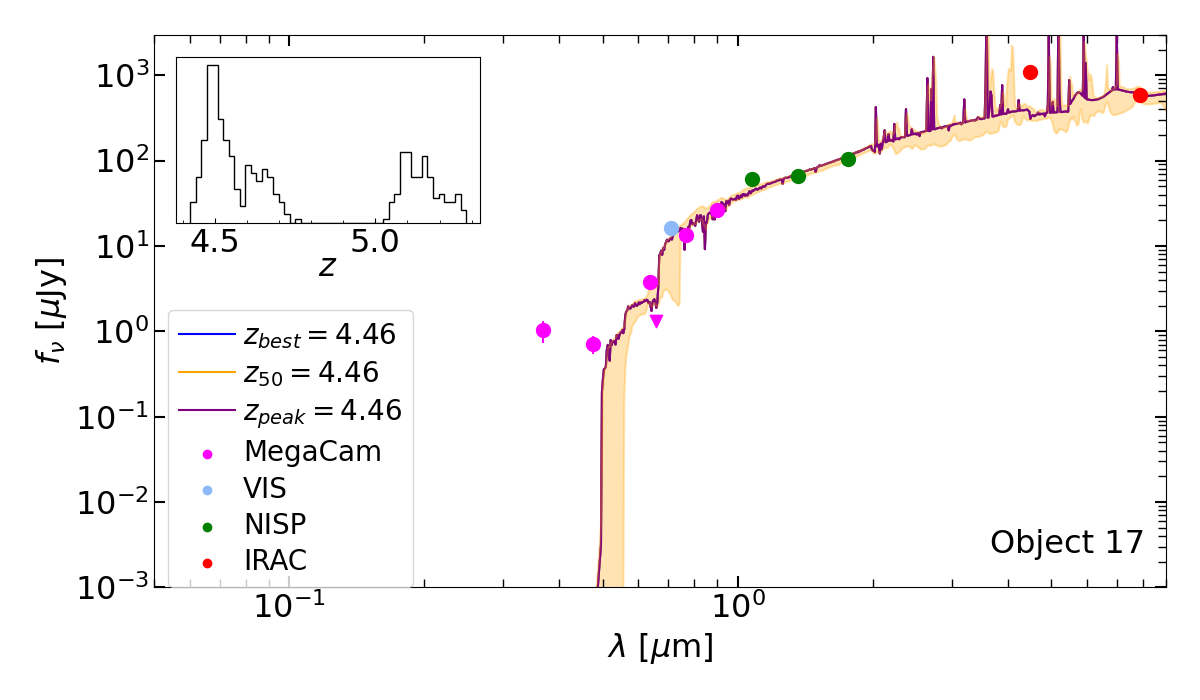}
         \includegraphics[width=0.49\linewidth, trim={0 25 0 25},clip, keepaspectratio]{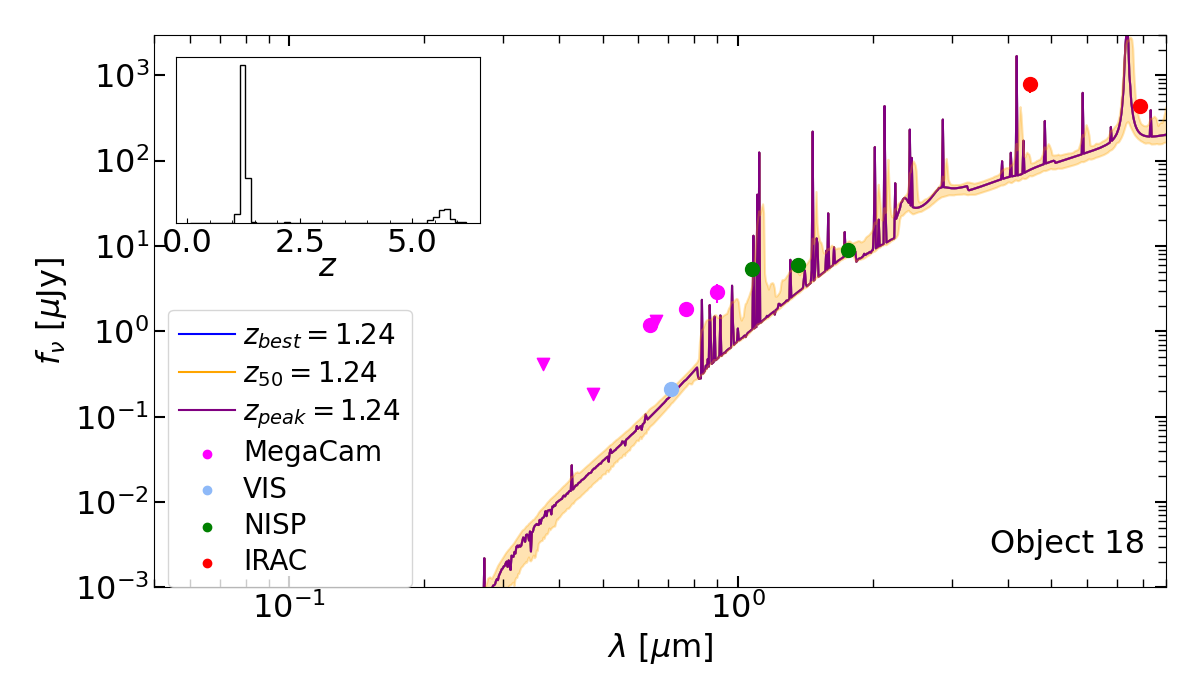}
\end{minipage}

    \vspace{0.0cm}

     \begin{minipage}{18cm}
        \centering
         \includegraphics[width=0.49\linewidth, trim={0 25 0 25},clip, keepaspectratio]{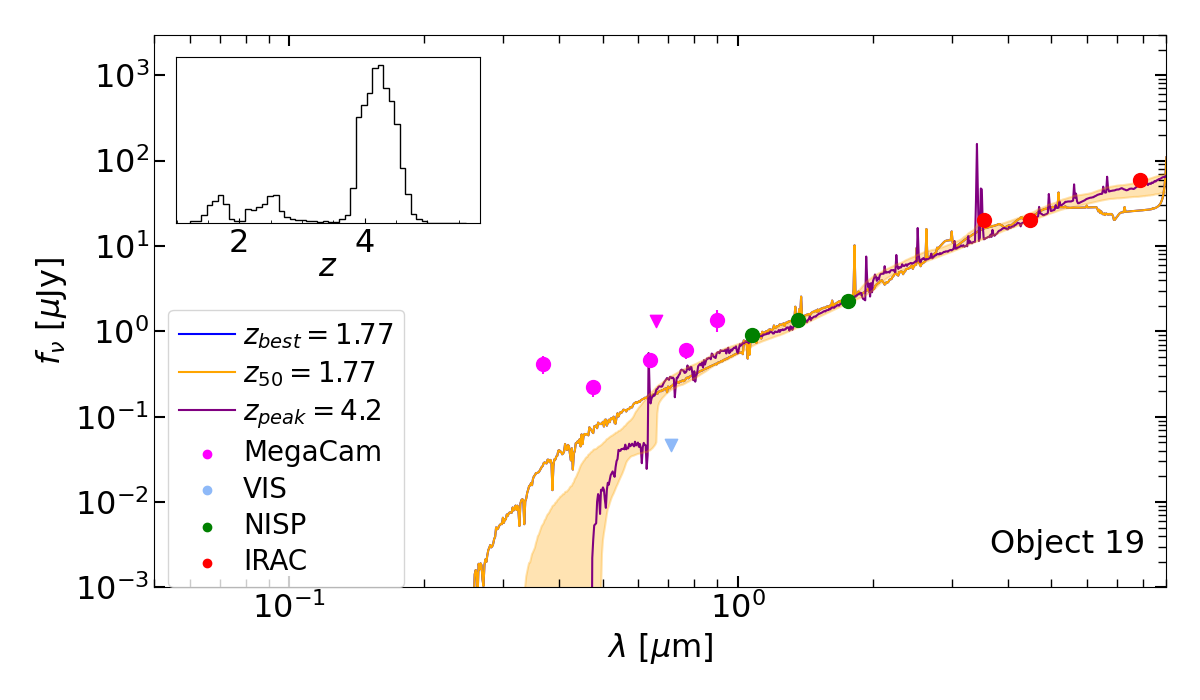}
         \includegraphics[width=0.49\linewidth, trim={0 25 0 25},clip, keepaspectratio]{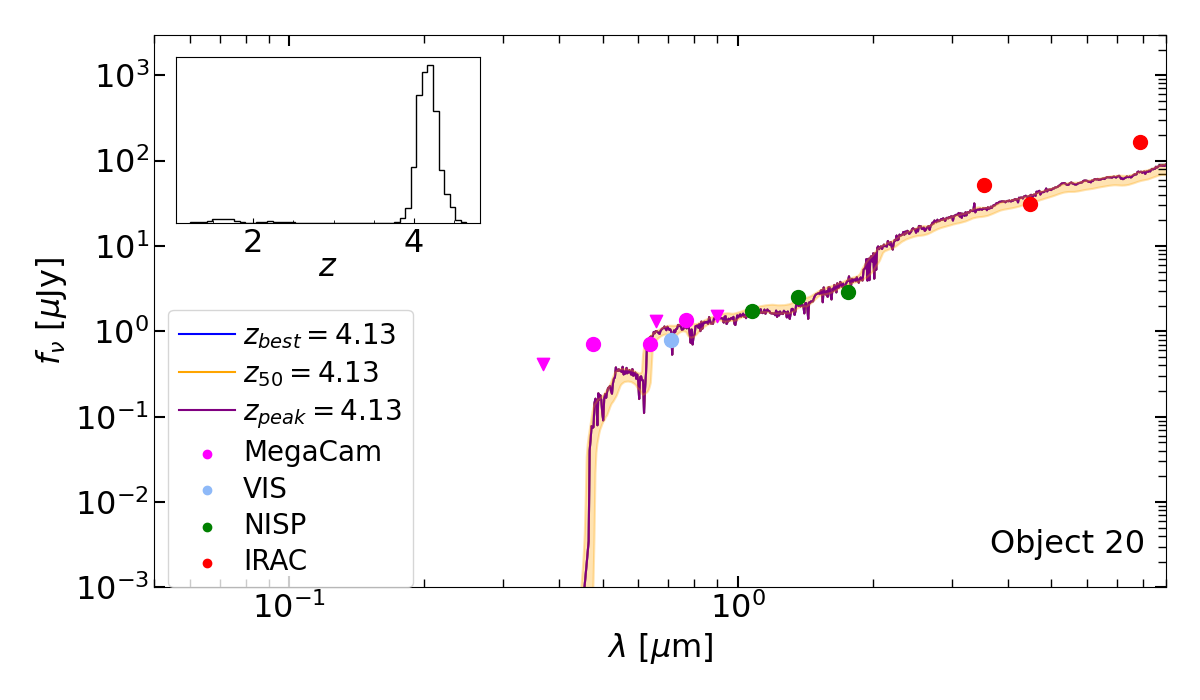}
\end{minipage}

\begin{minipage}{18cm}
        \centering
        \includegraphics[width=0.49\linewidth, trim={0 25 0 25},clip, keepaspectratio]{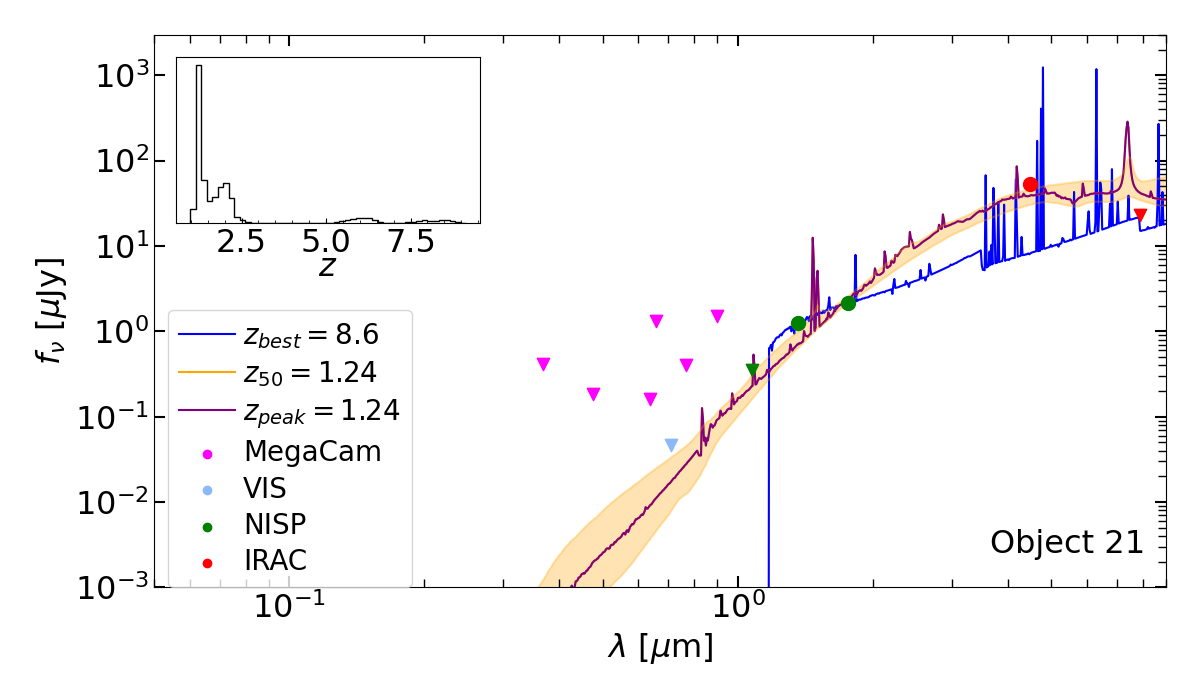}
         \includegraphics[width=0.49\linewidth, trim={0 25 0 25},clip, keepaspectratio]{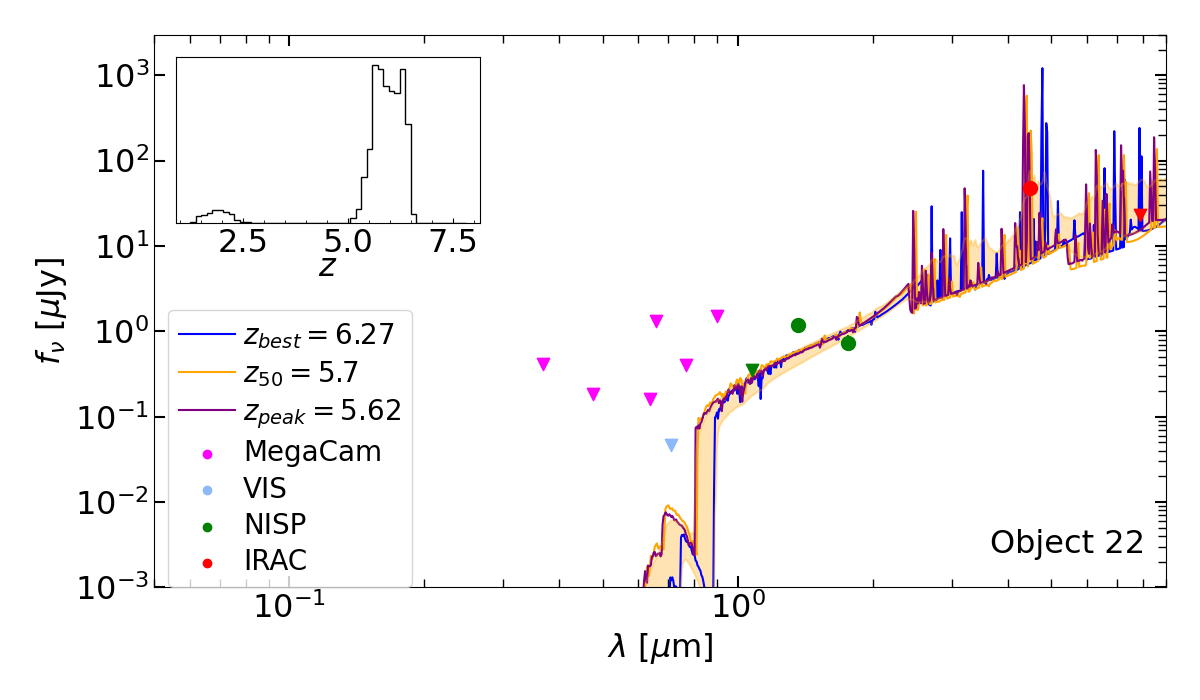}
\end{minipage}
\caption{Continued.}
\end{figure*}
\begin{figure*}
\addtocounter{figure}{-1}
     \begin{minipage}{18cm}
        \centering
        \includegraphics[width=0.49\linewidth, trim={0 25 0 25},clip, keepaspectratio]{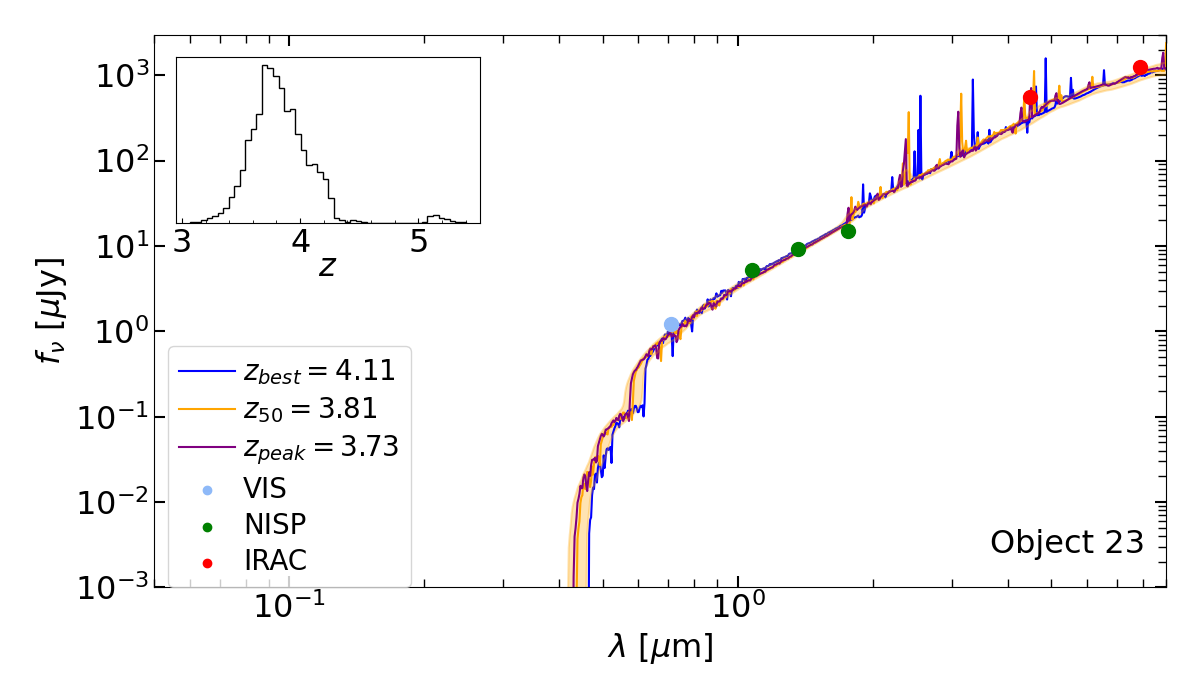}
         \includegraphics[width=0.49\linewidth, trim={0 25 0 25},clip, keepaspectratio]{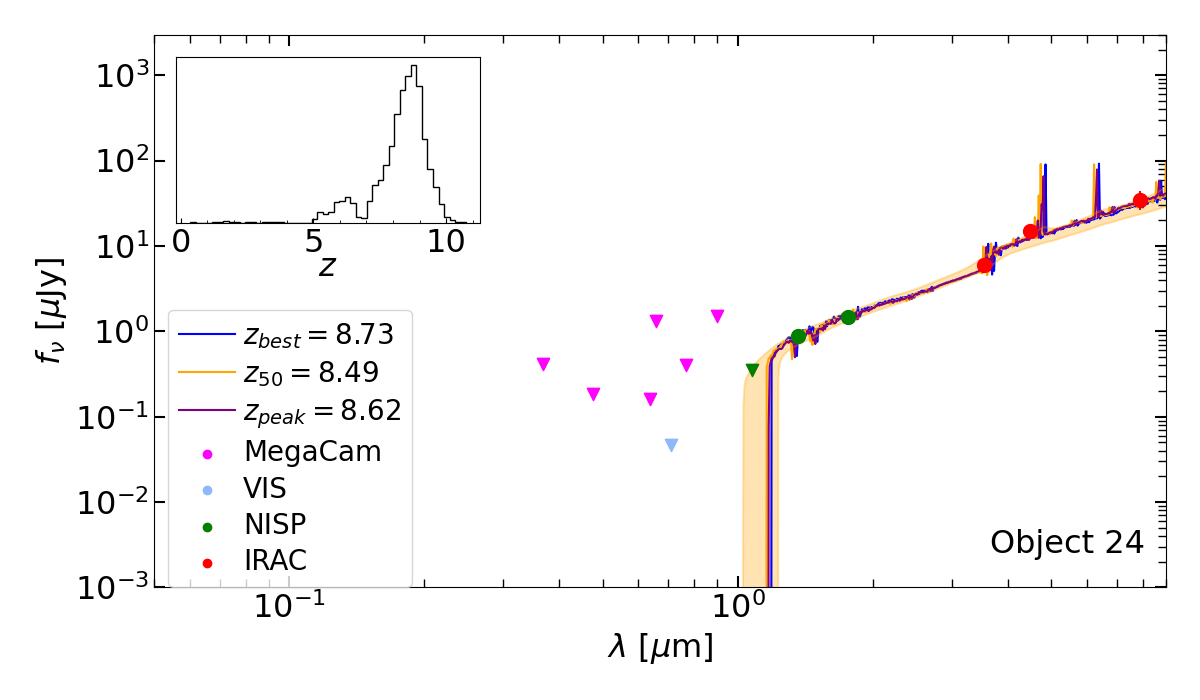}
\end{minipage}

    \vspace{0.0cm} 

     \begin{minipage}{18cm}
        \centering
       \includegraphics[width=0.49\linewidth, trim={0 25 0 25},clip, keepaspectratio]{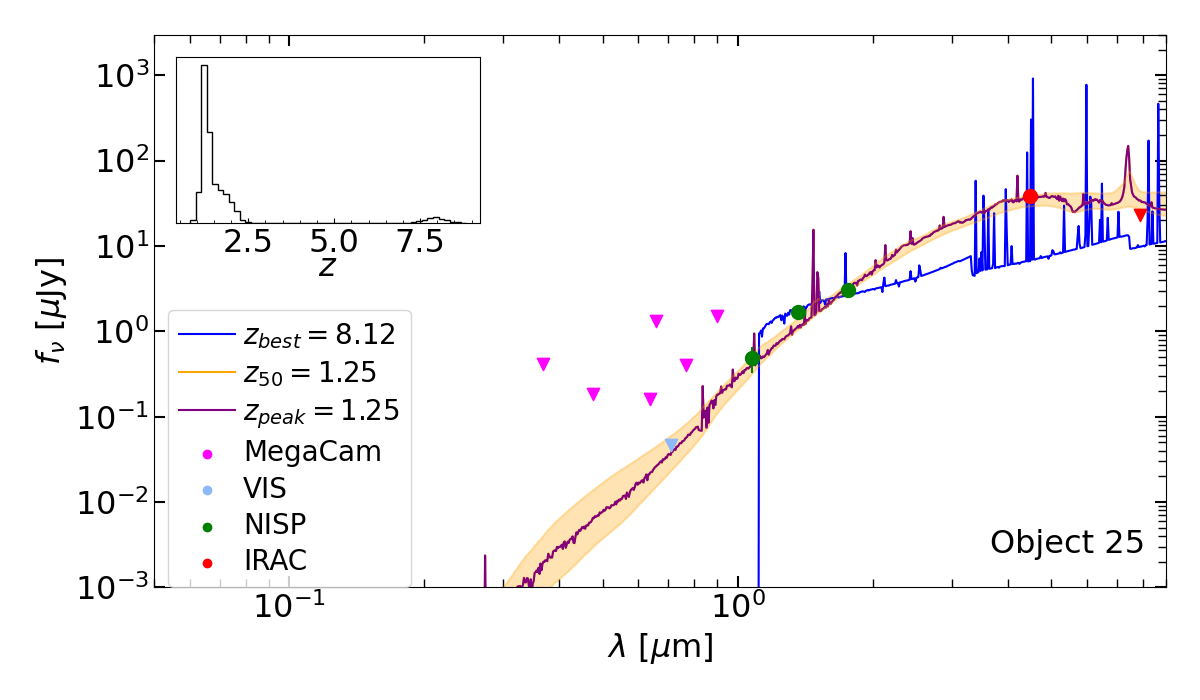}
         \includegraphics[width=0.49\linewidth, trim={0 25 0 25},clip, keepaspectratio]{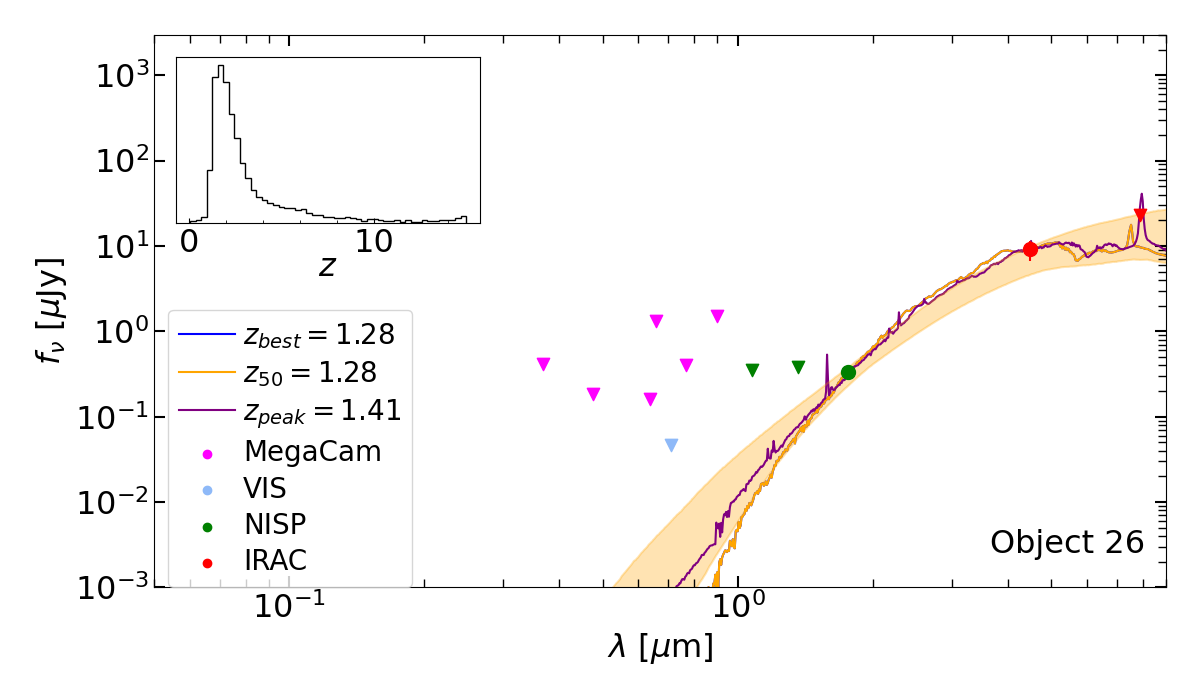}
\end{minipage}

    \vspace{0.0cm} 

     \begin{minipage}{18cm}
        \centering
        \includegraphics[width=0.49\linewidth, trim={0 25 0 25},clip, keepaspectratio]{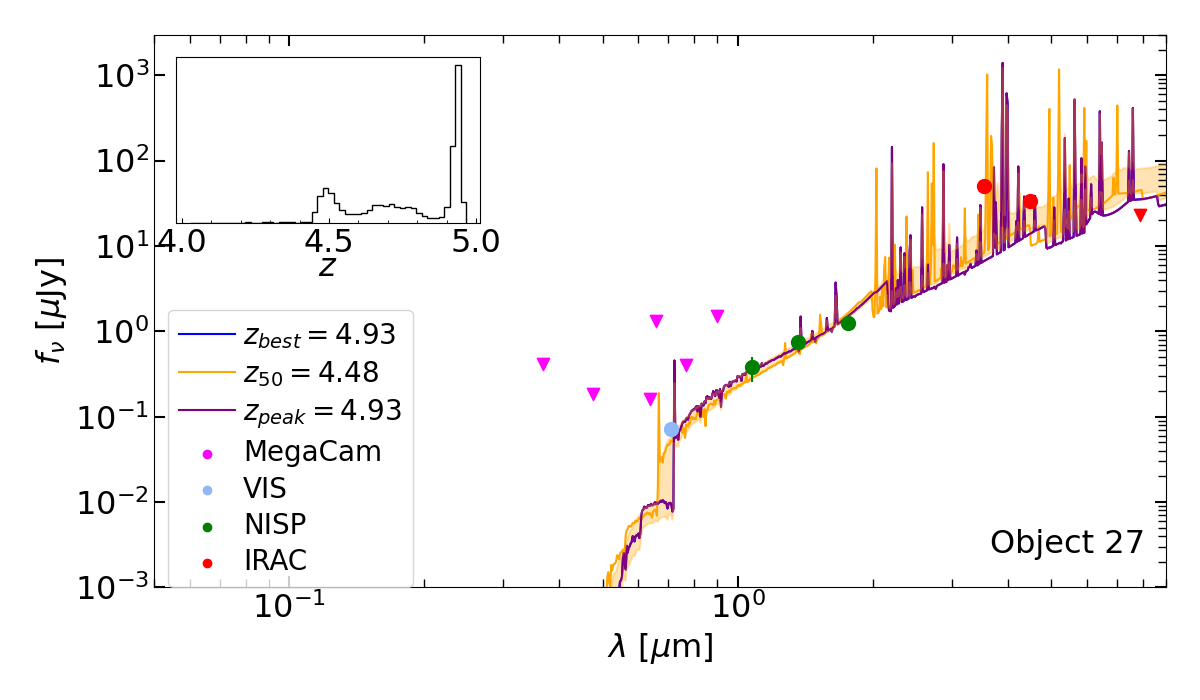}
         \includegraphics[width=0.49\linewidth, trim={0 25 0 25},clip, keepaspectratio]{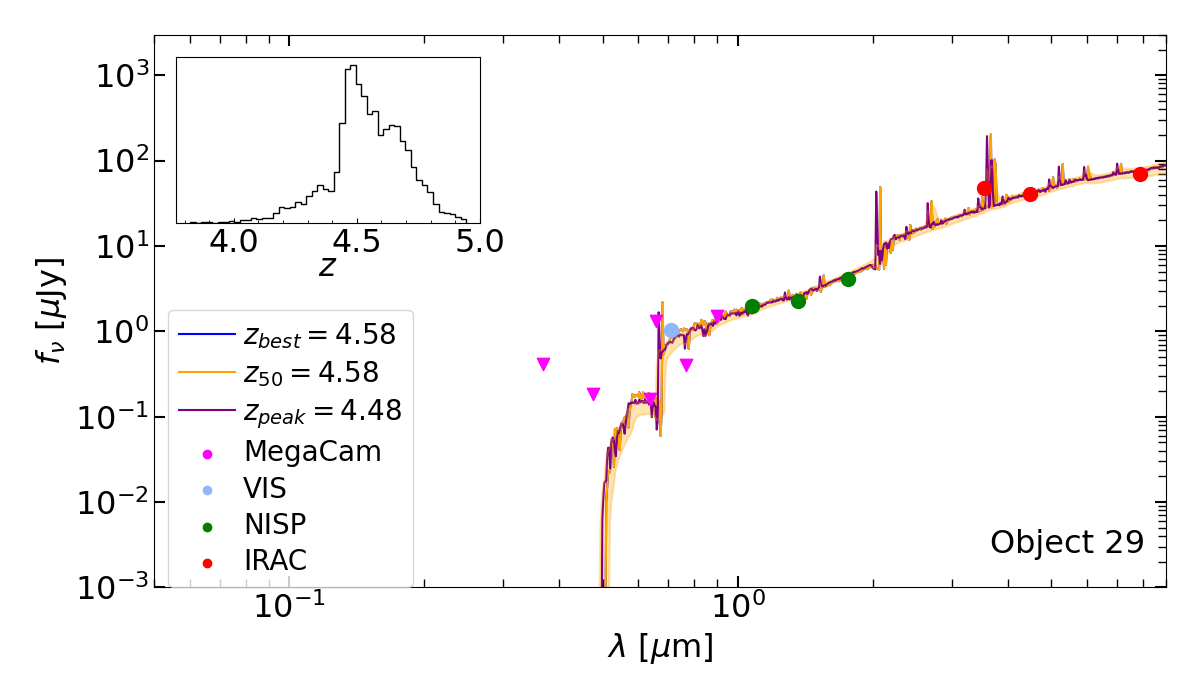}
\end{minipage}

    \vspace{0.0cm} 

     \begin{minipage}{18cm}
        \centering
        \includegraphics[width=0.49\linewidth, trim={0 25 0 25},clip, keepaspectratio]{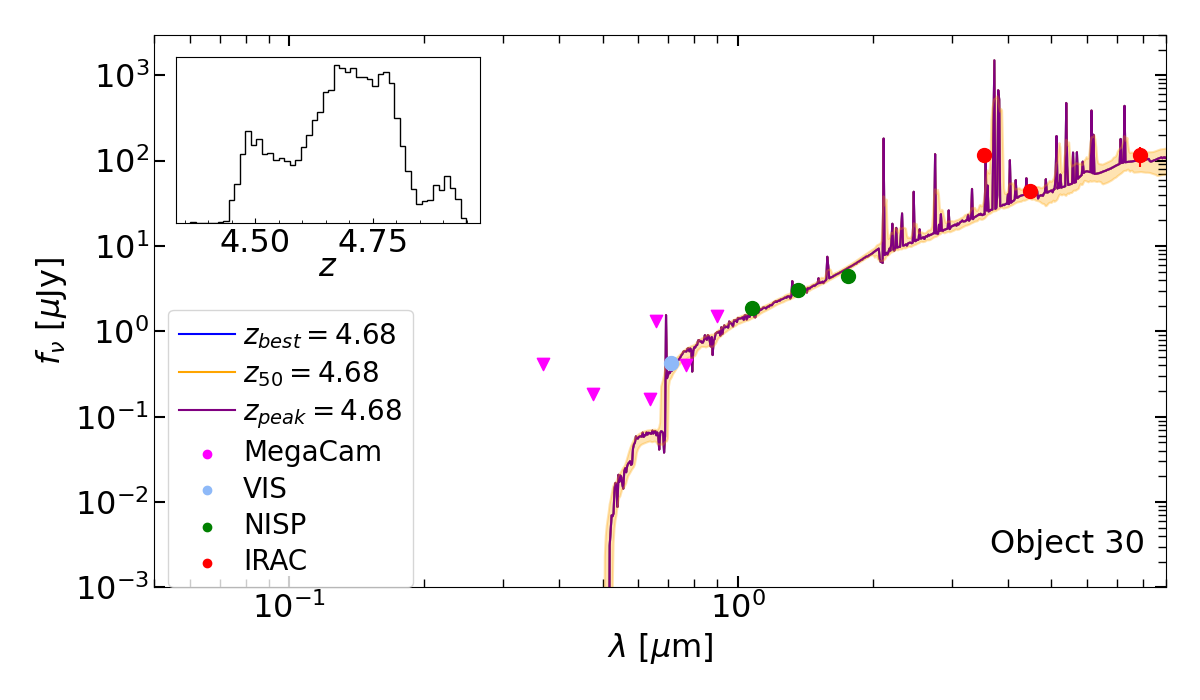}
         \includegraphics[width=0.49\linewidth, trim={0 25 0 25},clip, keepaspectratio]{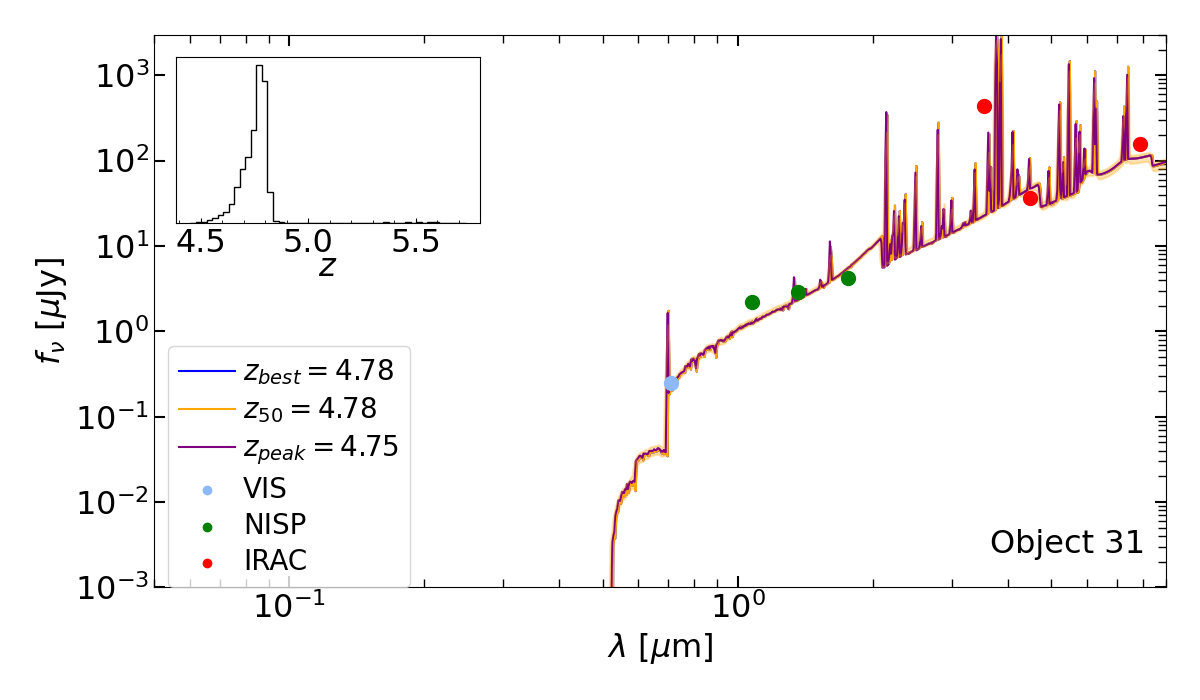}
\end{minipage}

    \vspace{0.0cm}

     \begin{minipage}{18cm}
        \centering
        \includegraphics[width=0.49\linewidth, trim={0 25 0 25},clip, keepaspectratio]{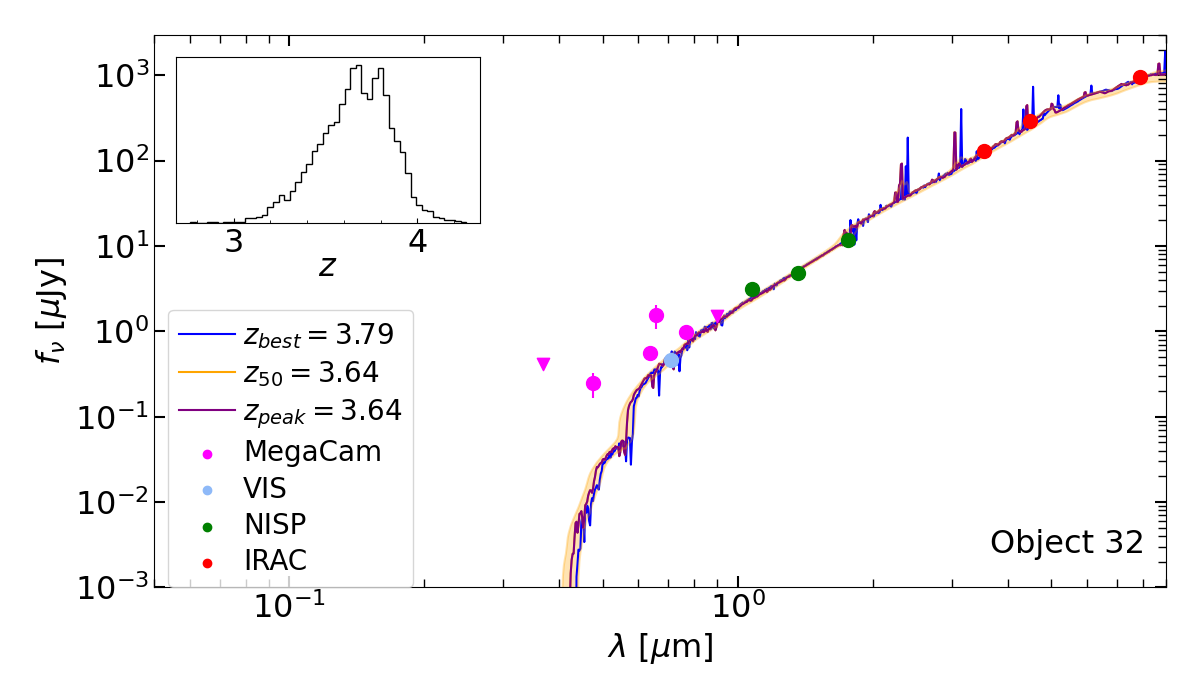}
         \includegraphics[width=0.49\linewidth, trim={0 25 0 25},clip, keepaspectratio]{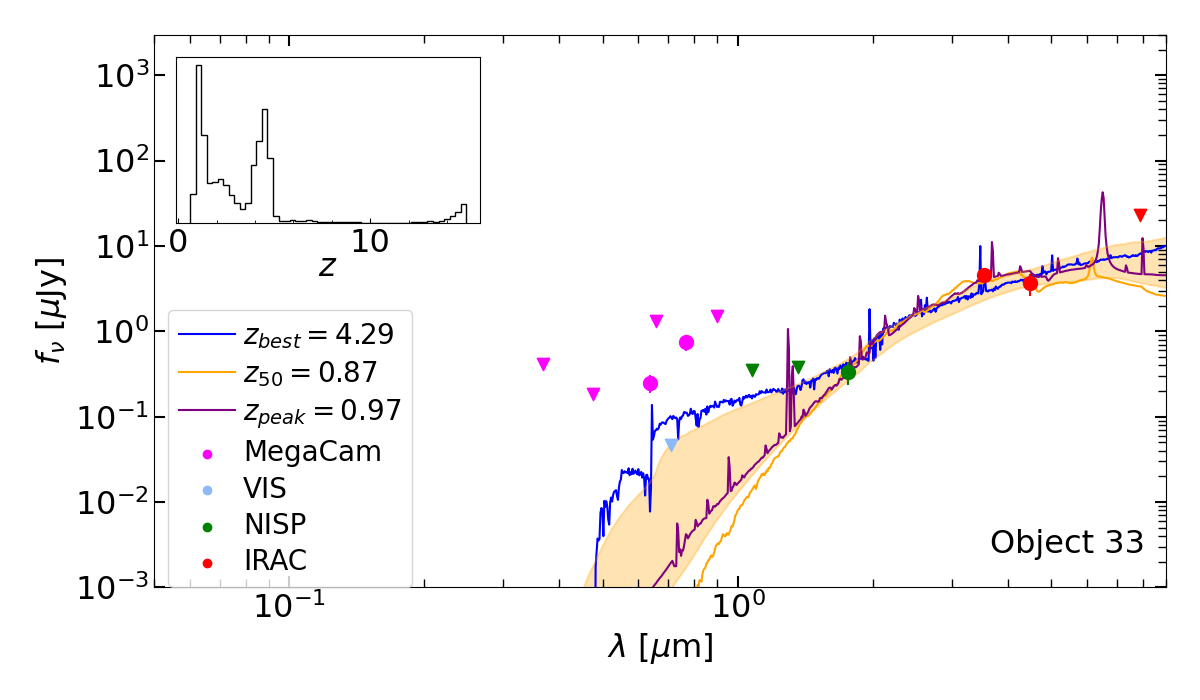}
\end{minipage}
\caption{Continued.}
\end{figure*}
\begin{figure*}
\addtocounter{figure}{-1}
     \begin{minipage}{18cm}
        \centering
        \includegraphics[width=0.49\linewidth, trim={0 25 0 25},clip, keepaspectratio]{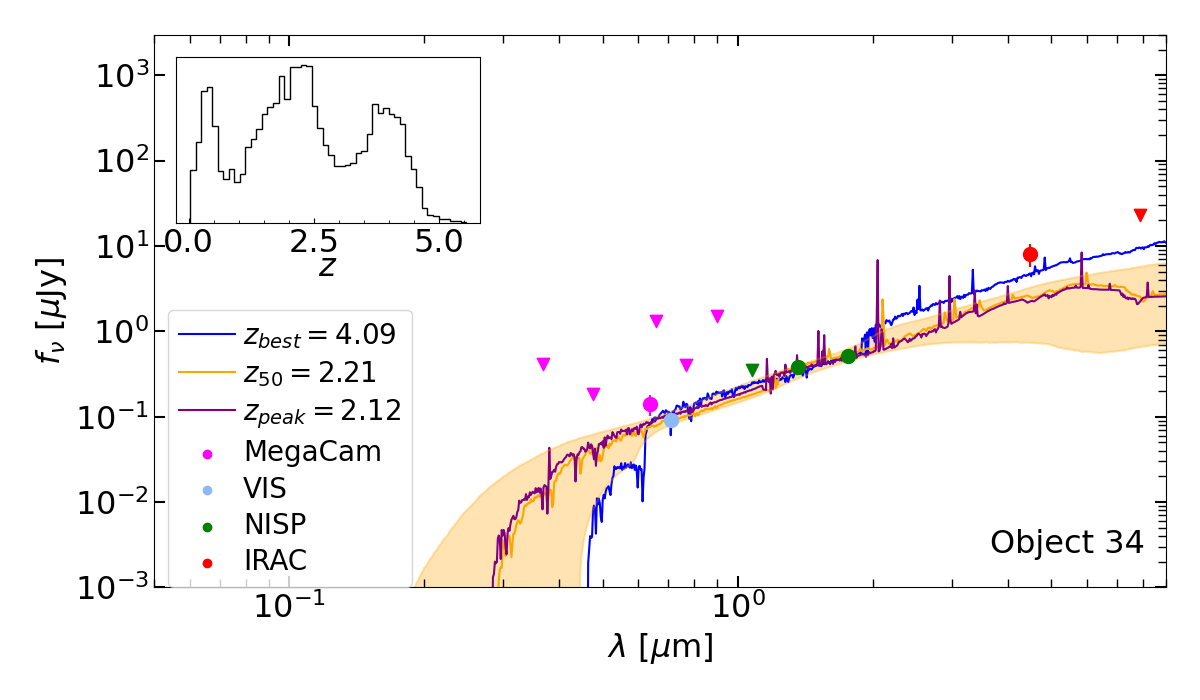}
         \includegraphics[width=0.49\linewidth, trim={0 25 0 25},clip, keepaspectratio]{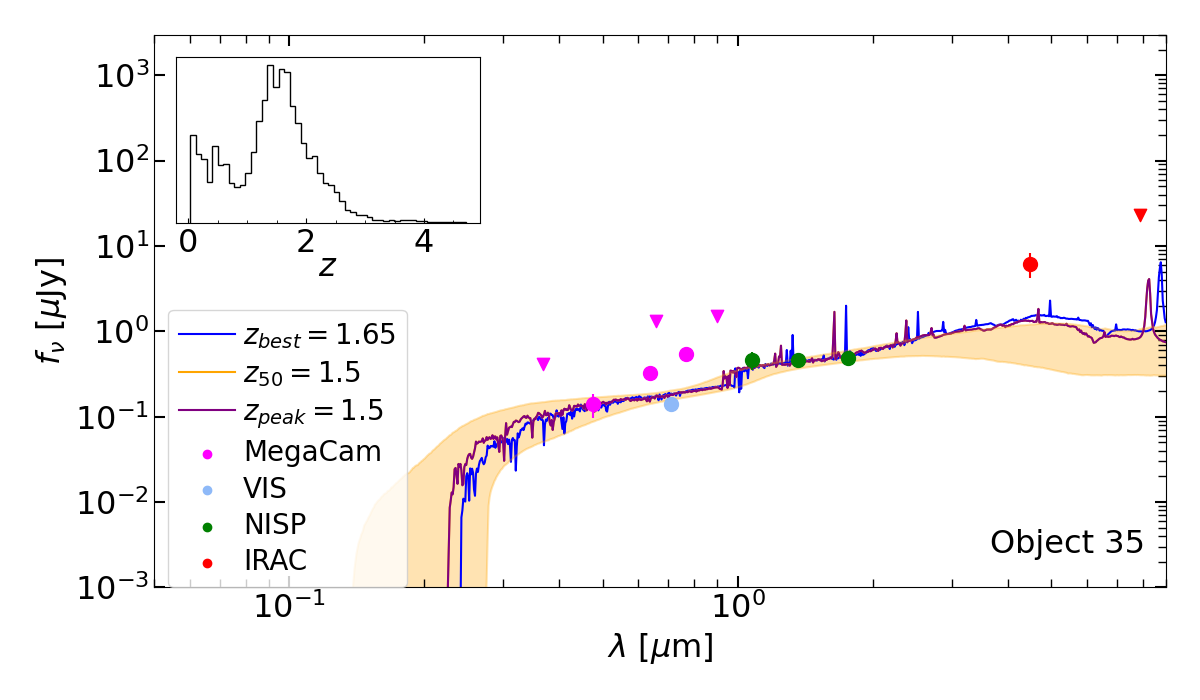}
\end{minipage}

     \begin{minipage}{18cm}
        \centering
       \includegraphics[width=0.49\linewidth, trim={0 25 0 25},clip, keepaspectratio]{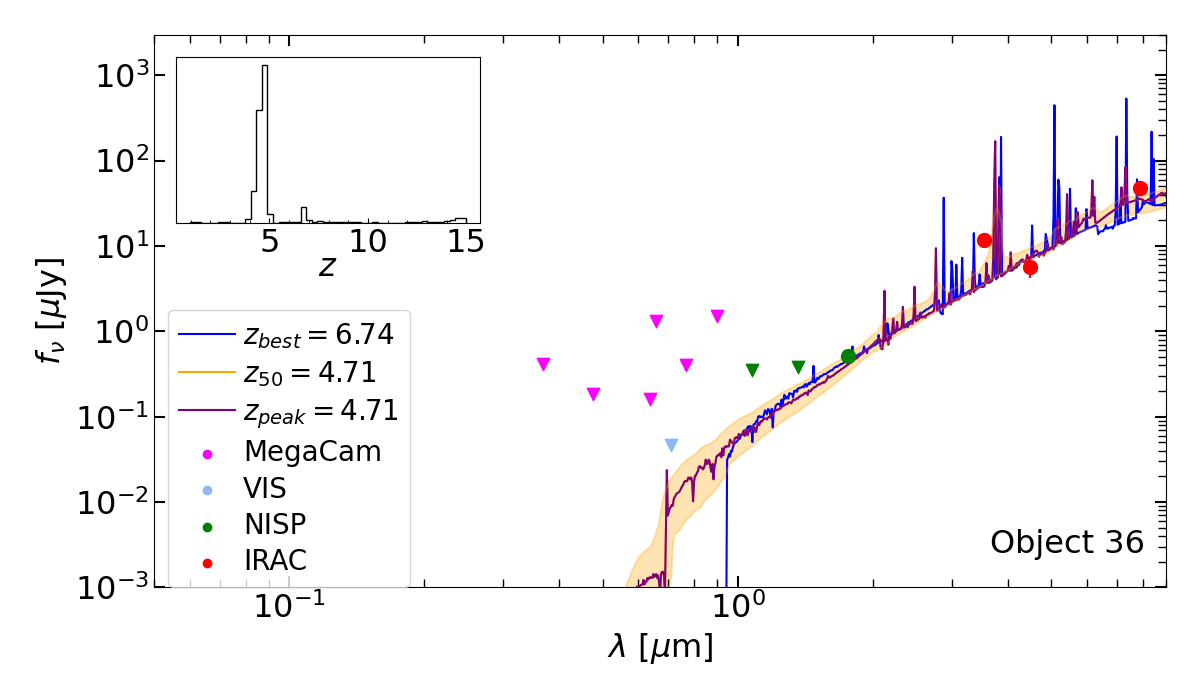}
         \includegraphics[width=0.49\linewidth, trim={0 25 0 25},clip, keepaspectratio]{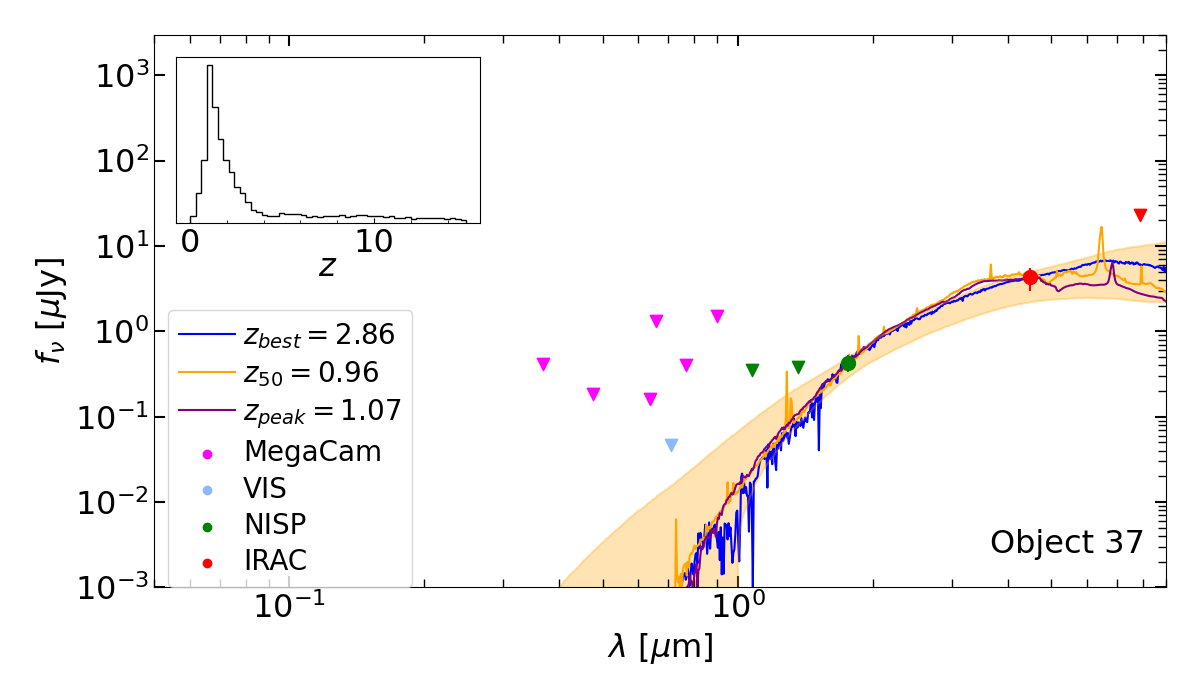}
\end{minipage}

    \vspace{0.0cm} 

     \begin{minipage}{18cm}
        \centering
        \includegraphics[width=0.49\linewidth, trim={0 25 0 25},clip, keepaspectratio]{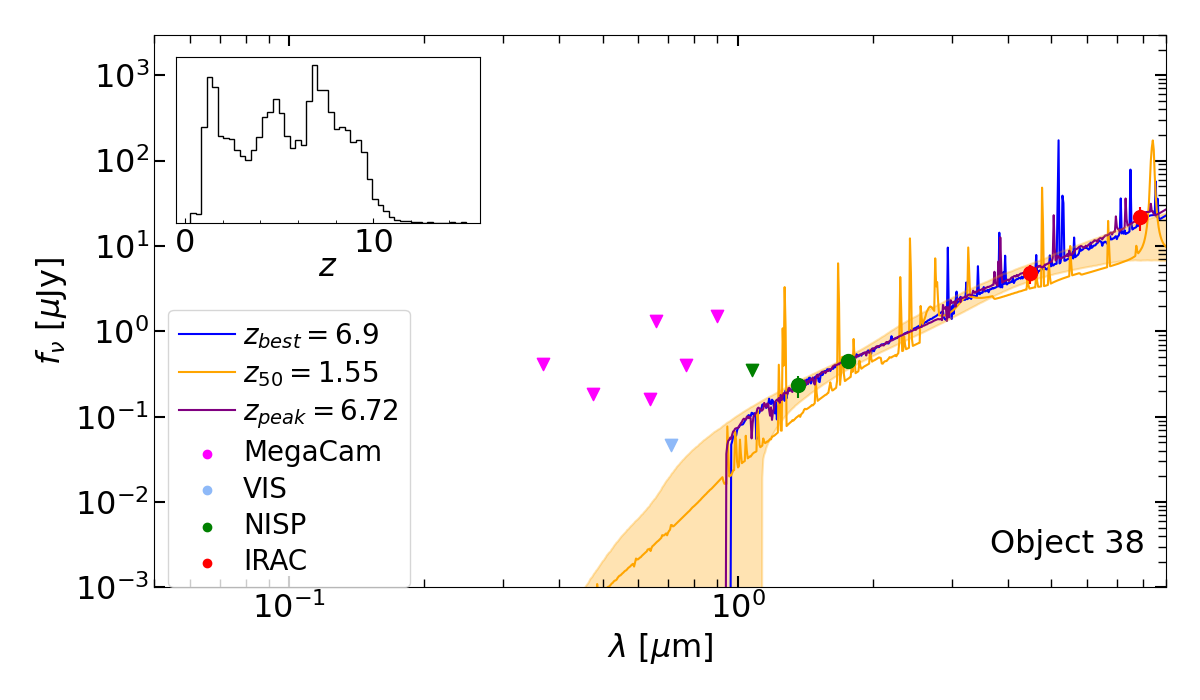}
         \includegraphics[width=0.49\linewidth, trim={0 25 0 25},clip, keepaspectratio]{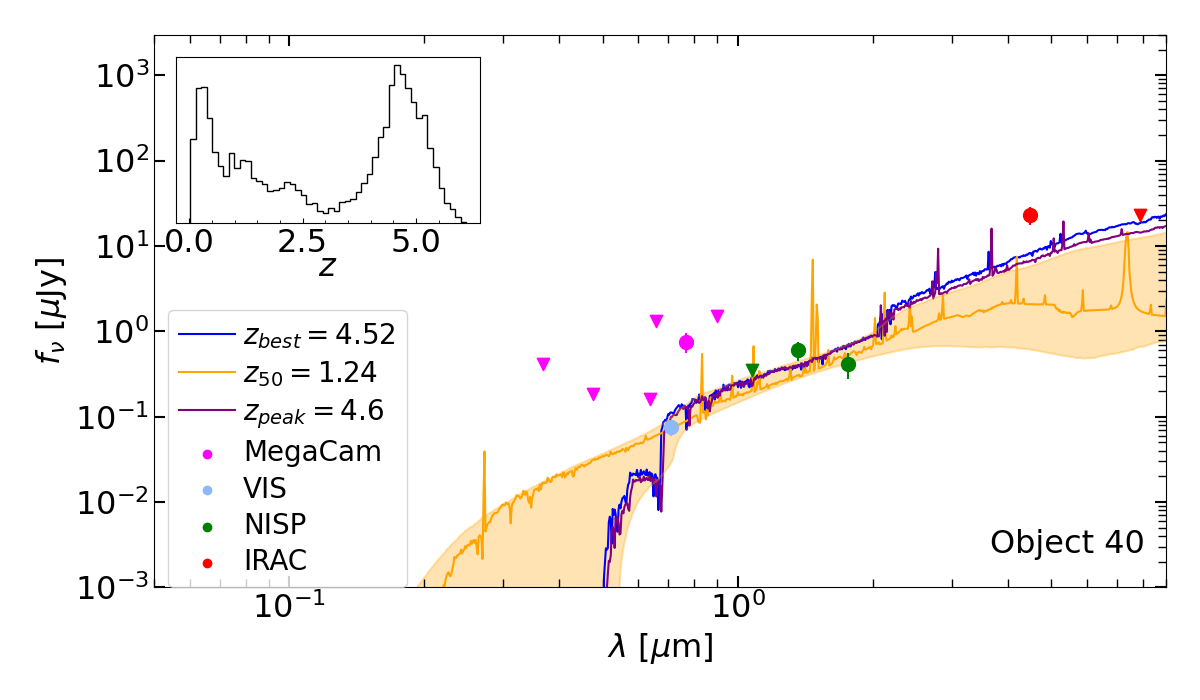}
\end{minipage}

    \vspace{0.0cm} 

     \begin{minipage}{18cm}
        \centering
        \includegraphics[width=0.49\linewidth, trim={0 25 0 25},clip, keepaspectratio]{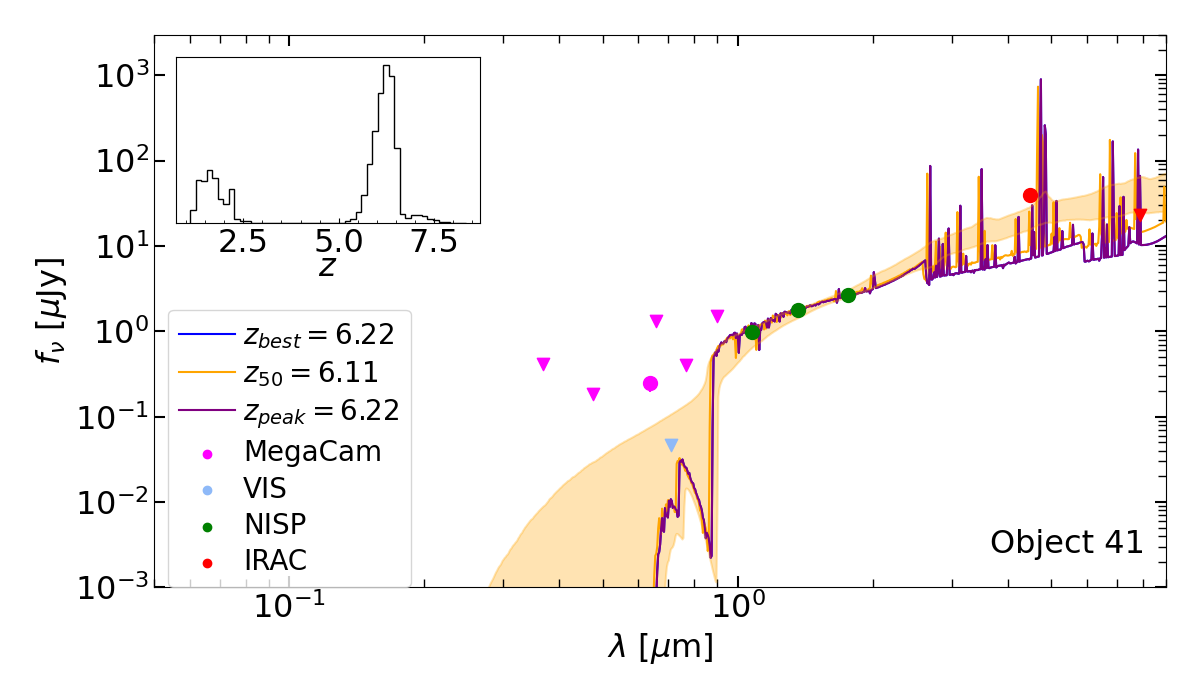}
         \includegraphics[width=0.49\linewidth, trim={0 25 0 25},clip, keepaspectratio]{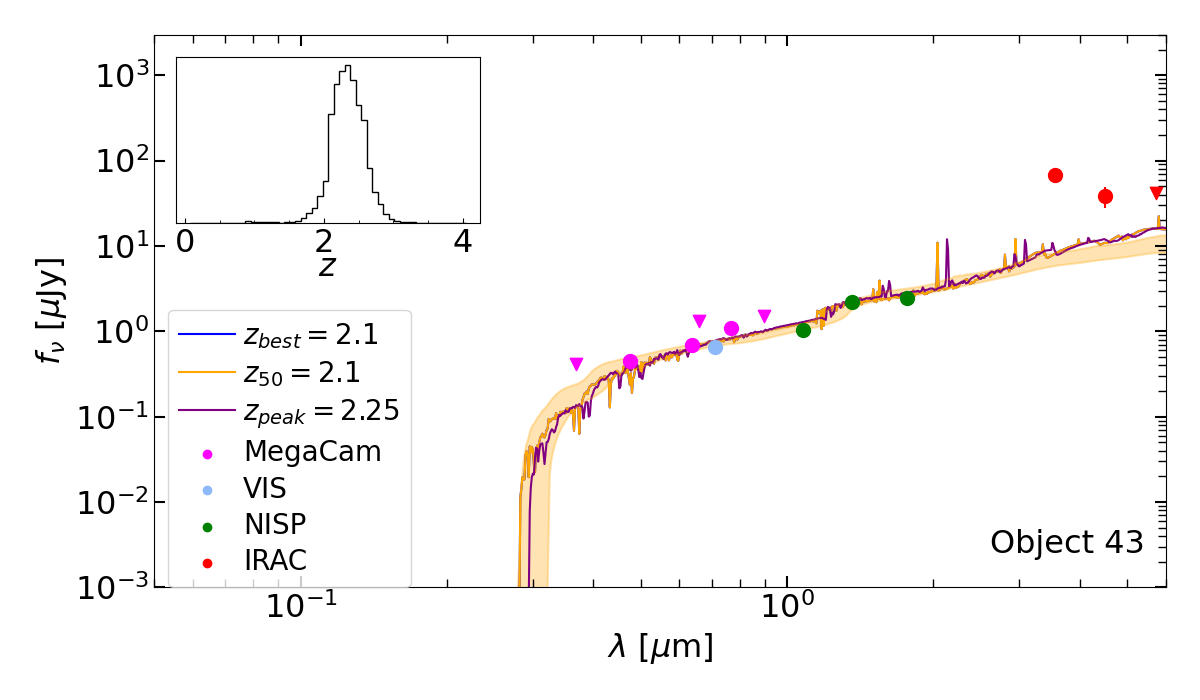}
\end{minipage}
\caption{Continued.}
\end{figure*}

\begin{table}[H]
    \centering
\caption{Main parameters output estimates retrieved by \texttt{Bagpipes} in our analyses, with the set-up presented in Table~\ref{tab:bagpipes}.}
\label{tab:bagpipes_results}
\renewcommand{\arraystretch}{1.3}
    \begin{tabular}{ccccccc} 
    \hline\hline
    ID & RA & Dec &$z$ & $\logten(M_*/M_\odot)$ & $A_V$ [mag] \\
\hline
    \noalign{\vskip 1pt}
$1$  & $\ra{3;19;50.15}$ & $\ang{+41;21;23.26}$ &  $5.11^{+0.26}_{-3.00}$  &  $11.30^{+0.26}_{-0.66}$  &  $1.94^{+0.39}_{-0.27}$ \\
$2$  & $\ra{3;19;47.59}$ & $\ang{+41;21;27.61}$ &  $8.20^{+0.84}_{-0.92}$  &  $12.15^{+0.23}_{-0.43}$  &  $2.92^{+0.37}_{-0.32}$ \\
$3$  & $\ra{3;19;46.80}$ & $\ang{+41;23;48.63}$ &  $4.70^{+0.13}_{-0.18}$  &  $11.27^{+0.09}_{-0.10}$  &  $3.69^{+0.20}_{-0.17}$ \\
$4$  & $\ra{3;19;57.77}$ & $\ang{+41;25;3.09}$ &  $1.10^{+1.30}_{-0.22}$  &  $10.29^{+0.58}_{-0.27}$  &  $4.97^{+0.90}_{-1.40}$ \\
$5$  & $\ra{3;19;48.42}$ & $\ang{+41;25;24.66}$ &  $6.50^{+0.41}_{-2.80}$  &  $12.16^{+0.15}_{-0.65}$  &  $2.22^{+0.51}_{-0.20}$ \\
$6$  & $\ra{3;19;47.53}$ & $\ang{+41;25;28.78}$ &  $6.51^{+0.09}_{-1.70}$  &  $11.55^{+0.19}_{-0.16}$  &  $2.72^{+0.41}_{-0.20}$ \\
$7$  & $\ra{3;19;53.70}$ & $\ang{+41;27;12.39}$ &  $6.61^{+0.13}_{-0.15}$  &  $11.43^{+0.36}_{-0.21}$  &  $2.36^{+0.21}_{-0.20}$ \\
$8$  & $\ra{3;19;34.83}$ & $\ang{+41;28;58.96}$ &  $1.30^{+5.30}_{-0.21}$  &  $10.94^{+1.20}_{-0.26}$  &  $3.89^{+0.75}_{-1.10}$ \\
$9$  & $\ra{3;19;44.49}$ & $\ang{+41;29;46.60}$ &  $4.10^{+0.12}_{-0.13}$  &  $12.00^{+0.07}_{-0.10}$  &  $0.95^{+0.23}_{-0.13}$ \\
$10$  & $\ra{3;19;43.22}$ & $\ang{+41;29;57.10}$ &  $4.50^{+0.16}_{-0.14}$  &  $10.09^{+0.47}_{-0.20}$  &  $1.98^{+0.17}_{-0.24}$ \\
$11$  & $\ra{3;19;45.23}$ & $\ang{+41;29;59.07}$ &  $6.02^{+0.06}_{-0.08}$  &  $12.29^{+0.08}_{-0.10}$  &  $2.78^{+0.13}_{-0.13}$ \\
$12$  & $\ra{3;19;43.43}$ & $\ang{+41;30;50.28}$ &  $4.28^{+0.02}_{-0.46}$  &  $12.26^{+0.06}_{-0.12}$  &  $2.03^{+0.18}_{-0.16}$ \\
$13$  & $\ra{3;19;38.48}$ & $\ang{+41;33;9.10}$ &  $14.60^{+0.79}_{-10.00}$  &  $12.10^{+0.37}_{-0.56}$  &  $1.24^{+1.40}_{-0.76}$ \\
$14$  & $\ra{3;20;3.65}$ & $\ang{+41;34;32.54}$ &  $3.70^{+0.55}_{-1.40}$  &  $10.57^{+0.44}_{-0.76}$  &  $1.66^{+0.34}_{-0.36}$ \\
$15$  & $\ra{3;20;0.51}$ & $\ang{+41;27;25.97}$ &  $2.11^{+3.40}_{-0.82}$  &  $10.96^{+0.75}_{-0.48}$  &  $4.39^{+1.20}_{-1.30}$ \\
$16$  & $\ra{3;20;2.04}$ & $\ang{+41;28;22.85}$ &  $0.30^{+1.20}_{-0.30}$  &  $7.42^{+0.55}_{-1.30}$  &  $0.54^{+0.34}_{-0.27}$ \\
$17$  & $\ra{3;19;22.93}$ & $\ang{+41;29;19.58}$ &  $4.50^{+0.53}_{-0.12}$  &  $12.41^{+0.12}_{-0.21}$  &  $2.49^{+0.16}_{-0.47}$ \\
$18$  & $\ra{3;19;24.28}$ & $\ang{+41;29;37.84}$ &  $1.23^{+0.15}_{-0.05}$  &  $10.49^{+0.17}_{-0.11}$  &  $5.10^{+0.14}_{-0.56}$ \\
$19$  & $\ra{3;20;3.62}$ & $\ang{+41;29;38.96}$ &  $4.20^{+0.29}_{-1.50}$  &  $11.64^{+0.15}_{-0.58}$  &  $2.18^{+0.39}_{-0.17}$ \\
$20$  & $\ra{3;19;44.45}$ & $\ang{+41;29;46.91}$ &  $4.10^{+0.12}_{-0.12}$  &  $11.95^{+0.06}_{-0.08}$  &  $0.83^{+0.23}_{-0.19}$ \\
$21$  & $\ra{3;19;24.07}$ & $\ang{+41;29;52.41}$ &  $1.21^{+0.75}_{-0.22}$  &  $10.77^{+0.39}_{-0.35}$  &  $5.29^{+0.90}_{-0.93}$ \\
$22$  & $\ra{3;19;23.54}$ & $\ang{+41;29;52.20}$ &  $5.61^{+0.40}_{-0.36}$  &  $11.01^{+0.27}_{-0.18}$  &  $2.97^{+0.31}_{-0.26}$ \\
$23$  & $\ra{3;19;19.17}$ & $\ang{+41;29;50.48}$ &  $3.70^{+0.25}_{-0.20}$  &  $12.35^{+0.04}_{-0.08}$  &  $3.38^{+0.13}_{-0.13}$ \\
$24$  & $\ra{3;19;58.23}$ & $\ang{+41;30;57.44}$ &  $8.80^{+0.61}_{-1.10}$  &  $12.06^{+0.19}_{-0.32}$  &  $1.99^{+0.20}_{-0.16}$ \\
$25$  & $\ra{3;19;24.32}$ & $\ang{+41;32;8.49}$ &  $1.21^{+0.56}_{-0.14}$  &  $10.77^{+0.30}_{-0.23}$  &  $4.36^{+0.58}_{-0.68}$ \\
$26$  & $\ra{3;19;15.46}$ & $\ang{+41;32;54.26}$ &  $1.40^{+2.60}_{-0.78}$  &  $10.31^{+0.64}_{-0.55}$  &  $5.41^{+1.10}_{-1.40}$ \\
$27$  & $\ra{3;19;46.67}$ & $\ang{+41;32;53.56}$ &  $4.93^{+0.15}_{-0.28}$  &  $11.16^{+0.26}_{-0.13}$  &  $3.26^{+0.18}_{-0.18}$ \\
$28$  & $\ra{3;20;10.19}$ & $\ang{+41;32;51.22}$ &  $1.70^{+2.20}_{-1.70}$  &  $11.12^{+0.35}_{-0.99}$  &  $5.06^{+2.20}_{-0.52}$ \\
$29$  & $\ra{3;19;45.61}$ & $\ang{+41;35;41.00}$ &  $4.50^{+0.15}_{-0.12}$  &  $11.99^{+0.10}_{-0.11}$  &  $1.96^{+0.18}_{-0.15}$ \\
$30$  & $\ra{3;19;37.44}$ & $\ang{+41;36;4.92}$ &  $4.70^{+0.10}_{-0.14}$  &  $11.43^{+0.13}_{-0.13}$  &  $2.72^{+0.11}_{-0.12}$ \\
$31$  & $\ra{3;19;59.41}$ & $\ang{+41;37;30.58}$ &  $4.71^{+0.04}_{-0.07}$  &  $11.47^{+0.05}_{-0.06}$  &  $3.06^{+0.08}_{-0.09}$ \\
$32$  & $\ra{3;19;37.48}$ & $\ang{+41;38;44.40}$ &  $3.61^{+0.18}_{-0.21}$  &  $12.32^{+0.05}_{-0.11}$  &  $3.56^{+0.10}_{-0.10}$ \\
$33$  & $\ra{3;19;45.58}$ & $\ang{+41;38;52.15}$ &  $1.00^{+1.80}_{-1.00}$  &  $9.84^{+0.47}_{-0.71}$  &  $5.47^{+2.20}_{-1.00}$ \\
$34$  & $\ra{3;20;2.27}$ & $\ang{+41;39;2.93}$ &  $2.30^{+1.60}_{-1.40}$  &  $9.83^{+0.81}_{-0.99}$  &  $1.42^{+0.37}_{-0.41}$ \\
$35$  & $\ra{3;20;2.85}$ & $\ang{+41;39;23.56}$ &  $1.50^{+0.59}_{-0.94}$  &  $9.14^{+0.37}_{-0.89}$  &  $0.52^{+0.37}_{-0.32}$ \\
$36$  & $\ra{3;19;23.56}$ & $\ang{+41;39;51.67}$ &  $4.60^{+0.23}_{-0.27}$  &  $11.14^{+0.40}_{-0.24}$  &  $3.48^{+0.41}_{-0.46}$ \\
$37$  & $\ra{3;19;7.79}$ & $\ang{+41;40;5.80}$ &  $1.10^{+5.40}_{-0.80}$  &  $9.83^{+1.00}_{-0.57}$  &  $4.85^{+1.70}_{-1.30}$ \\
$38$  & $\ra{3;19;36.73}$ & $\ang{+41;41;20.30}$ &  $6.80^{+2.80}_{-3.40}$  &  $11.62^{+0.50}_{-0.90}$  &  $2.45^{+0.93}_{-0.34}$ \\
$40$  & $\ra{3;19;54.63}$ & $\ang{+41;43;38.82}$ &  $4.60^{+1.10}_{-3.30}$  &  $11.08^{+0.86}_{-1.80}$  &  $1.91^{+0.54}_{-0.55}$ \\
$41$  & $\ra{3;19;55.10}$ & $\ang{+41;44;12.25}$ &  $6.20^{+0.34}_{-4.40}$  &  $11.79^{+0.41}_{-0.84}$  &  $2.03^{+0.85}_{-0.23}$ \\
$42$  & $\ra{3;19;52.92}$ & $\ang{+41;26;25.24}$ &  $0.12^{+0.26}_{-0.12}$  &  $6.80^{+0.40}_{-0.82}$  &  $0.59^{+0.39}_{-0.23}$ \\
$43$  & $\ra{3;19;29.95}$ & $\ang{+41;31;57.77}$ &  $2.30^{+0.23}_{-0.21}$  &  $10.44^{+0.20}_{-0.24}$  &  $1.15^{+0.31}_{-0.21}$ \\
 \hline
    \end{tabular}
    \tablefoot{The associated errors represent the $\pm 1\,\sigma$ confidence intervals. The coordinates of each object are also listed.}
\end{table}

\end{appendix}
    
    \label{LastPage}
    
\end{document}